\documentclass[aps,prl,reprint,twocolum,amsmath,superscriptaddress,amssymb,showpacs,longbibliography]{revtex4-1}
\usepackage{amsmath,braket,amssymb}
\usepackage[final]{graphicx}
\usepackage{subfigure}
\usepackage{xcolor}
\usepackage[english]{babel}
\usepackage[utf8]{inputenc}
\usepackage{color}
\usepackage{mathtools}
\usepackage{empheq}
\usepackage{tensor}
\usepackage{titlesec} 

\usepackage{hhline}

\usepackage[colorlinks,citecolor=darkBlue,linkcolor=darkBlue,
urlcolor=blue,hyperindex]{hyperref}

\usepackage[normalem]{ulem}
\normalfont

\usepackage[linesnumbered, ruled,vlined]{algorithm2e}

\graphicspath{{./images/},{./imagesAppendix/}}

\definecolor{forestgreen}{rgb}{0.08, 0.4, 0.13}
\definecolor{darkBlue}{rgb}{0.08, 0.13, 0.4}

\definecolor{THc}{rgb}{0.9,0.3,0.2}

\newcommand\numberthis{\addtocounter{equation}{1}\tag{\theequation}}

\usepackage{amsthm}
\theoremstyle{definition}

\theoremstyle{plain}

\newcommand{\idg}[1]{{\bfseries #1)}}

\newcommand{\prlsection}[1]{{\em {#1}.---~}}

\renewcommand{\eqref}[1]{Eq.~(\ref{#1})} 

\newcommand{\subfigimg}[3][,]{%
	\setbox1=\hbox{\includegraphics[#1]{#3}}
	\leavevmode\rlap{\usebox1}
	\rlap{\hspace*{2pt}\raisebox{\dimexpr\ht1-0.5\baselineskip}{{\bfseries \large\textsf{#2}}}}
	\phantom{\usebox1}
}

\newcommand{\revA}[1]{{#1}}

\newcommand{\revB}[1]{{#1}}

\newcommand{\revC}[1]{#1}
\begin{document}

\title{Efficient quantum algorithms for stabilizer entropies}

\author{Tobias Haug}
\email{tobias.haug@u.nus.edu}
\affiliation{Quantum Research Center, Technology Innovation Institute, Abu Dhabi, UAE}
\affiliation{Blackett Laboratory, Imperial College London SW7 2AZ, UK}

\author{Soovin Lee}
\email{soovinlee310@gmail.com}
\affiliation{Blackett Laboratory, Imperial College London SW7 2AZ, UK}

\author{M. S. Kim}
\affiliation{Blackett Laboratory, Imperial College London SW7 2AZ, UK}

\date{\today}
	
\begin{abstract}
Stabilizer entropies (SEs) are measures of nonstabilizerness or `magic' that quantify the degree to which a state is described by stabilizers. SEs are especially interesting due to their connections to  scrambling, localization and property testing.
However, applications have been limited so far as previously known measurement protocols for SEs scale exponentially with the number of qubits.
Here, we efficiently measure SEs for integer R\'enyi index $n>1$  via Bell measurements. The SE of $N$-qubit quantum states can be measured with $O(n)$ copies and $O(nN)$ classical computational time, where for even $n$ we additionally require the complex conjugate of the state.
We provide efficient bounds of various nonstabilizerness monotones which are intractable to compute beyond a few qubits. Using the IonQ quantum computer, we measure SEs of random Clifford circuits doped with non-Clifford gates and give bounds for the stabilizer fidelity, stabilizer extent and robustness of magic. We provide efficient algorithms to measure Clifford-averaged $4n$-point out-of-time-order correlators and multifractal flatness. With these measures we study the scrambling time of doped Clifford circuits and random Hamiltonian evolution depending on nonstabilizerness. Counter-intuitively, random Hamiltonian evolution becomes less scrambled at long times which we reveal with the multifractal flatness.
Our results open up the exploration of nonstabilizerness with quantum computers.
\end{abstract}
	
\maketitle


Stabilizer states and Clifford operations are essential to quantum information and quantum computing~\cite{gottesman1997stabilizer,shor1996fault,kitaev2003fault}. They are the cornerstone to run quantum algorithms on most fault-tolerant quantum computers, where Clifford operations are intertwined with non-Clifford gates~\cite{bravyi2005universal,campbell2017roads}.
To characterize the amount of non-Clifford resources needed to realize quantum states and operations the resource theory of nonstabilizerness has been put forward~\cite{campbell2011catalysis,veitch2014resource,howard2017application,wang2019quantifying,beverland2020lower,jiang2021lower, liu2022many,bu2022complexity,haug2022scalable}. 
Stabilizer entropies (SEs)~\cite{leone2022stabilizer}  are measures of nonstabilizerness with efficient algorithms for matrix product states~\cite{haug2022quantifying,haug2023stabilizer,lami2023quantum} which have enabled the study of nonstabilizerness in many-body systems~\cite{oliviero2022magic,odavic2022complexity,chen2022magic,haug2023stabilizer,haug2022quantifying,lami2023quantum,goto2022probing,piemontese2023entanglement,chen2023magic}. 

Recently, SEs have also been related to various important properties of quantum systems. SEs probe error-correction~\cite{niroula2023phase} and measurement-induced phase transitions~\cite{bejan2023dynamical,fux2023entanglementmagic}, as well as relate to the entanglement spectrum~\cite{tirrito2023quantifying} and property testing~\cite{leone2023phase,leone2023nonstabilizerness}.
SEs are also connected to the participation entropy~\cite{turkeshi2023measuring}, which are helpful to understand Anderson~\cite{castellani1986multifractal} and many-body localization~\cite{stephan2009shannon}. 
Further, recent works established a fruitful connection between out-of-time-order correlators (OTOCs) and nonstabilizerness~\cite{leone2021quantum,garcia2022resource,leone2023nonstabilizerness}. 
OTOCs describe scrambling in quantum systems~\cite{xu2022scrambling,dowling2023scrambling}. However, OTOCs are challenging to measure directly and often require an inverse of the time evolution~\cite{li2017measuring}. Higher-order OTOCs and nonstabilizerness have been related to quantum chaos~\cite{leone2021quantum} and state certification~\cite{leone2023nonstabilizerness}.

The aforementioned properties make SEs highly interesting for experimental studies of quantum computers and simulators. However, the progress has so far been limited as all previously known measurement protocols for SEs scale exponentially with the number of qubits~\cite{oliviero2022measuring,haug2022scalable}.

Here, we efficiently measure SEs with integer index $n>1$ on quantum computers and simulators via Bell measurements on two copies of $N$-qubit quantum states. \revC{Our algorithms are practical to implement with $O(n)$ copies and $O(nN)$ classical post-processing time, where even $n$ requires also access to the complex conjugate of the state.}
We devise an efficient protocol to measure Clifford-averaged  multifractal flatness and $4n$-point OTOCs where for odd $n$ we do not require an inverse time evolution. We study the interplay of nonstabilizerness and scrambling and show that the number of Clifford gates needed for OTOCs to converge depends on the number of T-gates. Further, we use the multifractal flatness to show that random Hamiltonian evolution stops being random for long evolution times.
We also provide efficiently computable bounds to other nonstabilizerness monotones, which are otherwise intractable beyond a few qubits. Finally, we measure the Tsallis SE on the IonQ quantum computer and demonstrate SEs as efficient bounds for the robustness of magic, stabilizer extent and stabilizer fidelity. Our work introduces methods to uncover the key features that characterize the power of quantum computers and simulators.

\prlsection{SE}
For an $N$-qubit state $\ket{\psi}$, the R\'enyi-$n$ SE is given by~\cite{leone2022stabilizer}
\begin{align*}
M_n(\ket{\psi})=&(1-n)^{-1} \ln (\sum_{\sigma \in \mathcal{P}} 2^{-N}\braket{\psi|\sigma|\psi}^{2n})\,.\numberthis\label{eq:SRE}
\end{align*}
where $n$ is the index of the SE and $\mathcal{P}$ is the set of $4^N$ Pauli strings. The Pauli strings are $N$-qubit tensor products $\sigma_{\boldsymbol{r}}=\bigotimes_{j=1}^N \sigma_{\boldsymbol{r}_{2j-1}\boldsymbol{r}_{2j}}$ with $\boldsymbol{r}\in\{0,1\}^{2N}$ where $\sigma_{00}=I_{1}$, $\sigma_{01}=\sigma^x$, $\sigma_{10}=\sigma^z$ and $\sigma_{11}=\sigma^y$ with $\ell$-qubit identity matrix $I_\ell$ and Pauli matrices $\sigma^{k}$, $k\in\{x,y,z\}$.
$M_n$ is a faithful measure of nonstabilizerness for pure states, i.e. $M_n(\ket{\psi_\text{STAB}})=0$ only for pure stabilizer states $\ket{\psi_\text{STAB}}$, and greater zero else~\cite{leone2022stabilizer}. Further, SEs are invariant under Clifford unitaries $U_\text{C}$ with \revA{$M_n(U_\text{C}\ket{\psi})=M_n(\ket{\psi})$, where Clifford unitaries map any Pauli string $\sigma$ to another Pauli string $\sigma'$ via $U_\text{C}\sigma U_\text{C}^\dagger=\sigma'$}. Further, $M_n$ is additive with $M_n(\ket{\psi}\otimes \ket{\phi})=M_n(\ket{\psi})+M_n(\ket{\phi})$. $M_n$ for $n<2$ is not a monotone under channels that can map a pure state to another pure state, while the case $n\ge2$ remains an open problem.

\revA{Evaluating~\eqref{eq:SRE} requires an efficient measurement protocol for the $n$-th moment of the Pauli spectrum
\begin{equation}\label{eq:A}
A_n(\ket{\psi})=2^{-N}\sum_{\sigma \in \mathcal{P}}\braket{\psi|\sigma|\psi}^{2n}\,,
\end{equation}
which on first glance appears challenging due to the summation over exponentially many Pauli strings.}

\prlsection{Algorithms}
We now provide two algorithms to efficiently measure $A_n(\ket{\psi})$. 
First, we introduce Algorithm~\ref{alg:SEmeas} which is efficient for odd $n>1$. We write $A_n$ as the expectation value of an observable $\Gamma_{n}^{\otimes N}$ acting on $2n$ copies of $\ket{\psi}$ via the replica trick~\cite{haug2022quantifying}
\begin{align}\label{eq:contraction2}
	 A_n=2^{-N}\sum_{\sigma \in \mathcal{P}} \braket{\psi|\sigma|\psi}^{2n}=\bra{\psi}^{\otimes 2n}\Gamma_{n}^{\otimes N} \ket{\psi}^{\otimes 2n}\,,
\end{align}
where $\Gamma_{n}=\frac{1}{2}\sum_{k=0}^3 (\sigma^k)^{\otimes 2n}$. For  even $n>1$,  $2^{-N}\Gamma_{n}^{\otimes N}$ is a projector with two possible eigenvalues $\omega\in\{0,2^N\}$ as shown in the Supplemental Material (SM)~\ref{sec:spectrum}.
In contrast, for odd $n>1$ it is unitary and hermitian with eigenvalues $\omega\in\{-1,1\}$. \revA{This fact was previously pointed out in Ref.~\cite{gross2021schur} for stabilizer testing.}

To measure the operator $\Gamma_{n}^{\otimes N}$ we transform the operator into a diagonal eigenbasis.
We first recall the Bell transformation 
acting on two qubits
$U_\text{Bell}=(H\otimes I_1) \text{CNOT}$,
where $H=\frac{1}{\sqrt{2}}(\sigma^x+\sigma^z)$ is the Hadamard gate, and $\text{CNOT}=\exp(i\frac{\pi}{4}(I_1-\sigma^z)\otimes(I_1-\sigma^x))$. 
It turns out $\Gamma_{n}$ is diagonalized  by $U_\text{Bell}$
\begin{align*}
&A_n=\bra{\psi}^{\otimes 2n}({U_\text{Bell}^{\otimes n}}^\dagger \frac{1}{2}( (I_1\otimes I_1)^{\otimes n}+(\sigma^z\otimes I_1)^{\otimes n}\numberthis\label{eq:meas}\\&
+(I_1\otimes \sigma^z)^{\otimes n}+(-1)^{n}(\sigma^z\otimes \sigma^z)^{\otimes n})U_\text{Bell}^{\otimes n})^{\otimes N}\ket{\psi}^{\otimes 2n}\,.
\end{align*}
Algorithm~\ref{alg:SEmeas} utilizes this fact to provide an unbiased estimator for $A_n$, where $\oplus$ denotes binary addition. 
While~\eqref{eq:meas} involves $2n$ copies of $\ket{\psi}$, the Bell transformation~\eqref{eq:meas} can be written as tensor products. Thus, $A_n$ is evaluated  using only Bell measurements on two copies of the quantum state, which requires only a  $2N$-qubit quantum computer. Then, via post-processing  $A_n$ is computed as the parity of odd and even qubit index Bell measurement outcomes as derived in SM~\ref{sec:parity}.

We apply Hoeffding's inequality to bound the number of copies as $C=O(n\Delta\omega_n\epsilon^{-2})$, where $\epsilon$ is the error and $\Delta\omega_n$ the range of eigenvalues of $\Gamma_n^{\otimes N}$. For odd $n>1$, we have $\Delta\omega_n=2$ and $C=O(n\epsilon^{-2})$.
For even $n$, the eigenvalue spectrum of $\Gamma_n^{\otimes N}$ diverges and we require an exponential number of measurements. In SM~\ref{sec:shift-rule} we extend our algorithm to get gradients $\partial_k A_n$ via the shift-rule for variational quantum algorithms.

\begin{algorithm}[h]
 \SetAlgoLined
 \LinesNumbered
  \SetKwInOut{Input}{Input}
  \SetKwInOut{Output}{Output}
   \Input{   Integer $n>1$; $L$ repetitions;
   }
    \Output{$A_n(\ket{\psi})$
    }

$A_n= 0$

 \SetKwRepeat{Do}{do}{while}
    \For{$k=1,\dots,L$}{

    \For{$j=1,\dots,n$}{
        Prepare $\ket{\eta}=U_\text{Bell}^{\otimes N}\ket{\psi}\otimes\ket{\psi}$
    
        Sample in computational basis $\boldsymbol{r}^{(j)}\sim \vert \braket{\boldsymbol{r}\vert\eta}\vert^2$
    }
    $b= 1$

        \For{$\ell=1,\dots,N$}{
            
            $\nu_1=\bigoplus_{j=1}^n r^{(j)}_{2\ell-1}$;             $\nu_2=\bigoplus_{j=1}^n r^{(j)}_{2\ell}$

             \uIf{$n$ $\mathrm{is\,\,odd}$}{
               $b= b\cdot (-2\nu_1 \cdot\nu_2+1)$
              }
              \Else{
                $b= b\cdot 2(\nu_1-1) \cdot(\nu_2-1)$
              }
        }
        
    $A_n= A_n+b/L$
    }

 \caption{SE without complex conjugate}
 \label{alg:SEmeas}
\end{algorithm}

\begin{figure}[htbp]
	\centering	
	\subfigimg[width=0.3\textwidth]{}{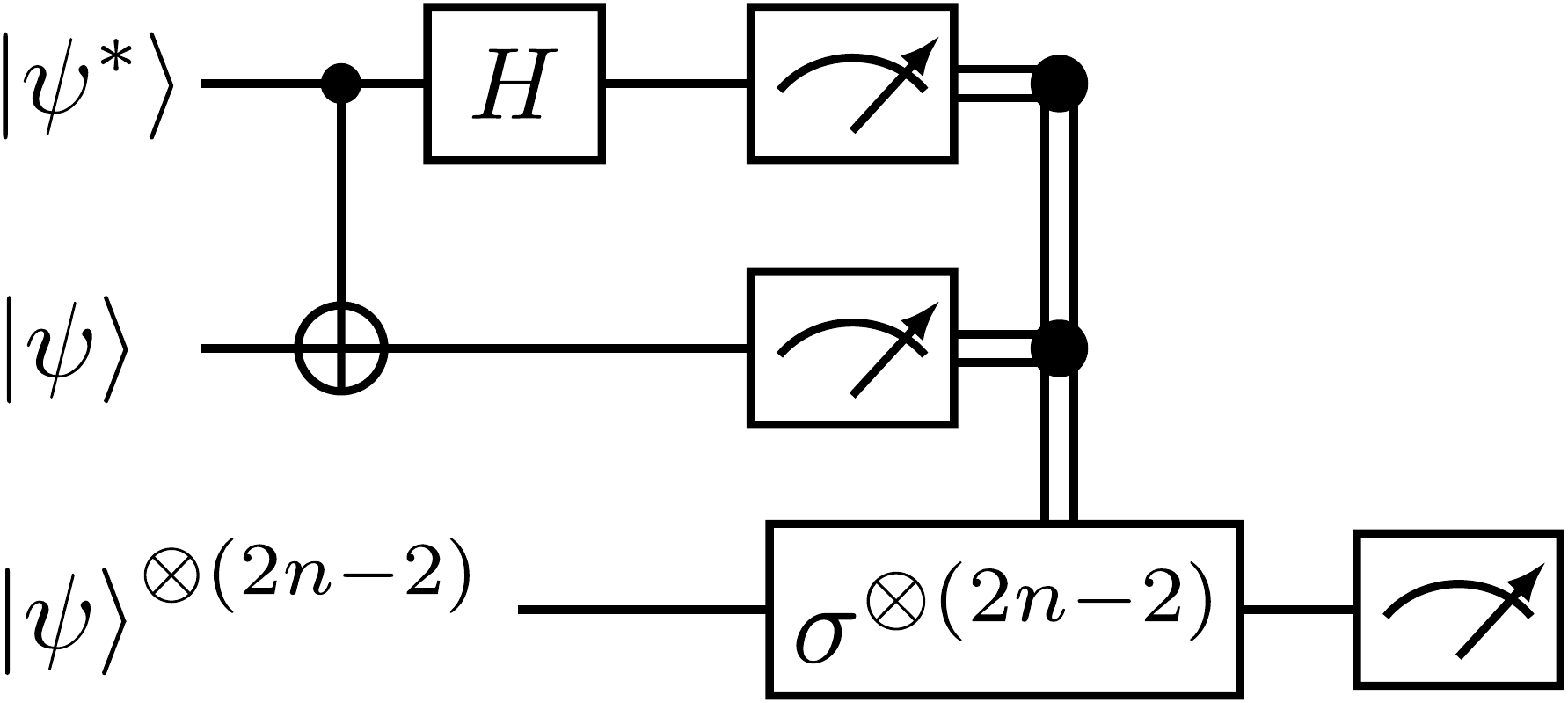}
	\caption{Measurement protocol for Algorithm~\ref{alg:SEmeaseven}.
	}
	\label{fig:seeven}
\end{figure}

\begin{algorithm}[ht]
 \SetAlgoLined
 \LinesNumbered
  \SetKwInOut{Input}{Input}
  \SetKwInOut{Output}{Output}
   \Input{   Integer $n>1$; $L$ repetitions
   }
    \Output{$A_n(\ket{\psi})$
    }

$A_n= 0$

 \SetKwRepeat{Do}{do}{while}
    \For{$k=1,\dots,L$}{
    Prepare $\ket{\eta}=U_\text{Bell}^{\otimes N}\ket{\psi^*}\otimes\ket{\psi}$

    Sample $\boldsymbol{r}\sim \vert \braket{\boldsymbol{r}\vert\eta}\vert^2$

    $b=1$
    
    \For{$\ell=1,\dots,2n-2$}{
    Prepare $\ket{\psi}$ and measure in eigenbasis of Paulistring $\sigma_{\boldsymbol{r}}$ for eigenvalue $\lambda\in\{+1,-1\}$

    $b= b\cdot\lambda$

    }

    $A_n= A_n+b/L$
    
    }

 \caption{SE with complex conjugate}
 \label{alg:SEmeaseven}
\end{algorithm}

Now, we provide Algorithm~\ref{alg:SEmeaseven} which is efficient for any integer $n>1$, however requires access to the complex conjugate $\ket{\psi^\ast}$.
We rewrite the SE as a sampling problem
\begin{equation}\label{eq:even}
A_n=\underset{\sigma\sim \Xi(\sigma)}{\mathbb{E}}[\bra{\psi}\sigma\ket{\psi}^{2n-2}]\,,
\end{equation}
where $\Xi(\sigma)=2^{-N}\braket{\psi|\sigma|\psi}^{2}$ is the probability distribution of Pauli strings $\sigma$.
The circuit for the algorithm is shown in Fig.\ref{fig:seeven}.
First, we prepare $\ket{\psi^\ast}\otimes\ket{\psi}$ on the quantum computer and transform into the Bell basis $\ket{\eta}=U_\text{Bell}^{\otimes N}\ket{\psi^*}\otimes\ket{\psi}$. 
Next, we sample from $2N$-qubit state $\ket{\eta}$ in the computational basis, gaining outcome $\boldsymbol{r}\in\{0,1\}^{2N}$.
As shown in~\cite{montanaro2017learning,lai2022learning}, we have $\Xi(\sigma_{\boldsymbol{r}})=\vert\braket{\boldsymbol{r}\vert\eta}\vert^2$
where $\ket{\boldsymbol{r}}$ is the computational basis state corresponding to bitstring $\boldsymbol{r}$. Thus, sampling $\boldsymbol{r}$ from $\ket{\eta}$ corresponds to sampling Pauli strings $\sigma_{\boldsymbol{r}}\sim \Xi(\sigma_{\boldsymbol{r}})$.
Then, we perform $2n-2$ measurements on $\ket{\psi}$ in the eigenbasis of the sampled $\sigma_{\boldsymbol{r}}$ and multiply the measured eigenvalues $\lambda_k$, gaining an unbiased estimator of $\bra{\psi}\sigma_{\boldsymbol{r}}\ket{\psi}^{2n-2}$.

The measured eigenvalues $\prod_{k=1}^{2n-2}\lambda_k\in\{+1,-1\}$ have a range $\Delta \omega_n=2$, thus according to Hoeffding's inequality we require at most $C=O(n\epsilon^{-2})$ copies of $\ket{\psi}$ and $O(\epsilon^{-2})$ copies of $\ket{\psi^\ast}$ for any integer $n>1$. 
Note that $\ket{\psi^{\ast}}$  cannot be efficiently prepared in general with only black-box access to $\ket{\psi}$~\cite{yang2014certifying,miyazaki2019complex,haug2023pseudorandom}. However, when we have a circuit description of the unitary preparing the state, $\ket{\psi^{\ast}}$ is constructed by an element-wise conjugation of the coefficients of the unitary~\cite{khatri2019quantum}.

\begin{figure}[htbp]
	\centering	
\subfigimg[width=0.238\textwidth]{a}{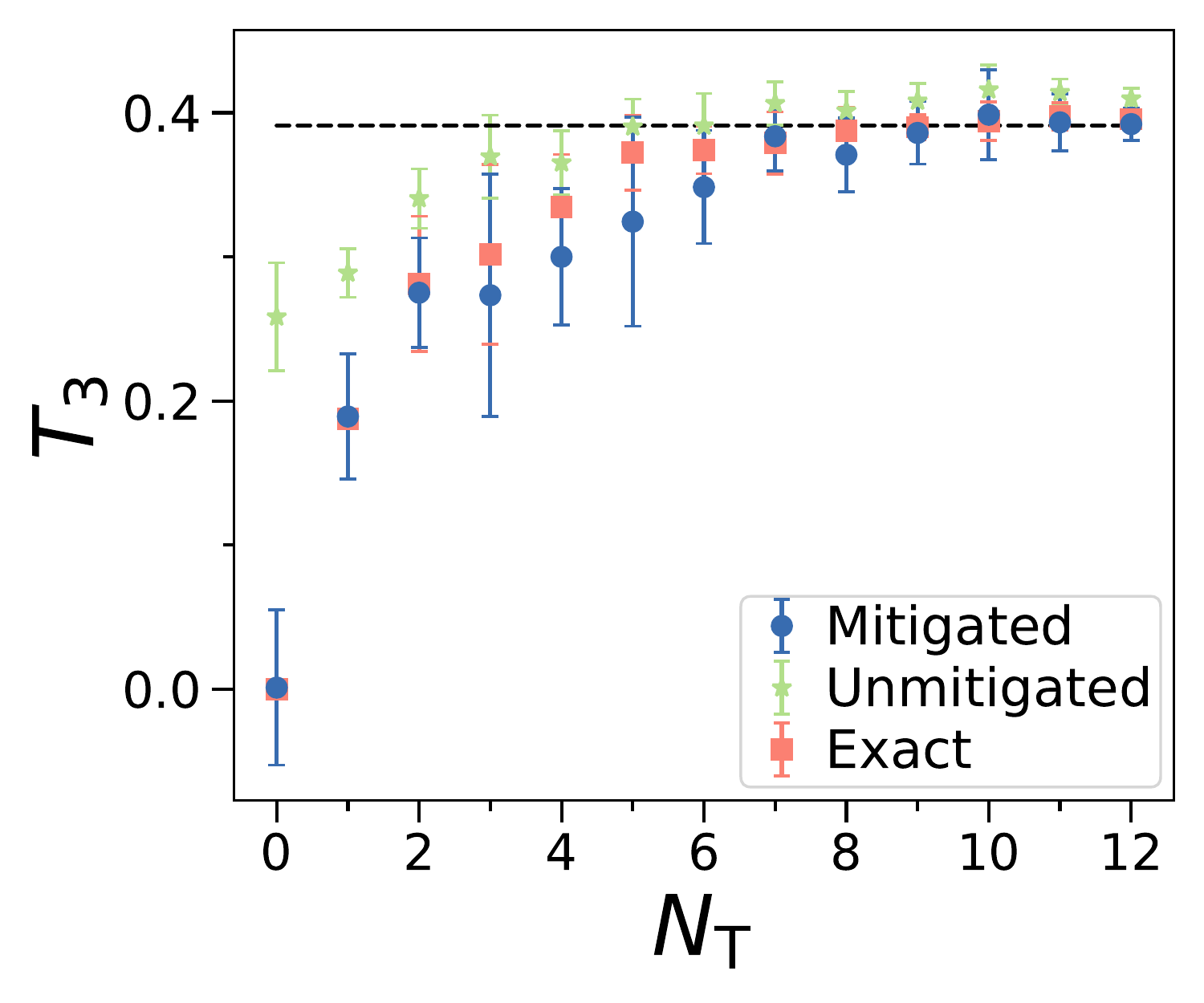}
 \subfigimg[width=0.238\textwidth]{b}{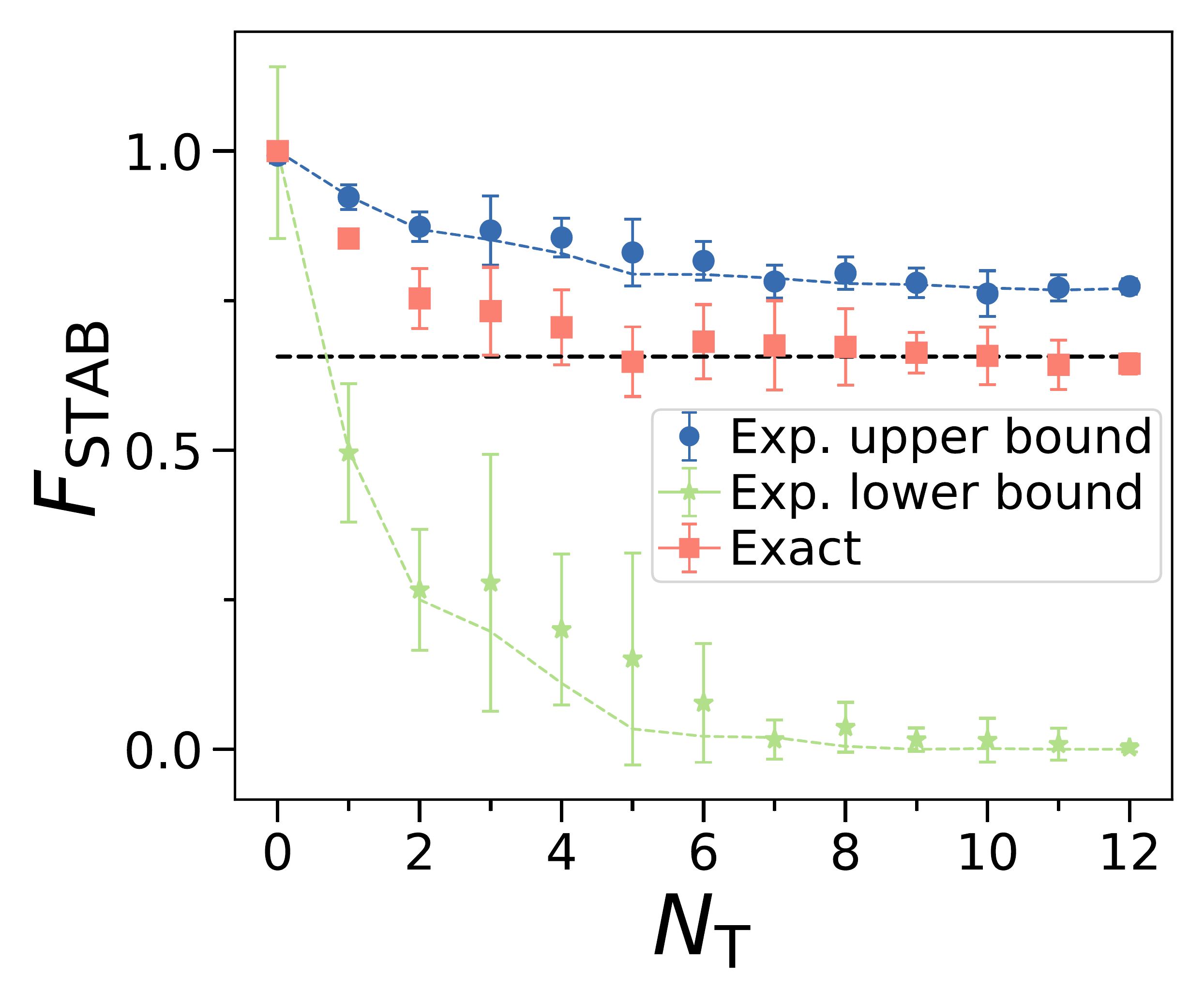}
	\caption{Measurement of nonstabilizerness for quantum states generated by~\eqref{eq:random_CliffTexp} with random Clifford circuits doped with $N_\text{T}$ T-gates on the IonQ quantum computer. 
 \idg{a} We show Tsallis SE $T_3$ with and without error mitigation as well as exact simulation. Dashed line is average value for Haar random states. Dots represent mean value and error bars the standard deviation taken over 6 random instances of the circuit. We have $N=3$ qubits, $10^3$ Bell measurements and a measured depolarization error of $p\approx0.1$.
 \idg{b} We show upper and lower bounds on $F_\text{STAB}$ via~\eqref{eq:stabbound} evaluated using the error mitigated $T_3$ as blue and green dots as well as simulation of the bound as dashed lines. The orange dots show simulations of $F_\text{STAB}$.
	}
	\label{fig:exp}
\end{figure}

\prlsection{Tsallis SE}
We now define a measure of nonstabilizerness which we call the Tsallis-$n$ SE~\cite{tsallis1988possible}
\begin{equation}\label{eq:Tsallis}
T_n(|\psi\rangle)=-(1-n)^{-1} (1-\sum_{\sigma \in \mathcal{P}}2^{-N}\braket{\psi|\sigma|\psi}^{2n})\,.
\end{equation}
They are a generalization of the linear SE $T_2$~\cite{leone2022stabilizer} and the von-Neumann SE $T_1=M_1=-2^{-N}\sum_{\sigma \in \mathcal{P}}\braket{\psi|\sigma|\psi}^{2}\ln(\braket{\psi|\sigma|\psi}^{2})$~\cite{haug2023stabilizer}. $T_n$ can be efficiently measured for integer $n>1$ using our protocols.
They are faithful measures of nonstabilizerness which are invariant under Clifford unitaries and related to R\'enyi SEs via $M_n=(1-n)^{-1}\ln(1+(1-n)T_n)$. Tsallis SEs lack the additive property of the R\'enyi SE, however our numerics suggest that Tsallis SEs may be a strong monotone which is a not necessary but highly desirable property for resource measures~\cite{chitambar2019quantum}.
Within extensive numerical optimization for $N\le6$ qubits we were unable to find states that could violate strong monotonicity for the Tsallis-$n$ SE for $n\ge2$ (see SM~\ref{sec:strong_mon}).

Note that measuring the R\'enyi SE $M_n\sim\ln(A_n)$ with precision $\epsilon_M$ requires $O(n\exp(M_n)\epsilon_M ^{-2} )$ samples due to the logarithm (see SM~\ref{sec:strong_mon}). Thus, $M_n$ is efficiently measurable as long as $M_n=O(\log(N))$.

\revA{\prlsection{Clifford-averaged OTOCs}
We now show how to efficiently measure $4n$-point OTOCs of unitary $U$ averaged over the Clifford group.}
The $4n$-point OTOC for $N$-qubit Pauli strings $\sigma$ and $\sigma'$ is given by~\cite{leone2023nonstabilizerness}
\begin{align*}
&\text{otoc}_{4n}(U,\sigma,\sigma')=\left(2^{-N}\text{tr}(\sigma U\sigma' U^\dagger)\right)^{2n}\numberthis
\end{align*}
We find that $\text{otoc}_{4n}$ averaged over the group of Clifford unitaries $\mathcal{C}_N$ can be related to SE of $U$, which we define via the Choi state $\ket{U}=I_N \otimes U \ket{\Phi}$, where $\ket{\Phi}=2^{-N/2}\sum_{i=0}^{2^N-1}\ket{i}\otimes \ket{i}$. In particular, we have
\begin{align*}
&\underset{U_\text{C},U_\text{C}'\in \mathcal{C}_N}{\mathbb{E}}[\text{otoc}_{4n}(U_\text{C}U U_\text{C}',\sigma,\sigma')]=\frac{A_n(\ket{U})4^{N}-1}{(4^{N}-1)^2}\numberthis\label{eq:OTOC_avg}
\end{align*}
where the Pauli strings $\sigma,\sigma'\in\mathcal{P}/\{I_N\}$ exclude the identity and the proof is found in SM~\ref{sec:MagicOTOC} using results of Ref.~\cite{leone2023nonstabilizerness}.
For odd $n>1$, we can efficiently measure~\eqref{eq:OTOC_avg} via Algorithm~\ref{alg:SEmeas}. For even $n>1$, we additionally require the complex conjugate $\ket{U^\ast}$ for Algorithm~\ref{alg:SEmeaseven}. The complex conjugate of the Choi state can be efficiently prepared with access to $U^\ast$ or $U^\dagger$ due to the ricochet property $\ket{U^\ast}= I_N \otimes U^\ast \ket{\Phi}=U^\dagger \otimes I_N \ket{\Phi}$~\cite{khatri2019quantum}.

\prlsection{Multifractal flatness}
The participation entropy is given by
    $\mathcal{I}_q(\ket{\psi})=\sum_{k} \vert\braket{k\vert \psi}\vert^{2q}$
where $\ket{k}$ are computational basis states, $q>0$ and $0\le \mathcal{I}_q\le 1$ ~\cite{castellani1986multifractal}. The participation entropy quantifies the spread of the wavefunction over basis states, i.e. $\mathcal{I}_q=1$ for computational basis states, while it is small when the state is delocalized over many computational basis states.
The multifractal flatness $\mathcal{F}(\ket{\psi})=\mathcal{I}_3(\ket{\psi})-\mathcal{I}_2^2(\ket{\psi})$ measures the flatness of the distribution $\vert\braket{k\vert \psi}\vert^2$, i.e. we have $\mathcal{F}=0$ when the distribution $\vert\braket{k\vert \psi}\vert^2$ is constant over its support, else we have $\mathcal{F}>0$. In particular, stabilizer states have $\mathcal{F}=0$~\cite{sierant2022universal}.

Recently, $\mathcal{F}$ averaged over $\mathcal{C}_N$ has been proposed as $\bar{\mathcal{F}}$~\cite{turkeshi2023measuring}. \revA{This quantity describes the participation ratio averaged over all possible choices of basis states. $\bar{\mathcal{F}}$ has been connected to SEs as follows~\cite{turkeshi2023measuring}}
\begin{equation}
\bar{\mathcal{F}}(\ket{\psi})=\underset{U_\text{C} \in \mathcal{C}_N}{\mathbb{E}}[\mathcal{F}(U_\text{C}\ket{\psi})]=\frac{2(1-A_2(\ket{\psi}))}{(2^N+1)(2^N+2)}\,.\label{eq:multi_avg}
\end{equation}
Thus, Algorithm~\ref{alg:SEmeaseven} allows us to efficiently measure $\bar{\mathcal{F}}(\ket{\psi})$ directly without the need of averaging over $\mathcal{C}_N$.

\prlsection{Bounds on nonstabilizerness}
We now provide efficient bounds on three magic monotones, namely the robustness of magic $R$~\cite{howard2017application}, stabilizer extent $\xi$~\cite{bravyi2019simulation} and the stabilizer fidelity $F_\text{STAB}=\max_{\ket{\phi}\in\text{STAB}}\vert \braket{\psi\vert\phi}\vert^2$~\cite{bravyi2019simulation}
(see SM~\ref{sec:upperbound}).
Computing $\text{R}$, $\xi$ and $F_\text{STAB}$ requires solving an optimization program over the set of pure $N$-qubit stabilizer states. As the number of stabilizer states scales as $O(2^{N^2})$, these three measures in general become numerically infeasible beyond $5$ qubits~\cite{aaronson2004improved,howard2017application}. 

Our algorithms provide efficient bounds for integer $n>1$  (see \cite{haug2023stabilizer} or SM~\ref{sec:upperbound}):
\begin{equation}\label{eq:hierarchy}
R\ge \xi \ge F_\text{STAB}^{-1}\ge A_n^{-\frac{1}{2n}}\,.
\end{equation}
The bound can be tightened for $n\ge\frac{1}{2}$ to
$\text{R}\ge A_n^{\frac{1}{2(1-n)}}$~\cite{howard2017application,leone2022stabilizer}. 
With methods from Ref.~\cite{gross2021schur,grewal2023improved}, we also prove a lower bound on $F_\text{STAB}$ for $n>1$ (see SM~\ref{sec:lowerbound})
\begin{equation}\label{eq:stabbound}
A_n^{\frac{1}{2n}} \ge F_\text{STAB}\ge \frac{A_n-2^{1-n}}{1-2^{1-n}}\,.
\end{equation}
\revA{The min-relative entropy of magic $D_\text{min}=-\ln(F_\text{STAB})$ can be seen as the distance to the nearest stabilizer state. We now argue that $D_\text{min}$, R\'enyi SEs with $n\ge2$ and the recently introduced additive Bell magic $\mathcal{B}_\text{a}$~\cite{haug2022scalable} are closely related. In particular, we find evidence for respective upper and lower bounds  independent of qubit number $N$ (see SM~\ref{sec:other_meas}). Via numerical optimization we find $1.7 M_2\gtrsim D_\text{min}\ge \frac{1}{4}M_2$ as well as $3.5 M_2 \gtrsim  \mathcal{B}_\text{a} \gtrsim 2.88 M_2 $ for at least $N\le4$, while similar bounds can also be found for larger $n$. Thus, $D_\text{min}$, $M_{n\ge2}$ and $\mathcal{B}_\text{a}$ can be seen as measures of nonstabilizerness that relate to the distance to the nearest stabilizer state.
In contrast, the robustness of magic $R$ and stabilizer extent $\xi$ relate to the degree a state can be approximated by a combination of stabilizer states. They belong to a different class of nonstabilizerness measures as they cannot be upper bounded with $D_\text{min}$ or $M_n$ for $n>1/2$~\cite{haug2023stabilizer}.  }

\prlsection{Demonstration}
We now study SEs with Bell measurements on the IonQ quantum computer~\cite{haug2022scalable} using Algorithm~\ref{alg:SEmeas} in Fig.~\ref{fig:exp}. 
We investigate random Clifford circuits $U_\text{C}$ doped with $N_\text{T}$ non-Clifford gates
\begin{equation}\label{eq:random_CliffTexp}
\ket{\psi(N_\text{T})}=U_\text{C}^{(0)}\prod_{k=1}^{N_\text{T}} V_\text{T}^{(k)}U_\text{C}^{(k)} \ket{0}\,,
\end{equation}
where $V_\text{T}^{(k)}=\exp(-i\frac{\pi}{8}\sigma^z_{g(k)})$ is the T-gate of the $k$-th layer acting on a randomly chosen qubit $g(k)$. With increasing $N_\text{T}$ these states transition from efficiently simulable stabilizer states to intractable quantum states~\cite{haferkamp2022random,leone2021quantum}.
To reduce noise, we compress the circuits into layered circuits composed of single-qubit operations and CNOT gates arranged in a nearest-neighbor chain configuration~\cite{haug2022scalable}.
The state prepared by the quantum computer is not pure but degraded by noise.
However,  SEs are faithful measures of nonstabilizerness only for pure states. Using measurements on the noisy state, we mitigate $A_n$ and $T_n$ from measurements on noisy states by assuming a global depolariziation error model (see SM~\ref{sec:mtg}).

In Fig.~\ref{fig:exp}a, we show $T_3$ with and without error mitigation for different $N_\text{T}$, where for each value we average over 6 random instances of~\eqref{eq:random_CliffTexp}. We find that the  results on the IonQ quantum computer with error mitigation closely match the simulated values. 
The Tsallis SE is zero for $N_\text{T}=0$, then increases with $N_\text{T}$ until it converges to the average value of Haar random states indicated as black dashed line.
In Fig.~\ref{fig:exp}b, we use the mitigated results for $A_3$ to compute upper and lower bounds for the stabilizer fidelity $F_\text{STAB}$ using~\eqref{eq:stabbound}. The measured result indeed gives valid bounds of the exactly simulated $F_\text{STAB}$. We find that the upper bound is relatively tight, while the lower bound is non-trivial only for small $N_\text{T}$. 
In SM~\ref{sec:exp_other}, we provide additional results for the IonQ quantum computer on measures of nonstabilizerness.
\revA{While our error mitigation scheme assumes global depolarization noise, it works well on actual quantum computers which have more complicated noise profiles. In SM~\ref{sec:noise}, we simulate our error mitigation scheme for various unital and non-unital noise models, and find very good performance. As SEs are moments of exponentially many Pauli strings, self-averaging effects may explain the good performance. }

\begin{figure}[htbp]
	\centering	
\subfigimg[width=0.238\textwidth]{a}{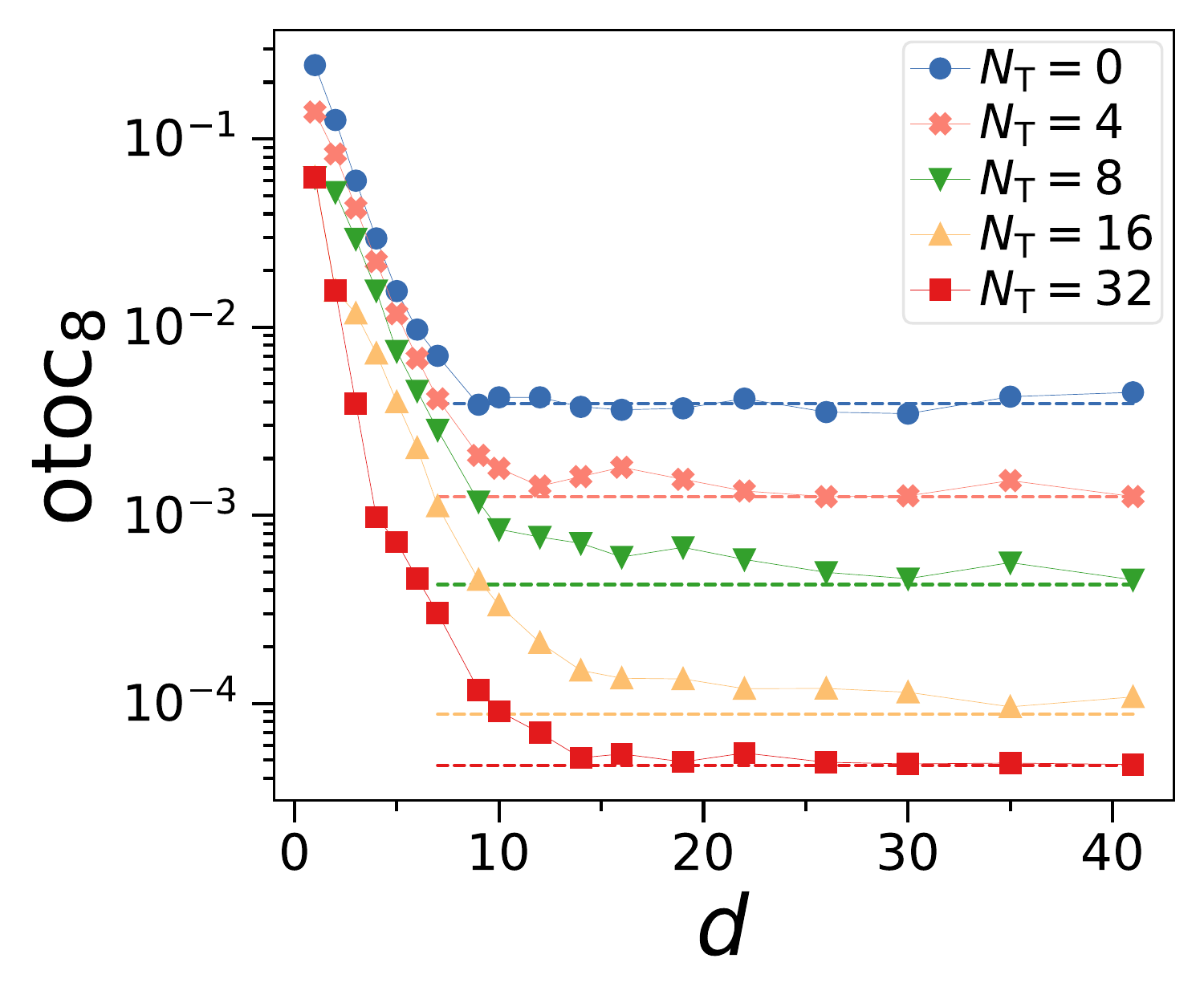}
 \subfigimg[width=0.238\textwidth]{b}{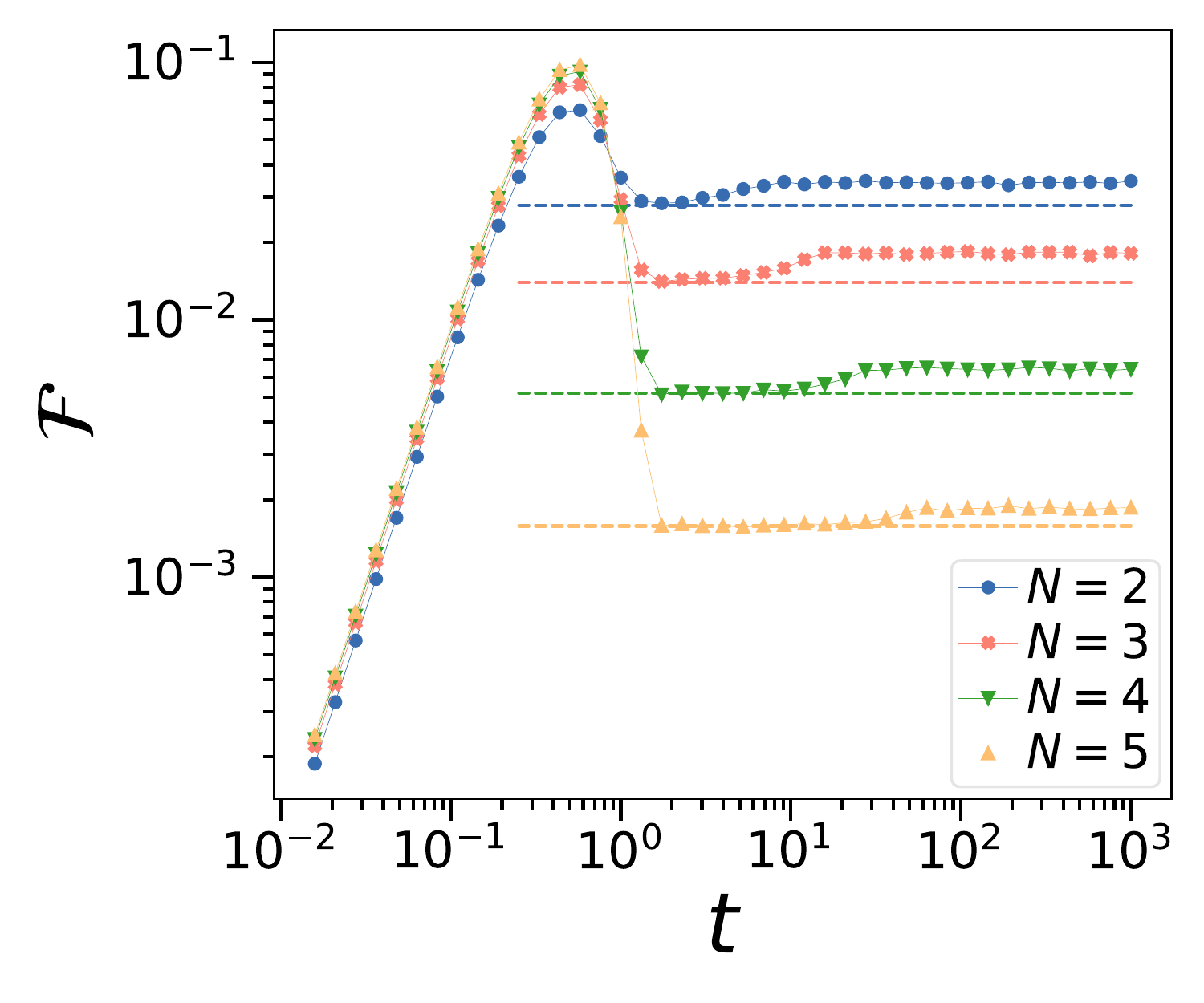}
	\caption{\idg{a} $\text{otoc}_{8}(U,\sigma^x_1,\sigma^x_1)$ against $d$ layers of single-qubit Clifford gates and CNOT gates arranged in a nearest-neighbor chain, doped with $N_\text{T}$ T-gates and $N=4$ qubits. Dashed line is the Clifford-averaged OTOC~\eqref{eq:OTOC_avg}.
 \idg{b} Multifractal flatness $\mathcal{F}$ for evolution in time $t$ with random Hamiltonian $\exp(-iH_\text{GUE}t)\ket{0}$. Dashed line is the Clifford-averaged multifractal flatness~\eqref{eq:multi_avg}.
	}
	\label{fig:sim}
\end{figure}

\revB{
\prlsection{Scrambling}
We now study scrambling using the multifractal flatness $\mathcal{F}$ and OTOCs. In Fig.~\ref{fig:sim}a, we show $\text{otoc}_{8}(U,\sigma^x_1,\sigma^x_1)$ against $d$ layers of Clifford gates doped with $N_\text{T}$ T-gates. We find that the OTOC decreases with $d$, converging to a minimum once the unitary is fully scrambled. This minimum is given by the Clifford averaged OTOC~\eqref{eq:OTOC_avg} and depends on $N_\text{T}$. The $d$ needed to converge depends on the number of T-gates, where for $N_\text{T}=0$ convergence is achieved for $d\sim10$, while higher $N_\text{T}$ requires larger $d$ to converge. We observe similar convergence for $\mathcal{F}$ and other OTOCs in SM~\ref{sec:MagicOTOC}.

In Fig.~\ref{fig:sim}b, we study $\mathcal{F}$ for the evolution of a state $\ket{\psi(t)}=\exp(-iH_\text{GUE}t)\ket{0}$ in time $t$ using a random Hamiltonian $H_\text{GUE}$ drawn from the Gaussian Unitary ensemble (GUE). We observe that $\mathcal{F}$ initially increases, reaching a maximum at $t\sim1$. This is followed by a sudden dip to the Clifford-averaged multifractal flatness~\eqref{eq:multi_avg}.  This is hallmark of reaching deep thermalization or unitary design, where the system is indistinguishable from Haar-random dynamics~\cite{cotler2017chaos}. Counter-intuitively, for longer (exponential) times $\mathcal{F}$ ramps up again, converging to a value above the Clifford-average. Here, the system stops being fully random due to dephasing of energy eigenvalues~\cite{cotler2017chaos}. In SM~\ref{sec:MagicOTOC}, we show how to measure $\mathcal{F}$ and approximate GUE Hamiltonians using a Hamiltonian of random Pauli strings which can be implemented in experiment.

}

\prlsection{Discussions}
\revA{We show how to efficiently measure SEs with a cost independent of qubit number $N$, which is an exponential improvement over previous protocols~\cite{haug2022scalable,oliviero2022measuring}.} For integer $n>1$, our protocol is asymptotically optimal with the number of copies scaling as $O(n\epsilon^{-2})$ and the classical post-processing time as $O(nN\epsilon^{-2})$ with additive error $\epsilon$. The protocol is easy to implement using Bell measurements which have been demonstrated for quantum computers and simulators~\cite{islam2015measuring,huang2021demonstrating,bluvstein2022quantum}. We note that our approach is distinct from the previously introduced Bell magic~\cite{haug2022scalable} as shown in SM~\ref{sec:Bell}.

Our measurement protocol allows for efficient experimental characterization of different important properties of quantum states.
We demonstrate an efficient bound on nonstabilizerness monotones which otherwise are hard to compute beyond a few qubits. These monotones serve as lower bounds on state preparation complexity and characterize the runtime of classical simulation algorithms~\cite{howard2017application,bravyi2019simulation}.
Further, we show how to efficiently measure Clifford-averaged $4n$-point OTOCs. Our protocol has the advantage that it does not require implementing time-reversal for odd $n>1$ which can be a challenge~\cite{xu2022scrambling}.
Our protocol can measure higher order OTOCs which promise to reveal more features compared to the usually considered $4$-point OTOCs~\cite{roberts2017chaos,garcia2021quantum}.
Our methods allow direct experimental study phase transitions in SE which have been found for purity testing~\cite{leone2023phase} and quantum error correction~\cite{niroula2023phase}.  \revA{Finally, we enable certification of magic gates in fault-tolerant quantum computers, where the SE could be directly evaluated from recent experimental data~\cite{bluvstein2023logical}.}

\revB{We use our methods to study scrambling in Clifford circuits doped with T-gates and random Hamiltonian evolution. OTOCs not only measure scrambling, but also depend on nonstabilizerness in a non-trivial way~\cite{leone2021isospectral}. We can disentangle these two effects by measuring the Clifford-averaged OTOC. We study when Clifford circuits doped with T-gates become fully scrambled, revealing that the depth depends on the number of T-gates.
We also study the scrambling with random Hamiltonians. Notably, random Hamiltonian evolution deep thermalizes at intermediate times, becoming indistinguishable from Haar random unitaries~\cite{cotler2017chaos} which we observe via the convergence of the multifractal flatness to its Clifford-averaged value. Counter-intuitively, the evolution becomes less random again for long times due to the dephasing of its energy eigenstates~\cite{cotler2017chaos}. While non-local OTOCs and SEs lack clear signatures of this effect  (see SE~L), we find that multifractal flatness and same-site OTOCs exhibit clear gaps to their Clifford-average.}

Future work could find efficient protocols for even $n$ without the need of complex conjugation and tighten the lower bound of SEs for the stabilizer fidelity.

The code for this work is available on Github~\cite{haug2023stabilizerentropy}.

\revC{\textit{Note added:} Before acceptance of this letter, the monotonicity of R\'enyi SE and strong monotonicity of Tsallis SE has been proven for $n\ge2$~\cite{leone2024stabilizer}.}

 \let\oldaddcontentsline\addcontentsline
\renewcommand{\addcontentsline}[3]{}

\begin{acknowledgements}
\prlsection{Acknowledgements}
We thank Hyukjoon Kwon, Ludovico Lami, Lorenzo Leone, Salvatore F.E. Oliviero, Adam Taylor and especially Lorenzo Piroli for inspiring discussions.
We thank IonQ for providing quantum computing resources.
This work is supported by a Samsung GRC project and the UK Hub in Quantum Computing and Simulation, part of the UK National Quantum Technologies Programme with funding from UKRI EPSRC grant EP/T001062/1. 
\end{acknowledgements}

\bibliography{bibliography}

\begin{thebibliography}{71}%
\makeatletter
\providecommand \@ifxundefined [1]{%
 \@ifx{#1\undefined}
}%
\providecommand \@ifnum [1]{%
 \ifnum #1\expandafter \@firstoftwo
 \else \expandafter \@secondoftwo
 \fi
}%
\providecommand \@ifx [1]{%
 \ifx #1\expandafter \@firstoftwo
 \else \expandafter \@secondoftwo
 \fi
}%
\providecommand \natexlab [1]{#1}%
\providecommand \enquote  [1]{``#1''}%
\providecommand \bibnamefont  [1]{#1}%
\providecommand \bibfnamefont [1]{#1}%
\providecommand \citenamefont [1]{#1}%
\providecommand \href@noop [0]{\@secondoftwo}%
\providecommand \href [0]{\begingroup \@sanitize@url \@href}%
\providecommand \@href[1]{\@@startlink{#1}\@@href}%
\providecommand \@@href[1]{\endgroup#1\@@endlink}%
\providecommand \@sanitize@url [0]{\catcode `\\12\catcode `\$12\catcode
  `\&12\catcode `\#12\catcode `\^12\catcode `\_12\catcode `\%12\relax}%
\providecommand \@@startlink[1]{}%
\providecommand \@@endlink[0]{}%
\providecommand \url  [0]{\begingroup\@sanitize@url \@url }%
\providecommand \@url [1]{\endgroup\@href {#1}{\urlprefix }}%
\providecommand \urlprefix  [0]{URL }%
\providecommand \Eprint [0]{\href }%
\providecommand \doibase [0]{http://dx.doi.org/}%
\providecommand \selectlanguage [0]{\@gobble}%
\providecommand \bibinfo  [0]{\@secondoftwo}%
\providecommand \bibfield  [0]{\@secondoftwo}%
\providecommand \translation [1]{[#1]}%
\providecommand \BibitemOpen [0]{}%
\providecommand \bibitemStop [0]{}%
\providecommand \bibitemNoStop [0]{.\EOS\space}%
\providecommand \EOS [0]{\spacefactor3000\relax}%
\providecommand \BibitemShut  [1]{\csname bibitem#1\endcsname}%
\let\auto@bib@innerbib\@empty
\bibitem [{\citenamefont {Gottesman}(1997)}]{gottesman1997stabilizer}%
  \BibitemOpen
  \bibfield  {author} {\bibinfo {author} {\bibfnamefont {Daniel}\ \bibnamefont
  {Gottesman}},\ }\emph {\bibinfo {title} {Stabilizer codes and quantum error
  correction. Caltech Ph. D}},\ \href@noop {} {Ph.D. thesis},\ \bibinfo
  {school} {Thesis, eprint: quant-ph/9705052} (\bibinfo {year}
  {1997})\BibitemShut {NoStop}%
\bibitem [{\citenamefont {Shor}(1996)}]{shor1996fault}%
  \BibitemOpen
  \bibfield  {author} {\bibinfo {author} {\bibfnamefont {Peter~W}\ \bibnamefont
  {Shor}},\ }\bibfield  {title} {\enquote {\bibinfo {title} {Fault-tolerant
  quantum computation},}\ }in\ \href@noop {} {\emph {\bibinfo {booktitle}
  {Proceedings of 37th conference on foundations of computer science}}}\
  (\bibinfo {organization} {IEEE},\ \bibinfo {year} {1996})\ pp.\ \bibinfo
  {pages} {56--65}\BibitemShut {NoStop}%
\bibitem [{\citenamefont {Kitaev}(2003)}]{kitaev2003fault}%
  \BibitemOpen
  \bibfield  {author} {\bibinfo {author} {\bibfnamefont {A~Yu}\ \bibnamefont
  {Kitaev}},\ }\bibfield  {title} {\enquote {\bibinfo {title} {Fault-tolerant
  quantum computation by anyons},}\ }\href {\doibase
  10.1016/S0003-4916(02)00018-0} {\bibfield  {journal} {\bibinfo  {journal}
  {Ann. Phys.}\ }\textbf {\bibinfo {volume} {303}},\ \bibinfo {pages} {2--30}
  (\bibinfo {year} {2003})}\BibitemShut {NoStop}%
\bibitem [{\citenamefont {Bravyi}\ and\ \citenamefont
  {Kitaev}(2005)}]{bravyi2005universal}%
  \BibitemOpen
  \bibfield  {author} {\bibinfo {author} {\bibfnamefont {Sergey}\ \bibnamefont
  {Bravyi}}\ and\ \bibinfo {author} {\bibfnamefont {Alexei}\ \bibnamefont
  {Kitaev}},\ }\bibfield  {title} {\enquote {\bibinfo {title} {Universal
  quantum computation with ideal clifford gates and noisy ancillas},}\ }\href
  {\doibase 10.1103/PhysRevA.71.022316} {\bibfield  {journal} {\bibinfo
  {journal} {Phys. Rev. A}\ }\textbf {\bibinfo {volume} {71}},\ \bibinfo
  {pages} {022316} (\bibinfo {year} {2005})}\BibitemShut {NoStop}%
\bibitem [{\citenamefont {Campbell}\ \emph {et~al.}(2017)\citenamefont
  {Campbell}, \citenamefont {Terhal},\ and\ \citenamefont
  {Vuillot}}]{campbell2017roads}%
  \BibitemOpen
  \bibfield  {author} {\bibinfo {author} {\bibfnamefont {Earl~T}\ \bibnamefont
  {Campbell}}, \bibinfo {author} {\bibfnamefont {Barbara~M}\ \bibnamefont
  {Terhal}}, \ and\ \bibinfo {author} {\bibfnamefont {Christophe}\ \bibnamefont
  {Vuillot}},\ }\bibfield  {title} {\enquote {\bibinfo {title} {Roads towards
  fault-tolerant universal quantum computation},}\ }\href {\doibase
  10.1038/nature23460} {\bibfield  {journal} {\bibinfo  {journal} {Nature}\
  }\textbf {\bibinfo {volume} {549}},\ \bibinfo {pages} {172--179} (\bibinfo
  {year} {2017})}\BibitemShut {NoStop}%
\bibitem [{\citenamefont {Campbell}(2011)}]{campbell2011catalysis}%
  \BibitemOpen
  \bibfield  {author} {\bibinfo {author} {\bibfnamefont {Earl~T.}\ \bibnamefont
  {Campbell}},\ }\bibfield  {title} {\enquote {\bibinfo {title} {Catalysis and
  activation of magic states in fault-tolerant architectures},}\ }\href
  {\doibase 10.1103/PhysRevA.83.032317} {\bibfield  {journal} {\bibinfo
  {journal} {Phys. Rev. A}\ }\textbf {\bibinfo {volume} {83}},\ \bibinfo
  {pages} {032317} (\bibinfo {year} {2011})}\BibitemShut {NoStop}%
\bibitem [{\citenamefont {Veitch}\ \emph {et~al.}(2014)\citenamefont {Veitch},
  \citenamefont {Mousavian}, \citenamefont {Gottesman},\ and\ \citenamefont
  {Emerson}}]{veitch2014resource}%
  \BibitemOpen
  \bibfield  {author} {\bibinfo {author} {\bibfnamefont {Victor}\ \bibnamefont
  {Veitch}}, \bibinfo {author} {\bibfnamefont {SA~Hamed}\ \bibnamefont
  {Mousavian}}, \bibinfo {author} {\bibfnamefont {Daniel}\ \bibnamefont
  {Gottesman}}, \ and\ \bibinfo {author} {\bibfnamefont {Joseph}\ \bibnamefont
  {Emerson}},\ }\bibfield  {title} {\enquote {\bibinfo {title} {The resource
  theory of stabilizer quantum computation},}\ }\href
  {https://doi.org/10.1088/1367-2630/16/1/013009} {\bibfield  {journal}
  {\bibinfo  {journal} {New J. Phys.}\ }\textbf {\bibinfo {volume} {16}},\
  \bibinfo {pages} {013009} (\bibinfo {year} {2014})}\BibitemShut {NoStop}%
\bibitem [{\citenamefont {Howard}\ and\ \citenamefont
  {Campbell}(2017)}]{howard2017application}%
  \BibitemOpen
  \bibfield  {author} {\bibinfo {author} {\bibfnamefont {Mark}\ \bibnamefont
  {Howard}}\ and\ \bibinfo {author} {\bibfnamefont {Earl}\ \bibnamefont
  {Campbell}},\ }\bibfield  {title} {\enquote {\bibinfo {title} {Application of
  a resource theory for magic states to fault-tolerant quantum computing},}\
  }\href {\doibase 10.1103/PhysRevLett.118.090501} {\bibfield  {journal}
  {\bibinfo  {journal} {Phys. Rev. Lett.}\ }\textbf {\bibinfo {volume} {118}},\
  \bibinfo {pages} {090501} (\bibinfo {year} {2017})}\BibitemShut {NoStop}%
\bibitem [{\citenamefont {Wang}\ \emph {et~al.}(2019)\citenamefont {Wang},
  \citenamefont {Wilde},\ and\ \citenamefont {Su}}]{wang2019quantifying}%
  \BibitemOpen
  \bibfield  {author} {\bibinfo {author} {\bibfnamefont {Xin}\ \bibnamefont
  {Wang}}, \bibinfo {author} {\bibfnamefont {Mark~M}\ \bibnamefont {Wilde}}, \
  and\ \bibinfo {author} {\bibfnamefont {Yuan}\ \bibnamefont {Su}},\ }\bibfield
   {title} {\enquote {\bibinfo {title} {Quantifying the magic of quantum
  channels},}\ }\href {https://doi.org/10.1088/1367-2630/ab451d} {\bibfield
  {journal} {\bibinfo  {journal} {New J. Phys.}\ }\textbf {\bibinfo {volume}
  {21}},\ \bibinfo {pages} {103002} (\bibinfo {year} {2019})}\BibitemShut
  {NoStop}%
\bibitem [{\citenamefont {Beverland}\ \emph {et~al.}(2020)\citenamefont
  {Beverland}, \citenamefont {Campbell}, \citenamefont {Howard},\ and\
  \citenamefont {Kliuchnikov}}]{beverland2020lower}%
  \BibitemOpen
  \bibfield  {author} {\bibinfo {author} {\bibfnamefont {Michael}\ \bibnamefont
  {Beverland}}, \bibinfo {author} {\bibfnamefont {Earl}\ \bibnamefont
  {Campbell}}, \bibinfo {author} {\bibfnamefont {Mark}\ \bibnamefont {Howard}},
  \ and\ \bibinfo {author} {\bibfnamefont {Vadym}\ \bibnamefont
  {Kliuchnikov}},\ }\bibfield  {title} {\enquote {\bibinfo {title} {Lower
  bounds on the non-clifford resources for quantum computations},}\ }\href
  {\doibase 10.1088/2058-9565/ab8963} {\bibfield  {journal} {\bibinfo
  {journal} {Quantum Science Tech.}\ }\textbf {\bibinfo {volume} {5}},\
  \bibinfo {pages} {035009} (\bibinfo {year} {2020})}\BibitemShut {NoStop}%
\bibitem [{\citenamefont {Jiang}\ and\ \citenamefont
  {Wang}(2023)}]{jiang2021lower}%
  \BibitemOpen
  \bibfield  {author} {\bibinfo {author} {\bibfnamefont {Jiaqing}\ \bibnamefont
  {Jiang}}\ and\ \bibinfo {author} {\bibfnamefont {Xin}\ \bibnamefont {Wang}},\
  }\bibfield  {title} {\enquote {\bibinfo {title} {Lower bound for the t count
  via unitary stabilizer nullity},}\ }\href {\doibase
  10.1103/PhysRevApplied.19.034052} {\bibfield  {journal} {\bibinfo  {journal}
  {Physical Review Applied}\ }\textbf {\bibinfo {volume} {19}},\ \bibinfo
  {pages} {034052} (\bibinfo {year} {2023})}\BibitemShut {NoStop}%
\bibitem [{\citenamefont {Liu}\ and\ \citenamefont
  {Winter}(2022)}]{liu2022many}%
  \BibitemOpen
  \bibfield  {author} {\bibinfo {author} {\bibfnamefont {Zi-Wen}\ \bibnamefont
  {Liu}}\ and\ \bibinfo {author} {\bibfnamefont {Andreas}\ \bibnamefont
  {Winter}},\ }\bibfield  {title} {\enquote {\bibinfo {title} {Many-body
  quantum magic},}\ }\href {\doibase 10.1103/PRXQuantum.3.020333} {\bibfield
  {journal} {\bibinfo  {journal} {PRX Quantum}\ }\textbf {\bibinfo {volume}
  {3}},\ \bibinfo {pages} {020333} (\bibinfo {year} {2022})}\BibitemShut
  {NoStop}%
\bibitem [{\citenamefont {Bu}\ \emph {et~al.}(2022)\citenamefont {Bu},
  \citenamefont {Garcia}, \citenamefont {Jaffe}, \citenamefont {Koh},\ and\
  \citenamefont {Li}}]{bu2022complexity}%
  \BibitemOpen
  \bibfield  {author} {\bibinfo {author} {\bibfnamefont {Kaifeng}\ \bibnamefont
  {Bu}}, \bibinfo {author} {\bibfnamefont {Roy~J}\ \bibnamefont {Garcia}},
  \bibinfo {author} {\bibfnamefont {Arthur}\ \bibnamefont {Jaffe}}, \bibinfo
  {author} {\bibfnamefont {Dax~Enshan}\ \bibnamefont {Koh}}, \ and\ \bibinfo
  {author} {\bibfnamefont {Lu}~\bibnamefont {Li}},\ }\bibfield  {title}
  {\enquote {\bibinfo {title} {Complexity of quantum circuits via sensitivity,
  magic, and coherence},}\ }\href {https://arxiv.org/abs/2204.12051} {\bibfield
   {journal} {\bibinfo  {journal} {arXiv:2204.12051}\ } (\bibinfo {year}
  {2022})}\BibitemShut {NoStop}%
\bibitem [{\citenamefont {Haug}\ and\ \citenamefont
  {Kim}(2023)}]{haug2022scalable}%
  \BibitemOpen
  \bibfield  {author} {\bibinfo {author} {\bibfnamefont {Tobias}\ \bibnamefont
  {Haug}}\ and\ \bibinfo {author} {\bibfnamefont {M.S.}\ \bibnamefont {Kim}},\
  }\bibfield  {title} {\enquote {\bibinfo {title} {Scalable measures of magic
  resource for quantum computers},}\ }\href {\doibase
  10.1103/PRXQuantum.4.010301} {\bibfield  {journal} {\bibinfo  {journal} {PRX
  Quantum}\ }\textbf {\bibinfo {volume} {4}},\ \bibinfo {pages} {010301}
  (\bibinfo {year} {2023})}\BibitemShut {NoStop}%
\bibitem [{\citenamefont {Leone}\ \emph {et~al.}(2022)\citenamefont {Leone},
  \citenamefont {Oliviero},\ and\ \citenamefont {Hamma}}]{leone2022stabilizer}%
  \BibitemOpen
  \bibfield  {author} {\bibinfo {author} {\bibfnamefont {Lorenzo}\ \bibnamefont
  {Leone}}, \bibinfo {author} {\bibfnamefont {Salvatore F.~E.}\ \bibnamefont
  {Oliviero}}, \ and\ \bibinfo {author} {\bibfnamefont {Alioscia}\ \bibnamefont
  {Hamma}},\ }\bibfield  {title} {\enquote {\bibinfo {title} {Stabilizer
  r\'enyi entropy},}\ }\href {\doibase 10.1103/PhysRevLett.128.050402}
  {\bibfield  {journal} {\bibinfo  {journal} {Phys. Rev. Lett.}\ }\textbf
  {\bibinfo {volume} {128}},\ \bibinfo {pages} {050402} (\bibinfo {year}
  {2022})}\BibitemShut {NoStop}%
\bibitem [{\citenamefont {Haug}\ and\ \citenamefont
  {Piroli}(2023{\natexlab{a}})}]{haug2022quantifying}%
  \BibitemOpen
  \bibfield  {author} {\bibinfo {author} {\bibfnamefont {Tobias}\ \bibnamefont
  {Haug}}\ and\ \bibinfo {author} {\bibfnamefont {Lorenzo}\ \bibnamefont
  {Piroli}},\ }\bibfield  {title} {\enquote {\bibinfo {title} {Quantifying
  nonstabilizerness of matrix product states},}\ }\href {\doibase
  10.1103/PhysRevB.107.035148} {\bibfield  {journal} {\bibinfo  {journal}
  {Phys. Rev. B}\ }\textbf {\bibinfo {volume} {107}},\ \bibinfo {pages}
  {035148} (\bibinfo {year} {2023}{\natexlab{a}})}\BibitemShut {NoStop}%
\bibitem [{\citenamefont {Haug}\ and\ \citenamefont
  {Piroli}(2023{\natexlab{b}})}]{haug2023stabilizer}%
  \BibitemOpen
  \bibfield  {author} {\bibinfo {author} {\bibfnamefont {Tobias}\ \bibnamefont
  {Haug}}\ and\ \bibinfo {author} {\bibfnamefont {Lorenzo}\ \bibnamefont
  {Piroli}},\ }\bibfield  {title} {\enquote {\bibinfo {title} {Stabilizer
  entropies and nonstabilizerness monotones},}\ }\href@noop {} {\bibfield
  {journal} {\bibinfo  {journal} {Quantum}\ }\textbf {\bibinfo {volume} {7}},\
  \bibinfo {pages} {1092} (\bibinfo {year} {2023}{\natexlab{b}})}\BibitemShut
  {NoStop}%
\bibitem [{\citenamefont {Lami}\ and\ \citenamefont
  {Collura}(2023)}]{lami2023quantum}%
  \BibitemOpen
  \bibfield  {author} {\bibinfo {author} {\bibfnamefont {Guglielmo}\
  \bibnamefont {Lami}}\ and\ \bibinfo {author} {\bibfnamefont {Mario}\
  \bibnamefont {Collura}},\ }\bibfield  {title} {\enquote {\bibinfo {title}
  {Nonstabilizerness via perfect pauli sampling of matrix product states},}\
  }\href {\doibase 10.1103/PhysRevLett.131.180401} {\bibfield  {journal}
  {\bibinfo  {journal} {Phys. Rev. Lett.}\ }\textbf {\bibinfo {volume} {131}},\
  \bibinfo {pages} {180401} (\bibinfo {year} {2023})}\BibitemShut {NoStop}%
\bibitem [{\citenamefont {Oliviero}\ \emph
  {et~al.}(2022{\natexlab{a}})\citenamefont {Oliviero}, \citenamefont {Leone},\
  and\ \citenamefont {Hamma}}]{oliviero2022magic}%
  \BibitemOpen
  \bibfield  {author} {\bibinfo {author} {\bibfnamefont {Salvatore F.~E.}\
  \bibnamefont {Oliviero}}, \bibinfo {author} {\bibfnamefont {Lorenzo}\
  \bibnamefont {Leone}}, \ and\ \bibinfo {author} {\bibfnamefont {Alioscia}\
  \bibnamefont {Hamma}},\ }\bibfield  {title} {\enquote {\bibinfo {title}
  {Magic-state resource theory for the ground state of the transverse-field
  ising model},}\ }\href {\doibase 10.1103/PhysRevA.106.042426} {\bibfield
  {journal} {\bibinfo  {journal} {Phys. Rev. A}\ }\textbf {\bibinfo {volume}
  {106}},\ \bibinfo {pages} {042426} (\bibinfo {year}
  {2022}{\natexlab{a}})}\BibitemShut {NoStop}%
\bibitem [{\citenamefont {Odavi{\'c}}\ \emph {et~al.}(2023)\citenamefont
  {Odavi{\'c}}, \citenamefont {Haug}, \citenamefont {Torre}, \citenamefont
  {Hamma}, \citenamefont {Franchini},\ and\ \citenamefont
  {Giampaolo}}]{odavic2022complexity}%
  \BibitemOpen
  \bibfield  {author} {\bibinfo {author} {\bibfnamefont {Jovan}\ \bibnamefont
  {Odavi{\'c}}}, \bibinfo {author} {\bibfnamefont {Tobias}\ \bibnamefont
  {Haug}}, \bibinfo {author} {\bibfnamefont {Gianpaolo}\ \bibnamefont {Torre}},
  \bibinfo {author} {\bibfnamefont {Alioscia}\ \bibnamefont {Hamma}}, \bibinfo
  {author} {\bibfnamefont {Fabio}\ \bibnamefont {Franchini}}, \ and\ \bibinfo
  {author} {\bibfnamefont {Salvatore~Marco}\ \bibnamefont {Giampaolo}},\
  }\bibfield  {title} {\enquote {\bibinfo {title} {Complexity of frustration: a
  new source of non-local non-stabilizerness},}\ }\href@noop {} {\bibfield
  {journal} {\bibinfo  {journal} {SciPost Physics}\ }\textbf {\bibinfo {volume}
  {15}},\ \bibinfo {pages} {131} (\bibinfo {year} {2023})}\BibitemShut
  {NoStop}%
\bibitem [{\citenamefont {Chen}\ \emph {et~al.}(2022)\citenamefont {Chen},
  \citenamefont {Garcia}, \citenamefont {Bu},\ and\ \citenamefont
  {Jaffe}}]{chen2022magic}%
  \BibitemOpen
  \bibfield  {author} {\bibinfo {author} {\bibfnamefont {Liyuan}\ \bibnamefont
  {Chen}}, \bibinfo {author} {\bibfnamefont {Roy~J}\ \bibnamefont {Garcia}},
  \bibinfo {author} {\bibfnamefont {Kaifeng}\ \bibnamefont {Bu}}, \ and\
  \bibinfo {author} {\bibfnamefont {Arthur}\ \bibnamefont {Jaffe}},\ }\bibfield
   {title} {\enquote {\bibinfo {title} {Magic of random matrix product
  states},}\ }\href {https://arxiv.org/abs/2211.10350} {\bibfield  {journal}
  {\bibinfo  {journal} {arXiv:2211.10350}\ } (\bibinfo {year}
  {2022})}\BibitemShut {NoStop}%
\bibitem [{\citenamefont {Goto}\ \emph {et~al.}(2022)\citenamefont {Goto},
  \citenamefont {Nosaka},\ and\ \citenamefont {Nozaki}}]{goto2022probing}%
  \BibitemOpen
  \bibfield  {author} {\bibinfo {author} {\bibfnamefont {Kanato}\ \bibnamefont
  {Goto}}, \bibinfo {author} {\bibfnamefont {Tomoki}\ \bibnamefont {Nosaka}}, \
  and\ \bibinfo {author} {\bibfnamefont {Masahiro}\ \bibnamefont {Nozaki}},\
  }\bibfield  {title} {\enquote {\bibinfo {title} {Probing chaos by magic
  monotones},}\ }\href@noop {} {\bibfield  {journal} {\bibinfo  {journal}
  {Physical Review D}\ }\textbf {\bibinfo {volume} {106}},\ \bibinfo {pages}
  {126009} (\bibinfo {year} {2022})}\BibitemShut {NoStop}%
\bibitem [{\citenamefont {Piemontese}\ \emph {et~al.}(2023)\citenamefont
  {Piemontese}, \citenamefont {Roscilde},\ and\ \citenamefont
  {Hamma}}]{piemontese2023entanglement}%
  \BibitemOpen
  \bibfield  {author} {\bibinfo {author} {\bibfnamefont {Stefano}\ \bibnamefont
  {Piemontese}}, \bibinfo {author} {\bibfnamefont {Tommaso}\ \bibnamefont
  {Roscilde}}, \ and\ \bibinfo {author} {\bibfnamefont {Alioscia}\ \bibnamefont
  {Hamma}},\ }\bibfield  {title} {\enquote {\bibinfo {title} {Entanglement
  complexity of the rokhsar-kivelson-sign wavefunctions},}\ }\href@noop {}
  {\bibfield  {journal} {\bibinfo  {journal} {Physical Review B}\ }\textbf
  {\bibinfo {volume} {107}},\ \bibinfo {pages} {134202} (\bibinfo {year}
  {2023})}\BibitemShut {NoStop}%
\bibitem [{\citenamefont {Chen}\ \emph {et~al.}(2023)\citenamefont {Chen},
  \citenamefont {Yan},\ and\ \citenamefont {Zhou}}]{chen2023magic}%
  \BibitemOpen
  \bibfield  {author} {\bibinfo {author} {\bibfnamefont {Junjie}\ \bibnamefont
  {Chen}}, \bibinfo {author} {\bibfnamefont {Yuxuan}\ \bibnamefont {Yan}}, \
  and\ \bibinfo {author} {\bibfnamefont {You}\ \bibnamefont {Zhou}},\
  }\bibfield  {title} {\enquote {\bibinfo {title} {Magic of quantum hypergraph
  states},}\ }\href@noop {} {\bibfield  {journal} {\bibinfo  {journal}
  {arXiv:2308.01886}\ } (\bibinfo {year} {2023})}\BibitemShut {NoStop}%
\bibitem [{\citenamefont {Niroula}\ \emph {et~al.}(2023)\citenamefont
  {Niroula}, \citenamefont {White}, \citenamefont {Wang}, \citenamefont
  {Johri}, \citenamefont {Zhu}, \citenamefont {Monroe}, \citenamefont {Noel},\
  and\ \citenamefont {Gullans}}]{niroula2023phase}%
  \BibitemOpen
  \bibfield  {author} {\bibinfo {author} {\bibfnamefont {Pradeep}\ \bibnamefont
  {Niroula}}, \bibinfo {author} {\bibfnamefont {Christopher~David}\
  \bibnamefont {White}}, \bibinfo {author} {\bibfnamefont {Qingfeng}\
  \bibnamefont {Wang}}, \bibinfo {author} {\bibfnamefont {Sonika}\ \bibnamefont
  {Johri}}, \bibinfo {author} {\bibfnamefont {Daiwei}\ \bibnamefont {Zhu}},
  \bibinfo {author} {\bibfnamefont {Christopher}\ \bibnamefont {Monroe}},
  \bibinfo {author} {\bibfnamefont {Crystal}\ \bibnamefont {Noel}}, \ and\
  \bibinfo {author} {\bibfnamefont {Michael~J.}\ \bibnamefont {Gullans}},\
  }\bibfield  {title} {\enquote {\bibinfo {title} {Phase transition in magic
  with random quantum circuits},}\ }\href@noop {} {\bibfield  {journal}
  {\bibinfo  {journal} {arXiv:2304.10481}\ } (\bibinfo {year}
  {2023})}\BibitemShut {NoStop}%
\bibitem [{\citenamefont {Bejan}\ \emph {et~al.}(2023)\citenamefont {Bejan},
  \citenamefont {McLauchlan},\ and\ \citenamefont
  {Béri}}]{bejan2023dynamical}%
  \BibitemOpen
  \bibfield  {author} {\bibinfo {author} {\bibfnamefont {M.}~\bibnamefont
  {Bejan}}, \bibinfo {author} {\bibfnamefont {C.}~\bibnamefont {McLauchlan}}, \
  and\ \bibinfo {author} {\bibfnamefont {B.}~\bibnamefont {Béri}},\ }\bibfield
   {title} {\enquote {\bibinfo {title} {Dynamical magic transitions in
  monitored clifford+t circuits},}\ }\href@noop {} {\bibfield  {journal}
  {\bibinfo  {journal} {arXiv:2312.00132}\ } (\bibinfo {year}
  {2023})}\BibitemShut {NoStop}%
\bibitem [{\citenamefont {Fux}\ \emph {et~al.}(2023)\citenamefont {Fux},
  \citenamefont {Tirrito}, \citenamefont {Dalmonte},\ and\ \citenamefont
  {Fazio}}]{fux2023entanglementmagic}%
  \BibitemOpen
  \bibfield  {author} {\bibinfo {author} {\bibfnamefont {Gerald~E.}\
  \bibnamefont {Fux}}, \bibinfo {author} {\bibfnamefont {Emanuele}\
  \bibnamefont {Tirrito}}, \bibinfo {author} {\bibfnamefont {Marcello}\
  \bibnamefont {Dalmonte}}, \ and\ \bibinfo {author} {\bibfnamefont {Rosario}\
  \bibnamefont {Fazio}},\ }\bibfield  {title} {\enquote {\bibinfo {title}
  {Entanglement-magic separation in hybrid quantum circuits},}\ }\href@noop {}
  {\bibfield  {journal} {\bibinfo  {journal} {arXiv:2312.02039}\ } (\bibinfo
  {year} {2023})}\BibitemShut {NoStop}%
\bibitem [{\citenamefont {Tirrito}\ \emph {et~al.}(2023)\citenamefont
  {Tirrito}, \citenamefont {Tarabunga}, \citenamefont {Lami}, \citenamefont
  {Chanda}, \citenamefont {Leone}, \citenamefont {Oliviero}, \citenamefont
  {Dalmonte}, \citenamefont {Collura},\ and\ \citenamefont
  {Hamma}}]{tirrito2023quantifying}%
  \BibitemOpen
  \bibfield  {author} {\bibinfo {author} {\bibfnamefont {Emanuele}\
  \bibnamefont {Tirrito}}, \bibinfo {author} {\bibfnamefont {Poetri~Sonya}\
  \bibnamefont {Tarabunga}}, \bibinfo {author} {\bibfnamefont {Gugliemo}\
  \bibnamefont {Lami}}, \bibinfo {author} {\bibfnamefont {Titas}\ \bibnamefont
  {Chanda}}, \bibinfo {author} {\bibfnamefont {Lorenzo}\ \bibnamefont {Leone}},
  \bibinfo {author} {\bibfnamefont {Salvatore F.~E.}\ \bibnamefont {Oliviero}},
  \bibinfo {author} {\bibfnamefont {Marcello}\ \bibnamefont {Dalmonte}},
  \bibinfo {author} {\bibfnamefont {Mario}\ \bibnamefont {Collura}}, \ and\
  \bibinfo {author} {\bibfnamefont {Alioscia}\ \bibnamefont {Hamma}},\
  }\bibfield  {title} {\enquote {\bibinfo {title} {Quantifying
  non-stabilizerness through entanglement spectrum flatness},}\ }\href@noop {}
  {\bibfield  {journal} {\bibinfo  {journal} {arXiv:2304.01175}\ } (\bibinfo
  {year} {2023})}\BibitemShut {NoStop}%
\bibitem [{\citenamefont {Leone}\ \emph
  {et~al.}(2023{\natexlab{a}})\citenamefont {Leone}, \citenamefont {Oliviero},
  \citenamefont {Esposito},\ and\ \citenamefont {Hamma}}]{leone2023phase}%
  \BibitemOpen
  \bibfield  {author} {\bibinfo {author} {\bibfnamefont {Lorenzo}\ \bibnamefont
  {Leone}}, \bibinfo {author} {\bibfnamefont {Salvatore F.~E.}\ \bibnamefont
  {Oliviero}}, \bibinfo {author} {\bibfnamefont {Gianluca}\ \bibnamefont
  {Esposito}}, \ and\ \bibinfo {author} {\bibfnamefont {Alioscia}\ \bibnamefont
  {Hamma}},\ }\bibfield  {title} {\enquote {\bibinfo {title} {Phase transition
  in stabilizer entropy and efficient purity estimation},}\ }\href@noop {}
  {\bibfield  {journal} {\bibinfo  {journal} {arXiv:2302.07895}\ } (\bibinfo
  {year} {2023}{\natexlab{a}})}\BibitemShut {NoStop}%
\bibitem [{\citenamefont {Leone}\ \emph
  {et~al.}(2023{\natexlab{b}})\citenamefont {Leone}, \citenamefont {Oliviero},\
  and\ \citenamefont {Hamma}}]{leone2023nonstabilizerness}%
  \BibitemOpen
  \bibfield  {author} {\bibinfo {author} {\bibfnamefont {Lorenzo}\ \bibnamefont
  {Leone}}, \bibinfo {author} {\bibfnamefont {Salvatore F.~E.}\ \bibnamefont
  {Oliviero}}, \ and\ \bibinfo {author} {\bibfnamefont {Alioscia}\ \bibnamefont
  {Hamma}},\ }\bibfield  {title} {\enquote {\bibinfo {title} {Nonstabilizerness
  determining the hardness of direct fidelity estimation},}\ }\href {\doibase
  10.1103/PhysRevA.107.022429} {\bibfield  {journal} {\bibinfo  {journal}
  {Phys. Rev. A}\ }\textbf {\bibinfo {volume} {107}},\ \bibinfo {pages}
  {022429} (\bibinfo {year} {2023}{\natexlab{b}})}\BibitemShut {NoStop}%
\bibitem [{\citenamefont {Turkeshi}\ \emph {et~al.}(2023)\citenamefont
  {Turkeshi}, \citenamefont {Schir\`o},\ and\ \citenamefont
  {Sierant}}]{turkeshi2023measuring}%
  \BibitemOpen
  \bibfield  {author} {\bibinfo {author} {\bibfnamefont {Xhek}\ \bibnamefont
  {Turkeshi}}, \bibinfo {author} {\bibfnamefont {Marco}\ \bibnamefont
  {Schir\`o}}, \ and\ \bibinfo {author} {\bibfnamefont {Piotr}\ \bibnamefont
  {Sierant}},\ }\bibfield  {title} {\enquote {\bibinfo {title} {Measuring
  nonstabilizerness via multifractal flatness},}\ }\href {\doibase
  10.1103/PhysRevA.108.042408} {\bibfield  {journal} {\bibinfo  {journal}
  {Phys. Rev. A}\ }\textbf {\bibinfo {volume} {108}},\ \bibinfo {pages}
  {042408} (\bibinfo {year} {2023})}\BibitemShut {NoStop}%
\bibitem [{\citenamefont {Castellani}\ and\ \citenamefont
  {Peliti}(1986)}]{castellani1986multifractal}%
  \BibitemOpen
  \bibfield  {author} {\bibinfo {author} {\bibfnamefont {C}~\bibnamefont
  {Castellani}}\ and\ \bibinfo {author} {\bibfnamefont {L}~\bibnamefont
  {Peliti}},\ }\bibfield  {title} {\enquote {\bibinfo {title} {Multifractal
  wavefunction at the localisation threshold},}\ }\href@noop {} {\bibfield
  {journal} {\bibinfo  {journal} {Journal of physics A: mathematical and
  general}\ }\textbf {\bibinfo {volume} {19}},\ \bibinfo {pages} {L429}
  (\bibinfo {year} {1986})}\BibitemShut {NoStop}%
\bibitem [{\citenamefont {St{\'e}phan}\ \emph {et~al.}(2009)\citenamefont
  {St{\'e}phan}, \citenamefont {Furukawa}, \citenamefont {Misguich},\ and\
  \citenamefont {Pasquier}}]{stephan2009shannon}%
  \BibitemOpen
  \bibfield  {author} {\bibinfo {author} {\bibfnamefont {Jean-Marie}\
  \bibnamefont {St{\'e}phan}}, \bibinfo {author} {\bibfnamefont {Shunsuke}\
  \bibnamefont {Furukawa}}, \bibinfo {author} {\bibfnamefont {Gr{\'e}goire}\
  \bibnamefont {Misguich}}, \ and\ \bibinfo {author} {\bibfnamefont {Vincent}\
  \bibnamefont {Pasquier}},\ }\bibfield  {title} {\enquote {\bibinfo {title}
  {Shannon and entanglement entropies of one-and two-dimensional critical wave
  functions},}\ }\href@noop {} {\bibfield  {journal} {\bibinfo  {journal}
  {Physical Review B}\ }\textbf {\bibinfo {volume} {80}},\ \bibinfo {pages}
  {184421} (\bibinfo {year} {2009})}\BibitemShut {NoStop}%
\bibitem [{\citenamefont {Leone}\ \emph
  {et~al.}(2021{\natexlab{a}})\citenamefont {Leone}, \citenamefont {Oliviero},
  \citenamefont {Zhou},\ and\ \citenamefont {Hamma}}]{leone2021quantum}%
  \BibitemOpen
  \bibfield  {author} {\bibinfo {author} {\bibfnamefont {Lorenzo}\ \bibnamefont
  {Leone}}, \bibinfo {author} {\bibfnamefont {Salvatore F.~E.}\ \bibnamefont
  {Oliviero}}, \bibinfo {author} {\bibfnamefont {You}\ \bibnamefont {Zhou}}, \
  and\ \bibinfo {author} {\bibfnamefont {Alioscia}\ \bibnamefont {Hamma}},\
  }\bibfield  {title} {\enquote {\bibinfo {title} {Quantum chaos is quantum},}\
  }\href {https://arxiv.org/abs/2102.08406} {\bibfield  {journal} {\bibinfo
  {journal} {Quantum}\ }\textbf {\bibinfo {volume} {5}},\ \bibinfo {pages}
  {453} (\bibinfo {year} {2021}{\natexlab{a}})}\BibitemShut {NoStop}%
\bibitem [{\citenamefont {Garcia}\ \emph {et~al.}(2023)\citenamefont {Garcia},
  \citenamefont {Bu},\ and\ \citenamefont {Jaffe}}]{garcia2022resource}%
  \BibitemOpen
  \bibfield  {author} {\bibinfo {author} {\bibfnamefont {Roy~J}\ \bibnamefont
  {Garcia}}, \bibinfo {author} {\bibfnamefont {Kaifeng}\ \bibnamefont {Bu}}, \
  and\ \bibinfo {author} {\bibfnamefont {Arthur}\ \bibnamefont {Jaffe}},\
  }\bibfield  {title} {\enquote {\bibinfo {title} {Resource theory of quantum
  scrambling},}\ }\href@noop {} {\bibfield  {journal} {\bibinfo  {journal}
  {Proceedings of the National Academy of Sciences}\ }\textbf {\bibinfo
  {volume} {120}},\ \bibinfo {pages} {e2217031120} (\bibinfo {year}
  {2023})}\BibitemShut {NoStop}%
\bibitem [{\citenamefont {Xu}\ and\ \citenamefont
  {Swingle}(2022)}]{xu2022scrambling}%
  \BibitemOpen
  \bibfield  {author} {\bibinfo {author} {\bibfnamefont {Shenglong}\
  \bibnamefont {Xu}}\ and\ \bibinfo {author} {\bibfnamefont {Brian}\
  \bibnamefont {Swingle}},\ }\bibfield  {title} {\enquote {\bibinfo {title}
  {Scrambling dynamics and out-of-time ordered correlators in quantum many-body
  systems: a tutorial},}\ }\href@noop {} {\bibfield  {journal} {\bibinfo
  {journal} {arXiv:2202.07060}\ } (\bibinfo {year} {2022})}\BibitemShut
  {NoStop}%
\bibitem [{\citenamefont {Dowling}\ \emph {et~al.}(2023)\citenamefont
  {Dowling}, \citenamefont {Kos},\ and\ \citenamefont
  {Modi}}]{dowling2023scrambling}%
  \BibitemOpen
  \bibfield  {author} {\bibinfo {author} {\bibfnamefont {Neil}\ \bibnamefont
  {Dowling}}, \bibinfo {author} {\bibfnamefont {Pavel}\ \bibnamefont {Kos}}, \
  and\ \bibinfo {author} {\bibfnamefont {Kavan}\ \bibnamefont {Modi}},\
  }\bibfield  {title} {\enquote {\bibinfo {title} {Scrambling is necessary but
  not sufficient for chaos},}\ }\href@noop {} {\bibfield  {journal} {\bibinfo
  {journal} {Physical Review Letters}\ }\textbf {\bibinfo {volume} {131}},\
  \bibinfo {pages} {180403} (\bibinfo {year} {2023})}\BibitemShut {NoStop}%
\bibitem [{\citenamefont {Li}\ \emph {et~al.}(2017)\citenamefont {Li},
  \citenamefont {Fan}, \citenamefont {Wang}, \citenamefont {Ye}, \citenamefont
  {Zeng}, \citenamefont {Zhai}, \citenamefont {Peng},\ and\ \citenamefont
  {Du}}]{li2017measuring}%
  \BibitemOpen
  \bibfield  {author} {\bibinfo {author} {\bibfnamefont {Jun}\ \bibnamefont
  {Li}}, \bibinfo {author} {\bibfnamefont {Ruihua}\ \bibnamefont {Fan}},
  \bibinfo {author} {\bibfnamefont {Hengyan}\ \bibnamefont {Wang}}, \bibinfo
  {author} {\bibfnamefont {Bingtian}\ \bibnamefont {Ye}}, \bibinfo {author}
  {\bibfnamefont {Bei}\ \bibnamefont {Zeng}}, \bibinfo {author} {\bibfnamefont
  {Hui}\ \bibnamefont {Zhai}}, \bibinfo {author} {\bibfnamefont {Xinhua}\
  \bibnamefont {Peng}}, \ and\ \bibinfo {author} {\bibfnamefont {Jiangfeng}\
  \bibnamefont {Du}},\ }\bibfield  {title} {\enquote {\bibinfo {title}
  {Measuring out-of-time-order correlators on a nuclear magnetic resonance
  quantum simulator},}\ }\href@noop {} {\bibfield  {journal} {\bibinfo
  {journal} {Physical Review X}\ }\textbf {\bibinfo {volume} {7}},\ \bibinfo
  {pages} {031011} (\bibinfo {year} {2017})}\BibitemShut {NoStop}%
\bibitem [{\citenamefont {Oliviero}\ \emph
  {et~al.}(2022{\natexlab{b}})\citenamefont {Oliviero}, \citenamefont {Leone},
  \citenamefont {Hamma},\ and\ \citenamefont {Lloyd}}]{oliviero2022measuring}%
  \BibitemOpen
  \bibfield  {author} {\bibinfo {author} {\bibfnamefont {Salvatore F.~E.}\
  \bibnamefont {Oliviero}}, \bibinfo {author} {\bibfnamefont {Lorenzo}\
  \bibnamefont {Leone}}, \bibinfo {author} {\bibfnamefont {Alioscia}\
  \bibnamefont {Hamma}}, \ and\ \bibinfo {author} {\bibfnamefont {Seth}\
  \bibnamefont {Lloyd}},\ }\bibfield  {title} {\enquote {\bibinfo {title}
  {Measuring magic on a quantum processor},}\ }\href {\doibase
  10.1038/s41534-022-00666-5} {\bibfield  {journal} {\bibinfo  {journal} {npj
  Quantum Information}\ }\textbf {\bibinfo {volume} {8}},\ \bibinfo {pages}
  {148} (\bibinfo {year} {2022}{\natexlab{b}})}\BibitemShut {NoStop}%
\bibitem [{\citenamefont {Gross}\ \emph {et~al.}(2021)\citenamefont {Gross},
  \citenamefont {Nezami},\ and\ \citenamefont {Walter}}]{gross2021schur}%
  \BibitemOpen
  \bibfield  {author} {\bibinfo {author} {\bibfnamefont {David}\ \bibnamefont
  {Gross}}, \bibinfo {author} {\bibfnamefont {Sepehr}\ \bibnamefont {Nezami}},
  \ and\ \bibinfo {author} {\bibfnamefont {Michael}\ \bibnamefont {Walter}},\
  }\bibfield  {title} {\enquote {\bibinfo {title} {Schur--weyl duality for the
  clifford group with applications: Property testing, a robust hudson theorem,
  and de finetti representations},}\ }\href@noop {} {\bibfield  {journal}
  {\bibinfo  {journal} {Communications in Mathematical Physics}\ }\textbf
  {\bibinfo {volume} {385}},\ \bibinfo {pages} {1325--1393} (\bibinfo {year}
  {2021})}\BibitemShut {NoStop}%
\bibitem [{\citenamefont {Montanaro}(2017)}]{montanaro2017learning}%
  \BibitemOpen
  \bibfield  {author} {\bibinfo {author} {\bibfnamefont {Ashley}\ \bibnamefont
  {Montanaro}},\ }\bibfield  {title} {\enquote {\bibinfo {title} {Learning
  stabilizer states by bell sampling},}\ }\href@noop {} {\bibfield  {journal}
  {\bibinfo  {journal} {arXiv:1707.04012}\ } (\bibinfo {year}
  {2017})}\BibitemShut {NoStop}%
\bibitem [{\citenamefont {Lai}\ and\ \citenamefont
  {Cheng}(2022)}]{lai2022learning}%
  \BibitemOpen
  \bibfield  {author} {\bibinfo {author} {\bibfnamefont {Ching-Yi}\
  \bibnamefont {Lai}}\ and\ \bibinfo {author} {\bibfnamefont {Hao-Chung}\
  \bibnamefont {Cheng}},\ }\bibfield  {title} {\enquote {\bibinfo {title}
  {Learning quantum circuits of some t gates},}\ }\href@noop {} {\bibfield
  {journal} {\bibinfo  {journal} {IEEE Transactions on Information Theory}\
  }\textbf {\bibinfo {volume} {68}},\ \bibinfo {pages} {3951--3964} (\bibinfo
  {year} {2022})}\BibitemShut {NoStop}%
\bibitem [{\citenamefont {Yang}\ \emph {et~al.}(2014)\citenamefont {Yang},
  \citenamefont {Chiribella},\ and\ \citenamefont
  {Adesso}}]{yang2014certifying}%
  \BibitemOpen
  \bibfield  {author} {\bibinfo {author} {\bibfnamefont {Yuxiang}\ \bibnamefont
  {Yang}}, \bibinfo {author} {\bibfnamefont {Giulio}\ \bibnamefont
  {Chiribella}}, \ and\ \bibinfo {author} {\bibfnamefont {Gerardo}\
  \bibnamefont {Adesso}},\ }\bibfield  {title} {\enquote {\bibinfo {title}
  {Certifying quantumness: Benchmarks for the optimal processing of generalized
  coherent and squeezed states},}\ }\href@noop {} {\bibfield  {journal}
  {\bibinfo  {journal} {Physical Review A}\ }\textbf {\bibinfo {volume} {90}},\
  \bibinfo {pages} {042319} (\bibinfo {year} {2014})}\BibitemShut {NoStop}%
\bibitem [{\citenamefont {Miyazaki}\ \emph {et~al.}(2019)\citenamefont
  {Miyazaki}, \citenamefont {Soeda},\ and\ \citenamefont
  {Murao}}]{miyazaki2019complex}%
  \BibitemOpen
  \bibfield  {author} {\bibinfo {author} {\bibfnamefont {Jisho}\ \bibnamefont
  {Miyazaki}}, \bibinfo {author} {\bibfnamefont {Akihito}\ \bibnamefont
  {Soeda}}, \ and\ \bibinfo {author} {\bibfnamefont {Mio}\ \bibnamefont
  {Murao}},\ }\bibfield  {title} {\enquote {\bibinfo {title} {Complex
  conjugation supermap of unitary quantum maps and its universal implementation
  protocol},}\ }\href@noop {} {\bibfield  {journal} {\bibinfo  {journal}
  {Physical Review Research}\ }\textbf {\bibinfo {volume} {1}},\ \bibinfo
  {pages} {013007} (\bibinfo {year} {2019})}\BibitemShut {NoStop}%
\bibitem [{\citenamefont {Haug}\ \emph {et~al.}(2023)\citenamefont {Haug},
  \citenamefont {Bharti},\ and\ \citenamefont {Koh}}]{haug2023pseudorandom}%
  \BibitemOpen
  \bibfield  {author} {\bibinfo {author} {\bibfnamefont {Tobias}\ \bibnamefont
  {Haug}}, \bibinfo {author} {\bibfnamefont {Kishor}\ \bibnamefont {Bharti}}, \
  and\ \bibinfo {author} {\bibfnamefont {Dax~Enshan}\ \bibnamefont {Koh}},\
  }\bibfield  {title} {\enquote {\bibinfo {title} {Pseudorandom unitaries are
  neither real nor sparse nor noise-robust},}\ }\href@noop {} {\bibfield
  {journal} {\bibinfo  {journal} {arXiv:2306.11677}\ } (\bibinfo {year}
  {2023})}\BibitemShut {NoStop}%
\bibitem [{\citenamefont {Khatri}\ \emph {et~al.}(2019)\citenamefont {Khatri},
  \citenamefont {LaRose}, \citenamefont {Poremba}, \citenamefont {Cincio},
  \citenamefont {Sornborger},\ and\ \citenamefont {Coles}}]{khatri2019quantum}%
  \BibitemOpen
  \bibfield  {author} {\bibinfo {author} {\bibfnamefont {Sumeet}\ \bibnamefont
  {Khatri}}, \bibinfo {author} {\bibfnamefont {Ryan}\ \bibnamefont {LaRose}},
  \bibinfo {author} {\bibfnamefont {Alexander}\ \bibnamefont {Poremba}},
  \bibinfo {author} {\bibfnamefont {Lukasz}\ \bibnamefont {Cincio}}, \bibinfo
  {author} {\bibfnamefont {Andrew~T}\ \bibnamefont {Sornborger}}, \ and\
  \bibinfo {author} {\bibfnamefont {Patrick~J}\ \bibnamefont {Coles}},\
  }\bibfield  {title} {\enquote {\bibinfo {title} {Quantum-assisted quantum
  compiling},}\ }\href@noop {} {\bibfield  {journal} {\bibinfo  {journal}
  {Quantum}\ }\textbf {\bibinfo {volume} {3}},\ \bibinfo {pages} {140}
  (\bibinfo {year} {2019})}\BibitemShut {NoStop}%
\bibitem [{\citenamefont {Tsallis}(1988)}]{tsallis1988possible}%
  \BibitemOpen
  \bibfield  {author} {\bibinfo {author} {\bibfnamefont {Constantino}\
  \bibnamefont {Tsallis}},\ }\bibfield  {title} {\enquote {\bibinfo {title}
  {Possible generalization of boltzmann-gibbs statistics},}\ }\href@noop {}
  {\bibfield  {journal} {\bibinfo  {journal} {Journal of statistical physics}\
  }\textbf {\bibinfo {volume} {52}},\ \bibinfo {pages} {479--487} (\bibinfo
  {year} {1988})}\BibitemShut {NoStop}%
\bibitem [{\citenamefont {Chitambar}\ and\ \citenamefont
  {Gour}(2019)}]{chitambar2019quantum}%
  \BibitemOpen
  \bibfield  {author} {\bibinfo {author} {\bibfnamefont {Eric}\ \bibnamefont
  {Chitambar}}\ and\ \bibinfo {author} {\bibfnamefont {Gilad}\ \bibnamefont
  {Gour}},\ }\bibfield  {title} {\enquote {\bibinfo {title} {Quantum resource
  theories},}\ }\href {\doibase 10.1103/RevModPhys.91.025001} {\bibfield
  {journal} {\bibinfo  {journal} {Rev. Mod. Phys.}\ }\textbf {\bibinfo {volume}
  {91}},\ \bibinfo {pages} {025001} (\bibinfo {year} {2019})}\BibitemShut
  {NoStop}%
\bibitem [{\citenamefont {Sierant}\ and\ \citenamefont
  {Turkeshi}(2022)}]{sierant2022universal}%
  \BibitemOpen
  \bibfield  {author} {\bibinfo {author} {\bibfnamefont {Piotr}\ \bibnamefont
  {Sierant}}\ and\ \bibinfo {author} {\bibfnamefont {Xhek}\ \bibnamefont
  {Turkeshi}},\ }\bibfield  {title} {\enquote {\bibinfo {title} {Universal
  behavior beyond multifractality of wave functions at measurement-induced
  phase transitions},}\ }\href@noop {} {\bibfield  {journal} {\bibinfo
  {journal} {Physical Review Letters}\ }\textbf {\bibinfo {volume} {128}},\
  \bibinfo {pages} {130605} (\bibinfo {year} {2022})}\BibitemShut {NoStop}%
\bibitem [{\citenamefont {Bravyi}\ \emph {et~al.}(2019)\citenamefont {Bravyi},
  \citenamefont {Browne}, \citenamefont {Calpin}, \citenamefont {Campbell},
  \citenamefont {Gosset},\ and\ \citenamefont {Howard}}]{bravyi2019simulation}%
  \BibitemOpen
  \bibfield  {author} {\bibinfo {author} {\bibfnamefont {Sergey}\ \bibnamefont
  {Bravyi}}, \bibinfo {author} {\bibfnamefont {Dan}\ \bibnamefont {Browne}},
  \bibinfo {author} {\bibfnamefont {Padraic}\ \bibnamefont {Calpin}}, \bibinfo
  {author} {\bibfnamefont {Earl}\ \bibnamefont {Campbell}}, \bibinfo {author}
  {\bibfnamefont {David}\ \bibnamefont {Gosset}}, \ and\ \bibinfo {author}
  {\bibfnamefont {Mark}\ \bibnamefont {Howard}},\ }\bibfield  {title} {\enquote
  {\bibinfo {title} {Simulation of quantum circuits by low-rank stabilizer
  decompositions},}\ }\href {\doibase 10.22331/q-2019-09-02-181} {\bibfield
  {journal} {\bibinfo  {journal} {Quantum}\ }\textbf {\bibinfo {volume} {3}},\
  \bibinfo {pages} {181} (\bibinfo {year} {2019})}\BibitemShut {NoStop}%
\bibitem [{\citenamefont {Aaronson}\ and\ \citenamefont
  {Gottesman}(2004)}]{aaronson2004improved}%
  \BibitemOpen
  \bibfield  {author} {\bibinfo {author} {\bibfnamefont {Scott}\ \bibnamefont
  {Aaronson}}\ and\ \bibinfo {author} {\bibfnamefont {Daniel}\ \bibnamefont
  {Gottesman}},\ }\bibfield  {title} {\enquote {\bibinfo {title} {Improved
  simulation of stabilizer circuits},}\ }\href {\doibase
  10.1103/PhysRevA.70.052328} {\bibfield  {journal} {\bibinfo  {journal} {Phys.
  Rev. A}\ }\textbf {\bibinfo {volume} {70}},\ \bibinfo {pages} {052328}
  (\bibinfo {year} {2004})}\BibitemShut {NoStop}%
\bibitem [{\citenamefont {Grewal}\ \emph {et~al.}(2023)\citenamefont {Grewal},
  \citenamefont {Iyer}, \citenamefont {Kretschmer},\ and\ \citenamefont
  {Liang}}]{grewal2023improved}%
  \BibitemOpen
  \bibfield  {author} {\bibinfo {author} {\bibfnamefont {Sabee}\ \bibnamefont
  {Grewal}}, \bibinfo {author} {\bibfnamefont {Vishnu}\ \bibnamefont {Iyer}},
  \bibinfo {author} {\bibfnamefont {William}\ \bibnamefont {Kretschmer}}, \
  and\ \bibinfo {author} {\bibfnamefont {Daniel}\ \bibnamefont {Liang}},\
  }\bibfield  {title} {\enquote {\bibinfo {title} {Improved stabilizer
  estimation via bell difference sampling},}\ }\href@noop {} {\bibfield
  {journal} {\bibinfo  {journal} {arXiv:2304.13915}\ } (\bibinfo {year}
  {2023})}\BibitemShut {NoStop}%
\bibitem [{\citenamefont {Haferkamp}(2022)}]{haferkamp2022random}%
  \BibitemOpen
  \bibfield  {author} {\bibinfo {author} {\bibfnamefont {Jonas}\ \bibnamefont
  {Haferkamp}},\ }\bibfield  {title} {\enquote {\bibinfo {title} {Random
  quantum circuits are approximate unitary $ t $-designs in depth $o(nt^{5+
  o(1)})$},}\ }\href@noop {} {\bibfield  {journal} {\bibinfo  {journal}
  {Quantum}\ }\textbf {\bibinfo {volume} {6}},\ \bibinfo {pages} {795}
  (\bibinfo {year} {2022})}\BibitemShut {NoStop}%
\bibitem [{\citenamefont {Cotler}\ \emph {et~al.}(2017)\citenamefont {Cotler},
  \citenamefont {Hunter-Jones}, \citenamefont {Liu},\ and\ \citenamefont
  {Yoshida}}]{cotler2017chaos}%
  \BibitemOpen
  \bibfield  {author} {\bibinfo {author} {\bibfnamefont {Jordan}\ \bibnamefont
  {Cotler}}, \bibinfo {author} {\bibfnamefont {Nicholas}\ \bibnamefont
  {Hunter-Jones}}, \bibinfo {author} {\bibfnamefont {Junyu}\ \bibnamefont
  {Liu}}, \ and\ \bibinfo {author} {\bibfnamefont {Beni}\ \bibnamefont
  {Yoshida}},\ }\bibfield  {title} {\enquote {\bibinfo {title} {Chaos,
  complexity, and random matrices},}\ }\href@noop {} {\bibfield  {journal}
  {\bibinfo  {journal} {Journal of High Energy Physics}\ }\textbf {\bibinfo
  {volume} {2017}},\ \bibinfo {pages} {1--60} (\bibinfo {year}
  {2017})}\BibitemShut {NoStop}%
\bibitem [{\citenamefont {Islam}\ \emph {et~al.}(2015)\citenamefont {Islam},
  \citenamefont {Ma}, \citenamefont {Preiss}, \citenamefont {Tai},
  \citenamefont {Lukin}, \citenamefont {Rispoli},\ and\ \citenamefont
  {Greiner}}]{islam2015measuring}%
  \BibitemOpen
  \bibfield  {author} {\bibinfo {author} {\bibfnamefont {Rajibul}\ \bibnamefont
  {Islam}}, \bibinfo {author} {\bibfnamefont {Ruichao}\ \bibnamefont {Ma}},
  \bibinfo {author} {\bibfnamefont {Philipp~M}\ \bibnamefont {Preiss}},
  \bibinfo {author} {\bibfnamefont {M~Eric}\ \bibnamefont {Tai}}, \bibinfo
  {author} {\bibfnamefont {Alexander}\ \bibnamefont {Lukin}}, \bibinfo {author}
  {\bibfnamefont {Matthew}\ \bibnamefont {Rispoli}}, \ and\ \bibinfo {author}
  {\bibfnamefont {Markus}\ \bibnamefont {Greiner}},\ }\bibfield  {title}
  {\enquote {\bibinfo {title} {Measuring entanglement entropy in a quantum
  many-body system},}\ }\href@noop {} {\bibfield  {journal} {\bibinfo
  {journal} {Nature}\ }\textbf {\bibinfo {volume} {528}},\ \bibinfo {pages}
  {77--83} (\bibinfo {year} {2015})}\BibitemShut {NoStop}%
\bibitem [{\citenamefont {Huang}\ \emph {et~al.}(2022)\citenamefont {Huang},
  \citenamefont {Broughton}, \citenamefont {Cotler}, \citenamefont {Chen},
  \citenamefont {Li}, \citenamefont {Mohseni}, \citenamefont {Neven},
  \citenamefont {Babbush}, \citenamefont {Kueng}, \citenamefont {Preskill}
  \emph {et~al.}}]{huang2021demonstrating}%
  \BibitemOpen
  \bibfield  {author} {\bibinfo {author} {\bibfnamefont {Hsin-Yuan}\
  \bibnamefont {Huang}}, \bibinfo {author} {\bibfnamefont {Michael}\
  \bibnamefont {Broughton}}, \bibinfo {author} {\bibfnamefont {Jordan}\
  \bibnamefont {Cotler}}, \bibinfo {author} {\bibfnamefont {Sitan}\
  \bibnamefont {Chen}}, \bibinfo {author} {\bibfnamefont {Jerry}\ \bibnamefont
  {Li}}, \bibinfo {author} {\bibfnamefont {Masoud}\ \bibnamefont {Mohseni}},
  \bibinfo {author} {\bibfnamefont {Hartmut}\ \bibnamefont {Neven}}, \bibinfo
  {author} {\bibfnamefont {Ryan}\ \bibnamefont {Babbush}}, \bibinfo {author}
  {\bibfnamefont {Richard}\ \bibnamefont {Kueng}}, \bibinfo {author}
  {\bibfnamefont {John}\ \bibnamefont {Preskill}},  \emph {et~al.},\ }\bibfield
   {title} {\enquote {\bibinfo {title} {Quantum advantage in learning from
  experiments},}\ }\href@noop {} {\bibfield  {journal} {\bibinfo  {journal}
  {Science}\ }\textbf {\bibinfo {volume} {376}},\ \bibinfo {pages} {1182--1186}
  (\bibinfo {year} {2022})}\BibitemShut {NoStop}%
\bibitem [{\citenamefont {Bluvstein}\ \emph {et~al.}(2022)\citenamefont
  {Bluvstein}, \citenamefont {Levine}, \citenamefont {Semeghini}, \citenamefont
  {Wang}, \citenamefont {Ebadi}, \citenamefont {Kalinowski}, \citenamefont
  {Keesling}, \citenamefont {Maskara}, \citenamefont {Pichler}, \citenamefont
  {Greiner} \emph {et~al.}}]{bluvstein2022quantum}%
  \BibitemOpen
  \bibfield  {author} {\bibinfo {author} {\bibfnamefont {Dolev}\ \bibnamefont
  {Bluvstein}}, \bibinfo {author} {\bibfnamefont {Harry}\ \bibnamefont
  {Levine}}, \bibinfo {author} {\bibfnamefont {Giulia}\ \bibnamefont
  {Semeghini}}, \bibinfo {author} {\bibfnamefont {Tout~T}\ \bibnamefont
  {Wang}}, \bibinfo {author} {\bibfnamefont {Sepehr}\ \bibnamefont {Ebadi}},
  \bibinfo {author} {\bibfnamefont {Marcin}\ \bibnamefont {Kalinowski}},
  \bibinfo {author} {\bibfnamefont {Alexander}\ \bibnamefont {Keesling}},
  \bibinfo {author} {\bibfnamefont {Nishad}\ \bibnamefont {Maskara}}, \bibinfo
  {author} {\bibfnamefont {Hannes}\ \bibnamefont {Pichler}}, \bibinfo {author}
  {\bibfnamefont {Markus}\ \bibnamefont {Greiner}},  \emph {et~al.},\
  }\bibfield  {title} {\enquote {\bibinfo {title} {A quantum processor based on
  coherent transport of entangled atom arrays},}\ }\href@noop {} {\bibfield
  {journal} {\bibinfo  {journal} {Nature}\ }\textbf {\bibinfo {volume} {604}},\
  \bibinfo {pages} {451--456} (\bibinfo {year} {2022})}\BibitemShut {NoStop}%
\bibitem [{\citenamefont {Roberts}\ and\ \citenamefont
  {Yoshida}(2017)}]{roberts2017chaos}%
  \BibitemOpen
  \bibfield  {author} {\bibinfo {author} {\bibfnamefont {Daniel~A}\
  \bibnamefont {Roberts}}\ and\ \bibinfo {author} {\bibfnamefont {Beni}\
  \bibnamefont {Yoshida}},\ }\bibfield  {title} {\enquote {\bibinfo {title}
  {Chaos and complexity by design},}\ }\href@noop {} {\bibfield  {journal}
  {\bibinfo  {journal} {Journal of High Energy Physics}\ }\textbf {\bibinfo
  {volume} {2017}},\ \bibinfo {pages} {1--64} (\bibinfo {year}
  {2017})}\BibitemShut {NoStop}%
\bibitem [{\citenamefont {Garcia}\ \emph {et~al.}(2021)\citenamefont {Garcia},
  \citenamefont {Zhou},\ and\ \citenamefont {Jaffe}}]{garcia2021quantum}%
  \BibitemOpen
  \bibfield  {author} {\bibinfo {author} {\bibfnamefont {Roy~J}\ \bibnamefont
  {Garcia}}, \bibinfo {author} {\bibfnamefont {You}\ \bibnamefont {Zhou}}, \
  and\ \bibinfo {author} {\bibfnamefont {Arthur}\ \bibnamefont {Jaffe}},\
  }\bibfield  {title} {\enquote {\bibinfo {title} {Quantum scrambling with
  classical shadows},}\ }\href@noop {} {\bibfield  {journal} {\bibinfo
  {journal} {Physical Review Research}\ }\textbf {\bibinfo {volume} {3}},\
  \bibinfo {pages} {033155} (\bibinfo {year} {2021})}\BibitemShut {NoStop}%
\bibitem [{\citenamefont {Bluvstein}\ \emph {et~al.}(2023)\citenamefont
  {Bluvstein}, \citenamefont {Evered}, \citenamefont {Geim}, \citenamefont
  {Li}, \citenamefont {Zhou}, \citenamefont {Manovitz}, \citenamefont {Ebadi},
  \citenamefont {Cain}, \citenamefont {Kalinowski}, \citenamefont {Hangleiter}
  \emph {et~al.}}]{bluvstein2023logical}%
  \BibitemOpen
  \bibfield  {author} {\bibinfo {author} {\bibfnamefont {Dolev}\ \bibnamefont
  {Bluvstein}}, \bibinfo {author} {\bibfnamefont {Simon~J}\ \bibnamefont
  {Evered}}, \bibinfo {author} {\bibfnamefont {Alexandra~A}\ \bibnamefont
  {Geim}}, \bibinfo {author} {\bibfnamefont {Sophie~H}\ \bibnamefont {Li}},
  \bibinfo {author} {\bibfnamefont {Hengyun}\ \bibnamefont {Zhou}}, \bibinfo
  {author} {\bibfnamefont {Tom}\ \bibnamefont {Manovitz}}, \bibinfo {author}
  {\bibfnamefont {Sepehr}\ \bibnamefont {Ebadi}}, \bibinfo {author}
  {\bibfnamefont {Madelyn}\ \bibnamefont {Cain}}, \bibinfo {author}
  {\bibfnamefont {Marcin}\ \bibnamefont {Kalinowski}}, \bibinfo {author}
  {\bibfnamefont {Dominik}\ \bibnamefont {Hangleiter}},  \emph {et~al.},\
  }\bibfield  {title} {\enquote {\bibinfo {title} {Logical quantum processor
  based on reconfigurable atom arrays},}\ }\href@noop {} {\bibfield  {journal}
  {\bibinfo  {journal} {Nature}\ ,\ \bibinfo {pages} {1--3}} (\bibinfo {year}
  {2023})}\BibitemShut {NoStop}%
\bibitem [{\citenamefont {Leone}\ \emph
  {et~al.}(2021{\natexlab{b}})\citenamefont {Leone}, \citenamefont {Oliviero},\
  and\ \citenamefont {Hamma}}]{leone2021isospectral}%
  \BibitemOpen
  \bibfield  {author} {\bibinfo {author} {\bibfnamefont {Lorenzo}\ \bibnamefont
  {Leone}}, \bibinfo {author} {\bibfnamefont {Salvatore~FE}\ \bibnamefont
  {Oliviero}}, \ and\ \bibinfo {author} {\bibfnamefont {Alioscia}\ \bibnamefont
  {Hamma}},\ }\bibfield  {title} {\enquote {\bibinfo {title} {Isospectral
  twirling and quantum chaos},}\ }\href@noop {} {\bibfield  {journal} {\bibinfo
   {journal} {Entropy}\ }\textbf {\bibinfo {volume} {23}},\ \bibinfo {pages}
  {1073} (\bibinfo {year} {2021}{\natexlab{b}})}\BibitemShut {NoStop}%
\bibitem [{\citenamefont {Haug}()}]{haug2023stabilizerentropy}%
  \BibitemOpen
  \bibfield  {author} {\bibinfo {author} {\bibfnamefont {Tobias}\ \bibnamefont
  {Haug}},\ }\href@noop {} {\enquote {\bibinfo {title} {Code for measuring
  stabilizer entropy on quantum computers},}\ }\bibinfo {howpublished}
  {\url{https://github.com/txhaug/stabilizer_entropy}}\BibitemShut {NoStop}%
\bibitem [{\citenamefont {Leone}\ and\ \citenamefont
  {Bittel}(2024)}]{leone2024stabilizer}%
  \BibitemOpen
  \bibfield  {author} {\bibinfo {author} {\bibfnamefont {Lorenzo}\ \bibnamefont
  {Leone}}\ and\ \bibinfo {author} {\bibfnamefont {Lennart}\ \bibnamefont
  {Bittel}},\ }\bibfield  {title} {\enquote {\bibinfo {title} {Stabilizer
  entropies are monotones for magic-state resource theory},}\ }\href
  {https://arxiv.org/abs/2404.11652} {\bibfield  {journal} {\bibinfo  {journal}
  {arXiv:2404.11652}\ } (\bibinfo {year} {2024})}\BibitemShut {NoStop}%
\bibitem [{\citenamefont {Mitarai}\ \emph {et~al.}(2018)\citenamefont
  {Mitarai}, \citenamefont {Negoro}, \citenamefont {Kitagawa},\ and\
  \citenamefont {Fujii}}]{mitarai2018quantum}%
  \BibitemOpen
  \bibfield  {author} {\bibinfo {author} {\bibfnamefont {Kosuke}\ \bibnamefont
  {Mitarai}}, \bibinfo {author} {\bibfnamefont {Makoto}\ \bibnamefont
  {Negoro}}, \bibinfo {author} {\bibfnamefont {Masahiro}\ \bibnamefont
  {Kitagawa}}, \ and\ \bibinfo {author} {\bibfnamefont {Keisuke}\ \bibnamefont
  {Fujii}},\ }\bibfield  {title} {\enquote {\bibinfo {title} {Quantum circuit
  learning},}\ }\href@noop {} {\bibfield  {journal} {\bibinfo  {journal}
  {Physical Review A}\ }\textbf {\bibinfo {volume} {98}},\ \bibinfo {pages}
  {032309} (\bibinfo {year} {2018})}\BibitemShut {NoStop}%
\bibitem [{\citenamefont {Garcia-Escartin}\ and\ \citenamefont
  {Chamorro-Posada}(2013)}]{garcia2013swap}%
  \BibitemOpen
  \bibfield  {author} {\bibinfo {author} {\bibfnamefont {Juan~Carlos}\
  \bibnamefont {Garcia-Escartin}}\ and\ \bibinfo {author} {\bibfnamefont
  {Pedro}\ \bibnamefont {Chamorro-Posada}},\ }\bibfield  {title} {\enquote
  {\bibinfo {title} {Swap test and hong-ou-mandel effect are equivalent},}\
  }\href@noop {} {\bibfield  {journal} {\bibinfo  {journal} {Physical Review
  A}\ }\textbf {\bibinfo {volume} {87}},\ \bibinfo {pages} {052330} (\bibinfo
  {year} {2013})}\BibitemShut {NoStop}%
\bibitem [{\citenamefont {Brandao}\ \emph {et~al.}(2016)\citenamefont
  {Brandao}, \citenamefont {Harrow},\ and\ \citenamefont
  {Horodecki}}]{brandao2016local}%
  \BibitemOpen
  \bibfield  {author} {\bibinfo {author} {\bibfnamefont {Fernando~GSL}\
  \bibnamefont {Brandao}}, \bibinfo {author} {\bibfnamefont {Aram~W}\
  \bibnamefont {Harrow}}, \ and\ \bibinfo {author} {\bibfnamefont {Micha{\l}}\
  \bibnamefont {Horodecki}},\ }\bibfield  {title} {\enquote {\bibinfo {title}
  {Local random quantum circuits are approximate polynomial-designs},}\
  }\href@noop {} {\bibfield  {journal} {\bibinfo  {journal} {Communications in
  Mathematical Physics}\ }\textbf {\bibinfo {volume} {346}},\ \bibinfo {pages}
  {397--434} (\bibinfo {year} {2016})}\BibitemShut {NoStop}%
\bibitem [{\citenamefont {Arute}\ \emph {et~al.}(2019)\citenamefont {Arute},
  \citenamefont {Arya}, \citenamefont {Babbush}, \citenamefont {Bacon},
  \citenamefont {Bardin}, \citenamefont {Barends}, \citenamefont {Biswas},
  \citenamefont {Boixo}, \citenamefont {Brandao}, \citenamefont {Buell} \emph
  {et~al.}}]{arute2019quantum}%
  \BibitemOpen
  \bibfield  {author} {\bibinfo {author} {\bibfnamefont {Frank}\ \bibnamefont
  {Arute}}, \bibinfo {author} {\bibfnamefont {Kunal}\ \bibnamefont {Arya}},
  \bibinfo {author} {\bibfnamefont {Ryan}\ \bibnamefont {Babbush}}, \bibinfo
  {author} {\bibfnamefont {Dave}\ \bibnamefont {Bacon}}, \bibinfo {author}
  {\bibfnamefont {Joseph~C}\ \bibnamefont {Bardin}}, \bibinfo {author}
  {\bibfnamefont {Rami}\ \bibnamefont {Barends}}, \bibinfo {author}
  {\bibfnamefont {Rupak}\ \bibnamefont {Biswas}}, \bibinfo {author}
  {\bibfnamefont {Sergio}\ \bibnamefont {Boixo}}, \bibinfo {author}
  {\bibfnamefont {Fernando~GSL}\ \bibnamefont {Brandao}}, \bibinfo {author}
  {\bibfnamefont {David~A}\ \bibnamefont {Buell}},  \emph {et~al.},\ }\bibfield
   {title} {\enquote {\bibinfo {title} {Quantum supremacy using a programmable
  superconducting processor},}\ }\href@noop {} {\bibfield  {journal} {\bibinfo
  {journal} {Nature}\ }\textbf {\bibinfo {volume} {574}},\ \bibinfo {pages}
  {505--510} (\bibinfo {year} {2019})}\BibitemShut {NoStop}%
\bibitem [{\citenamefont {{\v{Z}}nidari{\v{c}}}\ \emph
  {et~al.}(2008)\citenamefont {{\v{Z}}nidari{\v{c}}}, \citenamefont {Prosen},\
  and\ \citenamefont {Prelov{\v{s}}ek}}]{vznidarivc2008many}%
  \BibitemOpen
  \bibfield  {author} {\bibinfo {author} {\bibfnamefont {Marko}\ \bibnamefont
  {{\v{Z}}nidari{\v{c}}}}, \bibinfo {author} {\bibfnamefont {Toma{\v{z}}}\
  \bibnamefont {Prosen}}, \ and\ \bibinfo {author} {\bibfnamefont {Peter}\
  \bibnamefont {Prelov{\v{s}}ek}},\ }\bibfield  {title} {\enquote {\bibinfo
  {title} {Many-body localization in the heisenberg x x z magnet in a random
  field},}\ }\href@noop {} {\bibfield  {journal} {\bibinfo  {journal} {Physical
  Review B}\ }\textbf {\bibinfo {volume} {77}},\ \bibinfo {pages} {064426}
  (\bibinfo {year} {2008})}\BibitemShut {NoStop}%
\bibitem [{\citenamefont {G{\"a}rttner}\ \emph {et~al.}(2017)\citenamefont
  {G{\"a}rttner}, \citenamefont {Bohnet}, \citenamefont {Safavi-Naini},
  \citenamefont {Wall}, \citenamefont {Bollinger},\ and\ \citenamefont
  {Rey}}]{garttner2017measuring}%
  \BibitemOpen
  \bibfield  {author} {\bibinfo {author} {\bibfnamefont {Martin}\ \bibnamefont
  {G{\"a}rttner}}, \bibinfo {author} {\bibfnamefont {Justin~G}\ \bibnamefont
  {Bohnet}}, \bibinfo {author} {\bibfnamefont {Arghavan}\ \bibnamefont
  {Safavi-Naini}}, \bibinfo {author} {\bibfnamefont {Michael~L}\ \bibnamefont
  {Wall}}, \bibinfo {author} {\bibfnamefont {John~J}\ \bibnamefont
  {Bollinger}}, \ and\ \bibinfo {author} {\bibfnamefont {Ana~Maria}\
  \bibnamefont {Rey}},\ }\bibfield  {title} {\enquote {\bibinfo {title}
  {Measuring out-of-time-order correlations and multiple quantum spectra in a
  trapped-ion quantum magnet},}\ }\href@noop {} {\bibfield  {journal} {\bibinfo
   {journal} {Nature Physics}\ }\textbf {\bibinfo {volume} {13}},\ \bibinfo
  {pages} {781--786} (\bibinfo {year} {2017})}\BibitemShut {NoStop}%
\bibitem [{\citenamefont {Mi}\ \emph {et~al.}(2021)\citenamefont {Mi},
  \citenamefont {Roushan}, \citenamefont {Quintana}, \citenamefont {Mandra},
  \citenamefont {Marshall}, \citenamefont {Neill}, \citenamefont {Arute},
  \citenamefont {Arya}, \citenamefont {Atalaya}, \citenamefont {Babbush} \emph
  {et~al.}}]{mi2021information}%
  \BibitemOpen
  \bibfield  {author} {\bibinfo {author} {\bibfnamefont {Xiao}\ \bibnamefont
  {Mi}}, \bibinfo {author} {\bibfnamefont {Pedram}\ \bibnamefont {Roushan}},
  \bibinfo {author} {\bibfnamefont {Chris}\ \bibnamefont {Quintana}}, \bibinfo
  {author} {\bibfnamefont {Salvatore}\ \bibnamefont {Mandra}}, \bibinfo
  {author} {\bibfnamefont {Jeffrey}\ \bibnamefont {Marshall}}, \bibinfo
  {author} {\bibfnamefont {Charles}\ \bibnamefont {Neill}}, \bibinfo {author}
  {\bibfnamefont {Frank}\ \bibnamefont {Arute}}, \bibinfo {author}
  {\bibfnamefont {Kunal}\ \bibnamefont {Arya}}, \bibinfo {author}
  {\bibfnamefont {Juan}\ \bibnamefont {Atalaya}}, \bibinfo {author}
  {\bibfnamefont {Ryan}\ \bibnamefont {Babbush}},  \emph {et~al.},\ }\bibfield
  {title} {\enquote {\bibinfo {title} {Information scrambling in quantum
  circuits},}\ }\href@noop {} {\bibfield  {journal} {\bibinfo  {journal}
  {Science}\ }\textbf {\bibinfo {volume} {374}},\ \bibinfo {pages} {1479--1483}
  (\bibinfo {year} {2021})}\BibitemShut {NoStop}%
\bibitem [{\citenamefont {Blocher}\ \emph {et~al.}(2022)\citenamefont
  {Blocher}, \citenamefont {Asaad}, \citenamefont {Mourik}, \citenamefont
  {Johnson}, \citenamefont {Morello},\ and\ \citenamefont
  {M{\o}lmer}}]{blocher2022measuring}%
  \BibitemOpen
  \bibfield  {author} {\bibinfo {author} {\bibfnamefont {Philip~Daniel}\
  \bibnamefont {Blocher}}, \bibinfo {author} {\bibfnamefont {Serwan}\
  \bibnamefont {Asaad}}, \bibinfo {author} {\bibfnamefont {Vincent}\
  \bibnamefont {Mourik}}, \bibinfo {author} {\bibfnamefont {Mark~AI}\
  \bibnamefont {Johnson}}, \bibinfo {author} {\bibfnamefont {Andrea}\
  \bibnamefont {Morello}}, \ and\ \bibinfo {author} {\bibfnamefont {Klaus}\
  \bibnamefont {M{\o}lmer}},\ }\bibfield  {title} {\enquote {\bibinfo {title}
  {Measuring out-of-time-ordered correlation functions without reversing time
  evolution},}\ }\href@noop {} {\bibfield  {journal} {\bibinfo  {journal}
  {Physical Review A}\ }\textbf {\bibinfo {volume} {106}},\ \bibinfo {pages}
  {042429} (\bibinfo {year} {2022})}\BibitemShut {NoStop}%
\end{thebibliography}%

\let\addcontentsline\oldaddcontentsline

\onecolumngrid
\newpage 

\appendix
\setcounter{secnumdepth}{2}
\setcounter{equation}{0}
\setcounter{figure}{0}
\renewcommand{\thetable}{S\arabic{table}}
\renewcommand{\theequation}{S\arabic{equation}}
\renewcommand{\thefigure}{S\arabic{figure}}
\titleformat{\section}[hang]{\normalfont\bfseries}
{Supplemental Materials \thesection:}{0.5em}{\centering}

\clearpage
\begin{center}
	\textbf{\large Supplemental Materials}
\end{center}
\setcounter{equation}{0}
\setcounter{figure}{0}
\setcounter{table}{0}
\makeatletter
\renewcommand{\theequation}{S\arabic{equation}}
\renewcommand{\thefigure}{S\arabic{figure}}
\renewcommand{\bibnumfmt}[1]{[S#1]}
We provide additional details on the efficient measurement of stabilizer entropies and prove bounds on stabilizer fidelity. Further, we show how to compute gradients, mitigate errors and provide additional results for the IonQ quantum computer. Finally, we study nonstabilizerness and scrambling using Clifford-averaged multifractal flatness and OTOCs, as well as provide an efficient measurement scheme for multifractal flatness.
\tableofcontents

\section{Spectrum of SE observable} \label{sec:spectrum}

We now discuss the eigenvalue spectrum of the observable that measures the SE.
The SE is defined as a sum over powers of expectation value of the Pauli operator. By assuming access to $n$ copies, we can write it in terms of an SE observable 
\begin{equation}
\begin{aligned}
A_n=\sum_{\sigma \in \mathcal{P}} \frac{\left\langle\psi|\sigma| \psi\right\rangle^{2 n}}{2^N} = &\left\langle\psi\right|^{\otimes 2n} \Gamma_n^{\otimes N} \left|\psi\right\rangle^{\otimes 2n},
\end{aligned}
\end{equation}
where $\Gamma_n = \frac{1}{2}\sum_{\alpha=0}^3 (\sigma_\alpha)^{\otimes 2n}$. 
Using the definitions $\sigma_y = i\sigma_x\sigma_z$ and $(A\otimes B)(C\otimes D) = (AC)\otimes(BD)$, the following can be deduced:

\begin{equation}
    \sigma_y^{\otimes2n} = 
    \begin{cases}    +\sigma_x^{\otimes2n}\sigma_z^{\otimes2n}  &:  \text{$n=$   even,} \\ 
        -\sigma_x^{\otimes2n}\sigma_z^{\otimes2n}  &:  \text{$n=$   odd.}
    \end{cases}
\end{equation}
With the above considerations, $\Gamma_n^2$ can then be expanded in terms of $I_1, \sigma_x,$ and $ \sigma_z$. Finally, consider the commutation relations of tensor-product Paulis $[\sigma_x^{\otimes 2n}, \sigma_z^{\otimes 2n}] = 0$ to simplify the expanded terms and realise, for $n \geq 2$,

\begin{equation}
    \Gamma_n^2 = 
    \begin{cases}
    2\Gamma_n &:  \text{$n=$   even,} \\ 
    I_{2n} &: \text{$n=$   odd.}
    \end{cases}
\end{equation}
For odd $n$, we have $\Gamma_n^2=I_{2n}$. This implies that the eigenvalues of $\Gamma_n$ must $\lambda=\pm 1$. Thus, the SE operator $A_n=\Gamma_n^{\otimes N}$ has the same eigenvalue spectrum $\lambda=\pm 1$. In contrast for even $n$, we have $\Gamma_n^2=2\Gamma_n$.  One can check that the possible eigenvalues are $\lambda=0$ and $\lambda=2$. Thus, the spectrum of $\Gamma_n^{\otimes N}$ consists of eigenvalues $\lambda=0$ and $\lambda=2^N$.

\section{Parity check to evaluate SEs and Hoeffding's inequality}\label{sec:parity}
We now derive the parity check rules to evaluate $A_n$ using Algorithm 1.
To evaluate $A_n$, we note it is composed as a tensor product of $N$ operators which we evaluate one by one with a simple rule. In particular, for the operator $U_\text{Bell}^{\otimes n} \Gamma_{n} {U_\text{Bell}^{\otimes n}}^\dagger$ and the corresponding state $\ket{\psi}^{\otimes 2n}$, we reorder the position of the qubits $(r_1,r_2,\dots,r_{2n})\rightarrow (r_1,r_{3},\dots,r_{2n-1},r_2,r_{4},\dots,r_{2n})$ such that the odd-indexed qubits occupy the first half, and the even-indexed qubits the second half of the register. Then, we find for the transformed operator for odd $n$ the simple form
\begin{equation}
    [U_\text{Bell}^{\otimes n} \Gamma_{n} {U_\text{Bell}^{\otimes n}}^\dagger]_\text{reorder}=-\frac{1}{2} (1-{\sigma^z}^{\otimes n})\otimes (1-{\sigma^z}^{\otimes n})+1
\end{equation}
and for even $n$
\begin{equation}
    [U_\text{Bell}^{\otimes n} \Gamma_{n} {U_\text{Bell}^{\otimes n}}^\dagger]_\text{reorder}=\frac{1}{2} (1+{\sigma^z}^{\otimes n})\otimes (1+{\sigma^z}^{\otimes n})\,.
\end{equation}
Here, the expectation value $\langle x \rangle=\bra{x}(1-{\sigma^z}^{\otimes n})\ket{x}$ for $n$-dimensional computational basis state $\ket{x}$ is $\langle x \rangle=2$ when $x$ has odd parity, and $0$ otherwise. 
In contrast, $\bra{x}(1+{\sigma^z}^{\otimes n})\ket{x}$ is $2$ when $x$ has even parity, and is $0$ otherwise. 
Thus, we see that $U_\text{Bell}^{\otimes n} \Gamma_{n} {U_\text{Bell}^{\otimes n}}^\dagger$ can be evaluated by checking the parity of the odd-indexed and even-indexed qubits.

We bound the maximal number $L$ of measurement steps needed to measure SE with Hoeffding's inequality. The failure probability $\Delta$ to get an error $\epsilon$ between expectation value $A_n$ and estimation $\hat{A}_n$ is given by
\begin{equation}\label{eq:Hoeffding}
    P(\vert \hat{A}_n - A_n\vert\ge \epsilon)=\delta\le 2\exp(-\frac{2\epsilon^2 L}{\Delta\omega_n^2})\,,
\end{equation}
where $\Delta\omega_n$ is the range of eigenvalues of $\Gamma_n^{\otimes N}$.
To estimate $A_n$ within $\epsilon$ accuracy and $\delta$ failure probability we require at most
\begin{equation}
L\le\frac{\Delta\omega_n^2}{2\epsilon^2}\log(\frac{2}{\delta})\,
\end{equation}
measurement steps. Each step of the algorithm uses $2n$ copies of the state as seen in Algorithm 1, thus the total number of copies is $C=2Ln$. 
For odd $n>1$, we have $\Delta\omega_n=2$ and the number of copies of $\ket{\psi}$ scales as $C=O(n\epsilon^{-2})$.
For even $n$, the eigenvalue spectrum of $\Gamma_n^{\otimes N}$ diverges and Algorithm~1 of the main text requires in general an exponential number of measurements.

\section{Parameter-shift rule for gradient of SE}\label{sec:shift-rule}

We now derive the parameter-shift rule for observables that act on multiple copies of a state. First, we regard the generator $G$ of the unitary operator $\mathcal{G}(\mu)= e^{-i \mu G}$ has two distinct eigenvalues, we can shift the eigenvalues to $\pm r$. Note that any single qubit gate is of this form, which implies $G^2 = r^2 I$, where $I$ is the identity matrix. The Taylor series of $\mathcal{G}(\mu)$ shows
\begin{equation}
\begin{aligned}
\mathcal{G}(\mu)= & \exp (-i \mu G)=\sum_{k=0}^{\infty} \frac{(-i \mu)^k G^k}{k !} \\
= & \sum_{k=0}^{\infty} \frac{(-i \mu)^{2 k} G^{2 k}}{(2 k) !}+\sum_{k=0}^{\infty} \frac{(-i \mu)^{2 k+1} G^{2 k+1}}{(2 k+1) !} \\
= & I \sum_{k=0}^{\infty} \frac{(-1)^k(r \mu)^{2 k}}{(2 k) !} -i r^{-1} G \sum_{k=0}^{\infty} \frac{(-1)^k(r \mu)^{2 k+1}}{(2 k+1) !} \\
= & I \cos (r \mu)-i r^{-1} G \sin (r \mu).
\end{aligned}
\end{equation}
Thus the following identity holds
\begin{equation}
\mathcal{G}(\frac{\pi}{4r})= \frac{1}{\sqrt{2}} (I-ir^{-1} G).
\end{equation}

Now, we assume an $N$-qubit parameterized quantum circuit $\ket{\psi}\equiv\ket{\psi(\boldsymbol{\theta})} = \prod_{n=1}^d V_n(\boldsymbol{\theta}_n) W_n|0\rangle$ with $d$ layers, entangling gates $W_n$, parameters $\boldsymbol{\theta}$ and parameterized rotations $V_n(\boldsymbol{\theta}_n) = e^{-i\frac{\boldsymbol{\theta}_k}{2}\sigma_n}$ given by some Pauli strings $\sigma_n$.
For any operators $U$, $O$, and $V$ we have
\begin{equation}
\begin{aligned}
\langle\psi|U^{\dagger} O V| \psi\rangle+ \text { h.c. } =& \frac{1}{2}\left[\langle\psi|(U+V)^{\dagger} \hat{O}(U+V)| \psi\rangle \right. \left. - \langle\psi|(U-V)^{\dagger} \hat{O}(U-V)| \psi\rangle \right],
\end{aligned}
\end{equation}
where h.c. is the hermitian conjugate of the preceding terms~\cite{mitarai2018quantum}.

To measure SE, we calculate the expectation value 
\begin{equation}
\langle O^{(K)}(\boldsymbol{\theta})\rangle = \langle\psi(\boldsymbol{\theta})|^{\otimes K} O^{(K)} |\psi(\boldsymbol{\theta})\rangle^{\otimes K}, 
\end{equation}
over $K$ copies of state $\ket{\psi}\in\mathbb{C}^{2^N}$ with  in respect to an operator $O^{(K)}\in\mathbb{C}^{2^{NK}}$. For $K=1$, i.e. measurements on a single quantum state, the shift rule is given by $\partial_k \langle O^{(1)}(\boldsymbol{\theta})\rangle = \frac{1}{2} (\langle O^{(1)}(\boldsymbol{\theta} + \mathbf{e}_k \frac{\pi}{2})\rangle - \langle O^{(1)}(\boldsymbol{\theta}-\mathbf{e}_k\frac{\pi}{2})\rangle)$, where $e_\mathbf{k}$ is the $k$th unit vector~\cite{mitarai2018quantum}.

We now derive the shift-rule for general $K$.
The derivative of the quantum state $\ket{\psi}$ is given by
\begin{equation}
\begin{aligned}\label{eq:deriv_single}
\partial_k|\psi(\boldsymbol{\theta})\rangle &=\prod_{n=k+1}^d\left[V_n\left(\boldsymbol{\theta}_n\right) W_n\right]\left(-i \frac{1}{2} \sigma_k\right) \prod_{n=1}^k\left[V_n\left(\boldsymbol{\theta}_n\right) W_n\right]|0\rangle \\
& \equiv U_k\left(-i \frac{1}{2} \sigma_k\right)\left|\phi_k\right\rangle,
\end{aligned}
\end{equation}
where $U_k = \prod_{n=k+1}^d\left[V_n\left(\boldsymbol{\theta}_n\right) W_n\right]$ and $\left|\phi_k\right\rangle = \prod_{n=1}^k\left[V_n\left(\boldsymbol{\theta}_n\right) W_n\right]|0\rangle$. From the product rule, the $K$-copy derivative of $\langle O\rangle$ is given by
\begin{equation}\label{eq:deriv}
\partial_k \langle O^{(K)}\rangle=  K \left\langle\phi_k\right|^{\otimes K}U_k ^{\dagger \otimes K} O^{(K)} U_k ^{\otimes K}{\left[\left(-i \frac{1}{2} \sigma_k\right) \otimes I^{\otimes K-1}\right]\left|\phi_k\right\rangle ^{\otimes K}+\text { h.c. } }
\end{equation}
In the above, the states are freely rearranged while the factor of $K$ emerges from the product rule. We now define $O^\prime = U_k ^{\dagger \otimes K} O^{(K)} U_k ^{\otimes K}$. Then, applying~\eqref{eq:deriv} in the second row gives
\begin{equation}
\begin{aligned}
\partial_k \langle O\rangle =& \frac{K}{4}\langle\phi_k|^{\otimes K} O^{\prime}[(-i \sigma_k) \otimes I ^{\otimes K-1} ]| \phi_k\rangle ^{\otimes K}+\text { h.c. } \\
=&  \frac{K}{4} \left[ \langle\phi_k|^{\otimes K} [(I-i \sigma_k)^{\dagger} \otimes I^{\otimes K-1}] O^{\prime} [(I-i  \sigma_k) \otimes I^{\otimes K-1}]|\phi_k\rangle ^{\otimes N} \right.\\
& \left.- \langle\phi_k|^{\otimes K} [(I+i  \sigma_k)^{\dagger} \otimes I^{\otimes K-1}] O^{\prime} [(I+i  \sigma_k) \otimes I^{\otimes K-1}]|\phi_k\rangle^{\otimes K} \right].
\end{aligned}
\end{equation}
For any Pauli strings $\sigma_k$, we can use~\eqref{eq:deriv_single} to find

\begin{equation}
\begin{aligned}
\partial_k \langle O\rangle =& \frac{K}{2} \left[ \langle\phi_k|^{\otimes K}\left[e^{-i \frac{\pi}{2} \frac{1}{2} \sigma_k} \otimes I^{\otimes K-1}\right]^{\dagger} O^{\prime}\left[e^{-i \frac{\pi}{2} \frac{1}{2} \sigma_k} \otimes I^{\otimes K-1}\right]| \phi_k\rangle^{\otimes K} \right.\\
& \left. -\langle\phi_k|^{\otimes K}\left[e^{i \frac{\pi}{2} \frac{1}{2} \sigma_k} \otimes I^{\otimes K-1}\right]^{\dagger} O^{\prime}\left[e^{i \frac{\pi}{2} \frac{1}{2} \sigma_k} \otimes I^{\otimes K-1}\right]| \phi_k\rangle^{\otimes K} \right]\\
=& \frac{K}{2}\left[ \langle\psi(\boldsymbol{\theta}+\frac{\pi}{2} \mathbf{e}_k)|\langle\psi(\boldsymbol{\theta})|^{\otimes K-1}O^{(K)}| \psi(\boldsymbol{\theta}+\frac{\pi}{2} \mathbf{e}_k)\rangle| \psi(\boldsymbol{\theta})\rangle^{\otimes K-1} \right. \\
& \left. -\langle\psi(\boldsymbol{\theta}-\frac{\pi}{2} \mathbf{e}_k)|\langle\psi(\boldsymbol{\theta})|^{\otimes K-1}O^{(K)}| \psi(\boldsymbol{\theta}-\frac{\pi}{2} \mathbf{e}_k)\rangle| \psi(\boldsymbol{\theta})\rangle^{\otimes K-1} \right],
\end{aligned}
\end{equation}
where we have introduced the $k$th unit vector $e_\mathbf{k}$ and absorbed the exponential into the definition of the $k$th parameterized rotation: $U_k e^{-i \frac{\pi}{2} \frac{1}{2} \sigma_n} \ket{\phi_k} = | \psi(\boldsymbol{\theta}+\frac{\pi}{2} \mathbf{e}_k)\rangle$.
We can now get the $n$-th moment of the Pauli spectrum simply by setting $K=2n$ and performing Bell measurements.

To compute the gradient of the Tsallis SE in respect to the $k$th parameter, we have 
\begin{equation}
    \partial_k T_n=(1-n)^{-1}\partial_k  A_n\,,
\end{equation}
where $ A_n=\bra{\psi}^{\otimes 2n}\Gamma_{n}^{\otimes N} \ket{\psi}^{\otimes 2n}$.
For the R\'enyi SE, we have
\begin{equation}
    \partial_k M_n=(1-n)^{-1}\frac{1}{ A_n} \partial_k  A_n\,.
\end{equation}
Now, we present the algorithm to efficiently compute an unbiased estimator of the gradient of the SE on quantum computers for odd $n>1$. The algorithm is detailed in Algorithm~\ref{alg:gradientodd}. The algorithm computes the gradient of $A_n$ using the shift-rule. To calculate the gradient, one performs Bell measurements on a parameterized quantum state shifted by $\pm\pi/2$ with $\ket{\psi(\theta\pm\pi/2)}\ket{\psi(\theta)}$, as well as $n-1$ Bell measurements on states without shift $\ket{\psi(\theta)}\ket{\psi(\theta)}$. Then, use the same procedure as Algorithm 1 in the main text to estimate $K_\pm$, where its difference scaled by $n$ gives us $\partial_k A_n$.

\begin{figure}[htb]
  \centering
  \begin{minipage}{.6\linewidth}
\begin{algorithm}[H]
 \SetAlgoLined
 \LinesNumbered
  \SetKwInOut{Input}{Input}
  \SetKwInOut{Output}{Output}
   \Input{   Integer $n>1$\\
   $L$ repetitions\\
   Circuit parameter $\boldsymbol{\theta}$\\
   $N$-qubit parameterized unitary $U(\boldsymbol{\theta})$\\
   Gradient parameter index $k$
   }
    \Output{Gradient $\partial_k A_n$
    }

 \SetKwRepeat{Do}{do}{while}

    \For{$ s = -1, +1$}{
    
    $B_{s}= 0$
    
    \For{$q=1,\dots,L$}{

    Prepare $\ket{\eta_\text{shifted}}=U_\text{Bell}^{\otimes N}(U(\boldsymbol{\theta}+s\boldsymbol{e}_k)\otimes U(\boldsymbol{\theta}))\ket{0}^{\otimes 2N}$

    Sample in computational basis $\boldsymbol{r}^{(1)}\sim \vert \braket{\boldsymbol{r}\vert\eta_\text{shifted}}\vert^2$
    
    \For{$j=1,\dots,n-1$}{
        Prepare $\ket{\eta}=U_\text{Bell}^{\otimes N}(U(\boldsymbol{\theta})\otimes U(\boldsymbol{\theta}))\ket{0}^{\otimes 2N}$
    
        Sample in computational basis $\boldsymbol{r}^{(j+1)}\sim \vert \braket{\boldsymbol{r}\vert\eta}\vert^2$

    }

    $b= 1$

        \For{$\ell=1,\dots,N$}{
            
            $\nu_1=\bigoplus_{j=1}^n r^{(j)}_{2\ell-1}$
            
            $\nu_2=\bigoplus_{j=1}^n r^{(j)}_{2\ell}$

             \uIf{$n$ $\mathrm{is\,\,odd}$}{
               $b= b\cdot (-2\nu_1 \cdot\nu_2+1)$
              }
              \Else{
                $b= b\cdot 2(\nu_1-1) \cdot(\nu_2-1)$
              }
        }
        
    $B_{s}= B_{s}+b/L$
    }

    }
    
    $\partial_k A_n= n(B_{+} -B_{-})$
    
 \caption{Gradient of SE}
  \label{alg:gradientodd}
\end{algorithm}
  \end{minipage}
\end{figure}

\section{Strong monotonicity and measuring R\'enyi SEs}\label{sec:strong_mon}
A not necessary, but useful property of nonstabilizerness is  strong monotonicity where the measure is on average non-increasing under computational-basis measurements on a set of $k$ qubits~\cite{chitambar2019quantum}. This property demands that $T_n(\ket{\psi})\geq \sum_{\lambda}p_\lambda T_n[\ket{\psi_\lambda}]$ when using the projector $\Pi_\lambda=|\lambda\rangle \langle \lambda| \otimes \openone_{N\setminus k}$ onto the computational basis state $\lambda$ with corresponding probability $p_{\lambda}=\bra{\psi}\Pi_\lambda\ket{\psi}$ and post-measurement state $\ket{\psi_\lambda}$.
The R\'enyi SE $M_n$ is not a strong monotone for all $n$~\cite{haug2023stabilizer}. For $n<2$, we can show that the Tsallis SE $T_n$ is not a strong monotone using the counter-example from Ref.~\cite{haug2023stabilizer}. However, for $n\ge2$ we are unable to find any example where  strong monotonicity is violated. Using extensive numerical optimization for the strong monotonicity condition, we find numerical evidence that $T_n$ for $n\ge2$ is a strong monotone for at least 
 $N\le6$ qubits. 
We use numerical optimization of a general parameterized $N$-qubit state $\ket{\psi(\alpha,\beta)}=\sum_k \alpha_k+i\beta_k\ket{k}$. Here, we minimize the condition of strong monotonicity $\Delta T_n= T_n(\ket{\psi})-  \sum_{\lambda}p_\lambda T_n[\ket{\psi_\lambda}]$, where strong monotonicity is violated when $\Delta T_n<0$. Numerically, we do not find states that violate strong monotonicity for the Tsallis-$n$ SE for $n\ge2$, while we easily find counter-examples for the R\'enyi-$n$ SE.

\revA{
We now discuss the scaling of measuring the R\'enyi SE. We can estimate $A_n=2^{-N}\sum_{\sigma\in\mathcal{P}}\bra{\psi}\sigma\ket{\psi}^{2n}$ with additive precision using $O(n\epsilon^{-2})$ samples with our algorithms. 
The R\'enyi SE involves the logarithm of $A_n$, i.e. $M_n\sim \ln(A_n)$. 
Given $A_n$ estimated with additive precision $\epsilon$, we now consider the cost of estimating $M_n$ with additive precision $\epsilon_M$. 
Given exact $\bar{A}_n$ and $\bar{M}_n$, we assume from finite samples we have error  $\epsilon$ in estimating $A_n$, which gives us
\begin{equation}
    (1-n)^{-1} \ln(\bar{A}_n-\epsilon)=\bar{M}_n+\epsilon_M\,.
\end{equation}
We insert $A_n=\exp(M_n(1-n))$ to get
\begin{equation}
    \epsilon_M= (1-n)^{-1} \ln(1-e^{-\bar{M}_n(1-n)}\epsilon)
\end{equation}
We now assume $\epsilon\ll \bar{A}_n$ and get
\begin{equation}
    \epsilon_M\approx -(1-n)^{-1}e^{-\bar{M}_n(1-n)}\epsilon
\end{equation}
and finally
\begin{equation}
    \epsilon\approx -(1-n)e^{\bar{M}_n(1-n)}\epsilon_M\,.
\end{equation}
For $n>1$, to estimate $M_n$ with precision $\epsilon_M$, we require an accuracy $\epsilon=O(\exp(-M_n)\epsilon_M)$ exponentially small in $M_n$. With our algorithm this requires a number of samples $O(\exp(M_n)\epsilon_M^{-2})$ scaling in general exponentially in $M_n$. Assuming $M_n=O(\log(n))$, one can estimate $M_n$ with polynomial number of measurements.

Note that this challenge in estimating the logarithm of a function appears for any entropy involving logarithms, such as for the $2$-R\'enyi entanglement entropy where one has $\text{tr}(\rho^2)$ and $-\log(\text{tr}(\rho^2))$.
}

\section{Other nonstabilizerness monotones and known bounds}\label{sec:upperbound}

We introduce different magic monotones, and reproduce the proof of upper bound of stabilizer fidelity.

First, the robustness of magic for state $\rho$ is given by~\cite{howard2017application}
\begin{equation}
\text{R}(\rho)=\min_x (\sum_k \vert x_k \vert;\,\, \rho=\sum_k x_k \ket{\phi_k}\bra{\phi_k})\,,
\end{equation}
where $x_k\in\mathbb{R}$ and $\ket{\phi_k}\in\text{STAB}$ are stabilizer states. 
The stabilizer extent for a pure state $\ket{\psi}$ is given by~\cite{bravyi2019simulation}
\begin{equation}
\xi(\ket{\psi})=\min_{c}(\sum_k \vert c_k\vert;\,\,\ket{\psi}=\sum_{k} c_k\ket{\phi_k})^2\,,
\end{equation}
where $c_k\in\mathbb{C}$. 
Finally, the stabilizer fidelity measures the geometric distance to the closest stabilizer state~\cite{bravyi2019simulation}
\begin{equation}
F_\text{STAB}(\ket{\psi})=\max_{\ket{\phi}\in\text{STAB}}\vert\braket{\psi\vert\phi}\vert^2\,.
\end{equation}

The moment of the Pauli spectrum $A_n$ for $n>1$ upper bounds the stabilizer fidelity, which was previously proven in Ref.~\cite{haug2023stabilizer} via the R\'enyi SE. 
Here, we reproduce the proof for completeness and adapt it to our efficient measurement algorithms. First, we note that for any state $\ket{\psi}$ with stabilizer fidelity $F_\text{STAB}(\ket{\psi})$ one can find a Clifford unitary $U_C$ with
\begin{equation}\label{eq:dec}
\ket{\phi}=U_C\ket{\psi}=\sum_k a_k \ket{k}\,,
\end{equation}
 where 
\begin{equation}\label{eq:boundFmax}
	\vert a_0 \vert^2 = F_\text{STAB}(\ket{\psi})\,.
\end{equation}
One can see this immediately from $F_\text{STAB}=\vert \braket{\psi\vert\phi_\text{max}}\vert^2=\vert \braket{\psi\vert U_C^\text{max}\vert 0}\vert^2=\vert a_0\vert^2$. In~\eqref{eq:dec}, we denote $\{\ket{k}\}$ the the computational basis states. Next, we have
\begin{equation}\label{eq:inequality}
	2^{-N}\sum_{\sigma \in \mathcal{P}} \vert \bra{\phi}\sigma\ket{\phi}\vert^{2n}\ge 2^{-N}
	\sum_{\sigma \in \mathcal{P}_z} \vert \bra{\phi}\sigma\ket{\phi}\vert^{2n}\nonumber\ge
	2^{-2n N} \left(\sum_{\sigma \in \mathcal{P}_z} | \bra{\phi}\sigma\ket{\phi}|\right)^{2n}\,.\numberthis
\end{equation}
Here $\mathcal{P}$ is the set of all Pauli strings while $\mathcal{P}_z$ is the set of Pauli strings which contains strings with $\openone$ and $\sigma^z$ only.
We have used the convexity inequality
\begin{equation}\label{eq:powermean}
	\sum_{i=1}^m \vert a_i\vert^k\ge \frac{1}{m^{k-1}}\left(\sum_{i=1}^m \vert a_i \vert \right)^k\,.
\end{equation}
Now, we apply
\begin{equation}
	\left(\sum_{\sigma \in \mathcal{P}_z} | \bra{\phi}\sigma\ket{\phi}|\right)^{2n}\geq \left|  \sum_{\sigma \in \mathcal{P}_z} \bra{\phi}\sigma\ket{\phi}\right|^{2n},
\end{equation}
and the fact that
\begin{equation}
	\sum_{\sigma\in \mathcal{P}_z}\sigma=2^{N}\ket{0}\bra{0}\,,
\end{equation}
together with~\eqref{eq:inequality} to get
\begin{equation}\label{eq:almost_there}
2^{-N}\sum_{\sigma \in \mathcal{P}} \vert \bra{\phi}\sigma\ket{\phi}\vert^{2n}\geq |a_0|^{4n}=F_{\rm STAB}(\ket{\psi})^{2n}\,.
\end{equation}
Then, we use the fact that $A_n$ is invariant under unitary operations to get $\sum_{\sigma \in \mathcal{P}} \vert \bra{\phi}\sigma\ket{\phi}\vert^{2n}=\sum_{\sigma \in \mathcal{P}} \vert \bra{\psi}\sigma\ket{\psi}\vert^{2n}$. We insert the definition of $T_n$ to finally get
\begin{equation}
F_\text{STAB}(\ket{\psi})\le A_n(\ket{\psi}))^{\frac{1}{2n}}\,,
\end{equation}
which concludes the proof of the upper bound. Furthermore, we note the well known relationship between $R$, $\xi$ and $F_\text{STAB}$ ~\cite{bravyi2019simulation,liu2022many}
\begin{equation}
    R\ge \xi \ge F_\text{STAB}^{-1}\,,
\end{equation}
which combined gives us the final bound:
\begin{equation}\label{eq:hierarchy_sup}
R\ge \xi \ge F_\text{STAB}^{-1}\ge A_n^{-\frac{1}{2n}}\,.
\end{equation}

\section{Proof of lower bound of stabilizer fidelity}\label{sec:lowerbound}

Here, we prove that the $n$th moment of the Pauli spectrum $A_n$ is a lower bound of the stabilizer fidelity $F_\text{STAB}$.

The main ingredient is the relationship proven in~\cite{gross2021schur,grewal2023improved}
\begin{equation}
    F_\text{STAB}(\ket{\psi})\ge 2^{-N}\sum_{\sigma \in \mathcal{Q}} \bra{\psi}\sigma\ket{\psi}^2,
\end{equation}
where $\mathcal{Q}=\{\sigma \in \mathcal{P}: \bra{\psi}\sigma\ket{\psi}^2>\frac{1}{2}\}$.
We now have 
\begin{align*}
F_\text{STAB}&\ge 2^{-N}\sum_{\sigma \in \mathcal{Q}} \bra{\psi}\sigma\ket{\psi}^2=\underset{\sigma\sim \Xi(\sigma)}{\text{Pr}}[\bra{\psi}\sigma\ket{\psi}^2>\frac{1}{2}]=\underset{\sigma\sim \Xi(\sigma)}{\text{Pr}}[\bra{\psi}\sigma\ket{\psi}^{2(n-1)}>2^{1-n}]\\
&=1-\underset{\sigma\sim \Xi(\sigma)}{\text{Pr}}[\bra{\psi}\sigma\ket{\psi}^{2(n-1)}\le2^{1-n}]=1-\underset{\sigma\sim \Xi(\sigma)}{\text{Pr}}[1-\bra{\psi}\sigma\ket{\psi}^{2(n-1)}\ge1-2^{1-n}]\\
&\ge1-\frac{1}{1-2^{1-n}}(1-\underset{\sigma\sim \Xi(\sigma)}{\mathbb{E}}[\bra{\psi}\sigma\ket{\psi}^{2(n-1)}])=1-\frac{1}{1-2^{1-n}}(1-\sum_{\sigma\in\mathcal{P}}\frac{\bra{\psi}\sigma\ket{\psi}^{2n}}{2^{N}})\\
&=\frac{A_n-2^{1-n}}{1-2^{1-n}}. \numberthis
\end{align*}
In the first line we used the definition of $\mathcal{Q}$, and in the second line we used Markov's inequality which holds for $n>1$.

\revA{
\section{Relationship of measures of nonstabilizerness}\label{sec:other_meas}
Here we study the relationship of different measures of nonstabilizerness. In particular, we study the R\'enyi SE $M_n$, the min-relative entropy of magic $D_\text{min}=-\ln(F_\text{STAB})$, the log-free robustness of magic $\text{LR}=\ln(R)$ and the max-relative entropy of magic $\ln(\xi)$. We also consider the additive Bell magic~\cite{haug2022scalable} which is defined and further discussed in SM~\ref{sec:Bell}.

We now study the $N$-qubit product state
\begin{equation}\label{eq:Trel}
    \ket{\psi(s)}=2^{-N/2}(\ket{0}+\exp(i\pi s/4)\ket{1})^{\otimes N}\,.
\end{equation}
We study the scaling of this state with $s$. We show the scaling in Fig.~\ref{fig:rel}.

For $D_\text{min}$, $\mathcal{B}_\text{a}$ and $M_n$ with $n\ge2$ we find that these measures of nonstabilizerness scale as $\propto s^2N$. Defining $\theta=s\sqrt{N}$, we find for all these measures $\propto \theta^2$. This indicates that these measure are closely related. 
We study the relationship between these measures further by finding respective upper and lower bounds. We numerical maximize and minimize the ratios $D_\text{min}/M_n$ and $\mathcal{B}_\text{a}/M_n$ over all pure states for a given qubit number $N$. This gives us a lower and upper bound between these measures. Due to numerical complexity of computing  $D_\text{min}$ and $\mathcal{B}_\text{a}$ exactly, we can optimize up to $N\le4$. We find that the bounds barely change with $N$, indicating that they are likely valid even for higher qubit numbers. In particular, we find $1.7 M_2\gtrsim D_\text{min}\ge \frac{1}{4}M_2$ as well as $3.5 M_2 \gtrsim  \mathcal{B}_\text{a} \gtrsim 2.88 M_2 $ for at least $N\le4$.

\begin{figure*}[htbp]
	\centering	
 \subfigimg[width=0.3\textwidth]{}{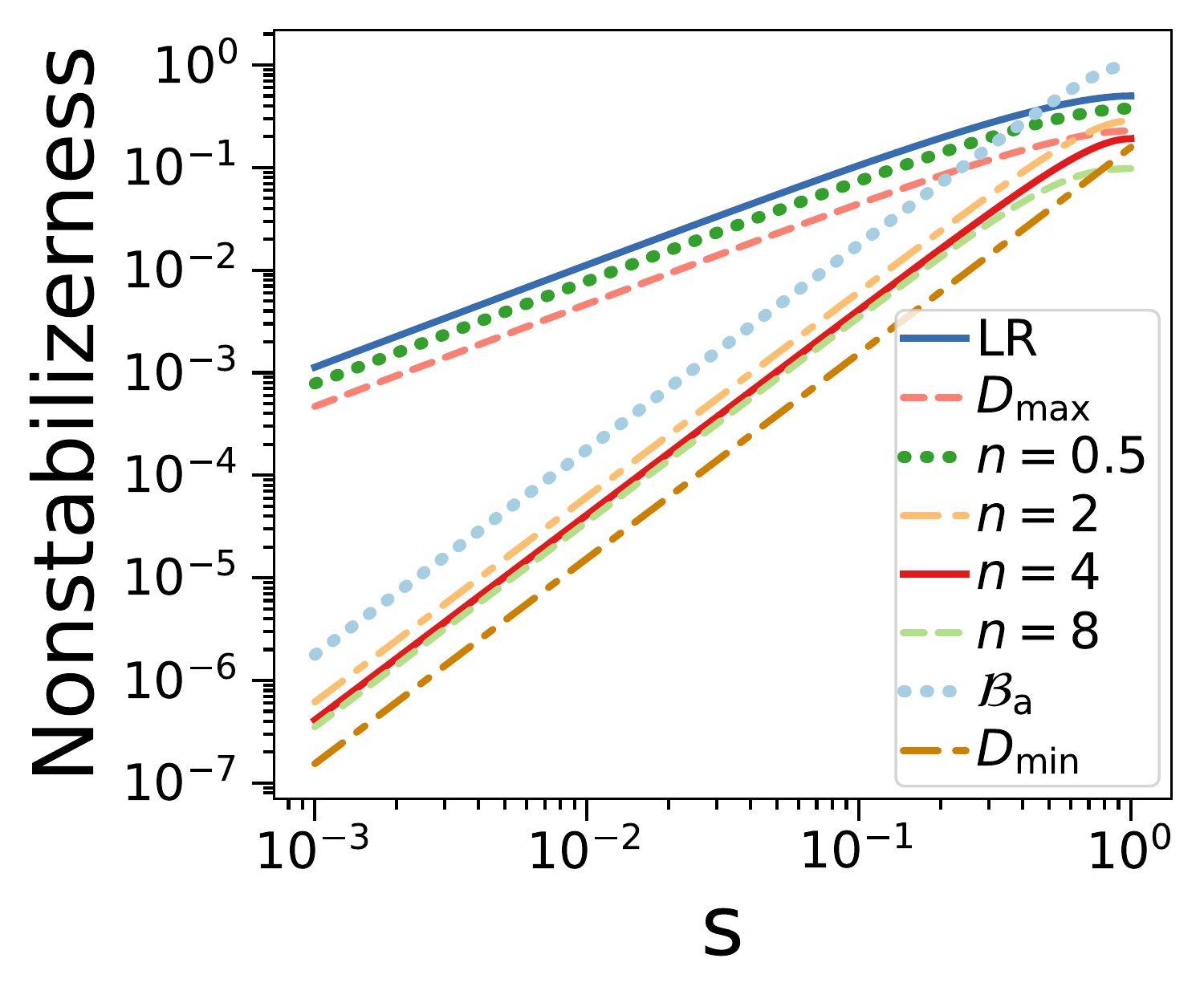}
	\caption{Scaling of different measures of nonstabilizerness for state~\eqref{eq:Trel} as function of $s$. 
	}
	\label{fig:rel}
\end{figure*}

In contrast, for $\text{LR}$, $D_\text{max}$ and $M_{1/2}$ we find that these measures scale as $\propto sN$ and thus $\propto \theta/\sqrt{N}$. These measures show a different scaling with $N$ compared to the measures that relate to geometric distance such as $D_\text{min}$. This shows that $\text{LR}$, $D_\text{max}$ and $M_{1/2}$ do not support upper bounds in respect to $M_n$ for $n>1/2$. As there are no known efficient measurement protocols for $M_{1/2}$, it is likely that one cannot find efficiently measurable upper bounds for $D_\text{max}$ and $\text{LR}$. 
}

\section{Error mitigation}\label{sec:mtg}
Noise in the quantum computer distorts the true value of the SE. We now propose how to mitigate noise without requiring additional measurements. 
Here, we assume that a pure state $\ket{\psi}$ is subject to a global depolarizing channel with probability $p$, resulting in the noisy state $\rho_\text{dp} = (1-p)|\psi\rangle\langle\psi| +pI_N2^{-N}$.
The depolarising probability $p$ can be deduced by the measurement of purity $\text{tr}(\rho_\text{dp}^2)$ as shown in Ref.~\cite{haug2022scalable}
\begin{equation}
    p = 1 - \frac{\sqrt{(2^N-1)(2^N \text{tr}(\rho^2_\text{dp})-1)}}{2^N-1}.
\end{equation}
Note that $\text{tr}(\rho^2_\text{dp})$ can be efficiently measured with Bell measurements~\cite{garcia2013swap}. In particular, the purity can be measured via  $\text{tr}(\rho^2_\text{dp})= A_1$ using Algorithm 1 in an efficient manner.
SEs for depolarized state $\rho_\text{dp}$ can be expressed using the expectation value of operator $\Gamma^{\otimes N}_{n} = 2^{-N}\sum_{\sigma\in \mathcal{P}}\sigma^{\otimes 2n}$ acting on $2n$ copies of the mixed state $\rho_\text{dp}$
\begin{equation}
\begin{aligned}
    A_n^\text{dp} &= \text{tr}(\rho_\text{dp}^{\otimes 2n} \Gamma^{\otimes N}_{n}) = \frac{1}{2^N} \sum_{\sigma\in\mathcal{P}}[\text{tr}(\rho_\text{dp}\sigma)]^{2n}= \frac{1}{2^N} \sum_{\sigma\in\mathcal{P}}\left[(1-p)\langle\psi|\sigma|\psi\rangle + p\delta_{\sigma,I_N}\right]^{2n}\\
    &= \frac{(1-p)^{2n}}{2^N}\sum_{\sigma\in\mathcal{P}}\langle\psi|\sigma|\psi\rangle^{2n} +\frac{1-(1-p)^{2n}}{2^N}
    = (1-p)^{2n} A_n^\text{mtg} + \frac{1-(1-p)^{2n}}{2^N}\,,
\end{aligned}
\end{equation}
where $ A_n^\text{mtg}$ is the mitigated expectation value using the measurement result affected by depolarizing noise $A_n^\text{dp}$. We find
\begin{equation}
 A_n^\text{mtg}=\frac{ A_n^\text{dp}}{(1-p)^{2n}}-\frac{(1-p)^{-2n}-1}{2^N}\,.
\end{equation}
We then substitute the above expression to calculate $T_n = -(1- A_n)/(1-n)$, to realize the expression of mitigated SE $T_n^\text{mtg}$ in terms of depolarized SE $T_n^\text{dp}$
\begin{equation}
    T_n^\text{mtg} = \frac{1}{(1-p)^{2n}} \left[T_n^\text{dp} - \frac{(1-(1-p)^{2n})(2^{-N}-1)}{(1-n)}\right].
\end{equation}
Similarly, we have for the R\'enyi SE
\begin{equation}
M_n^\text{mtg} =(1-n)^{-1} \ln(A_n^\text{mtg})\,.
\end{equation}

\section{Further IonQ quantum computer results}\label{sec:exp_other}
We present here further experimental results on nonstabilizerness.

\begin{figure*}[htbp]
	\centering	
  \subfigimg[width=0.3\textwidth]{c}{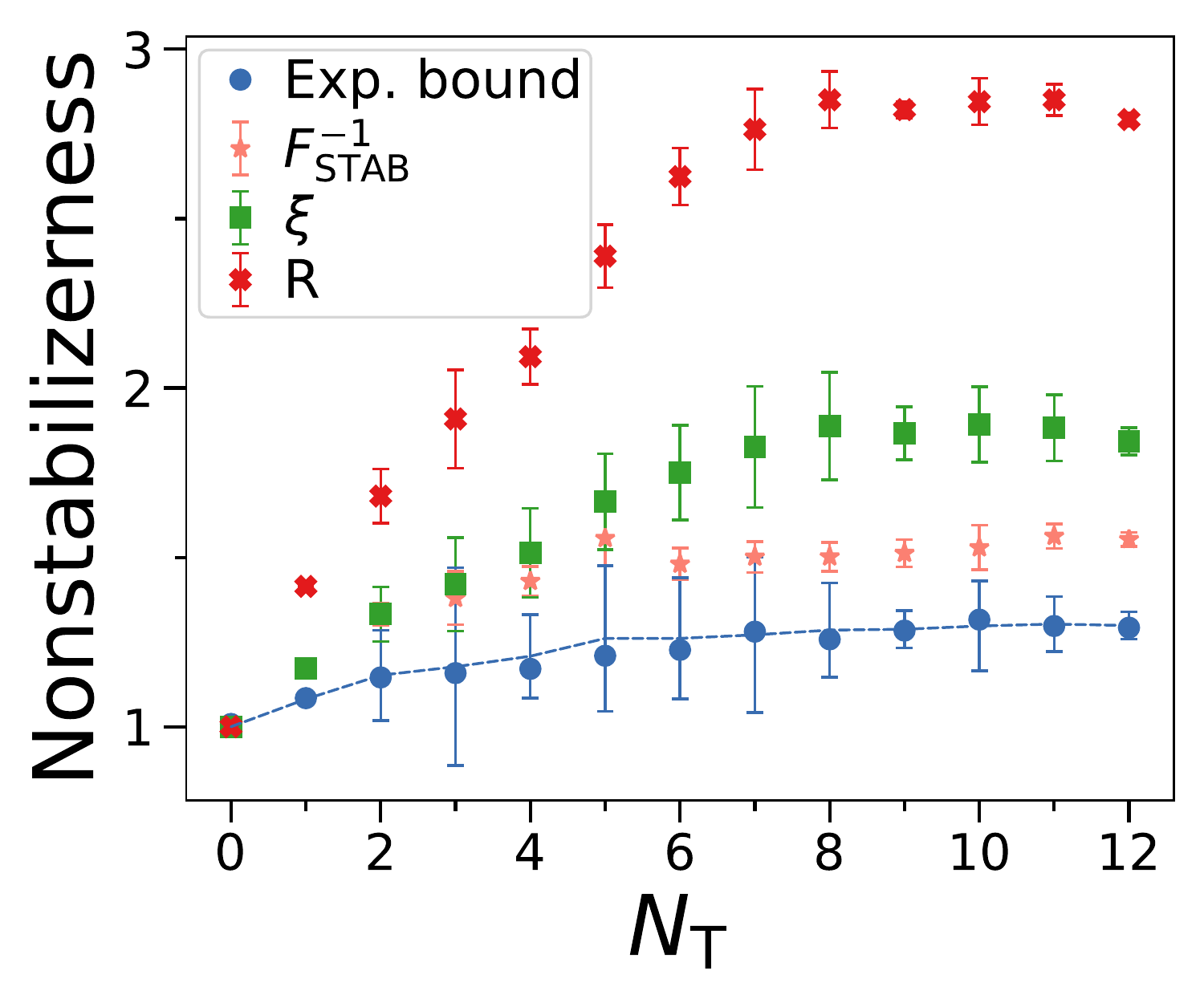}
	\caption{ We demonstrate the hierarchy of bounds~\eqref{eq:hierarchy_sup} for different measures of nonstabilizerness using error mitigated $A_3$ from IonQ experiment as well as simulations of $F_\text{STAB}^{-1}$, stabilizer extent $\xi$ and robustness of magic $R$. Quantum states generated by random Clifford circuits doped with $N_\text{T}$ T-gates. 
	}
	\label{fig:exp_sup}
\end{figure*}

In Fig.~\ref{fig:exp_sup}, we demonstrate the hierarchy of bounds between the measures of nonstabilizerness, where $A_3$ is measured in experiment.

\begin{figure*}[htbp]
	\centering	
  	\subfigimg[width=0.3\textwidth]{a}{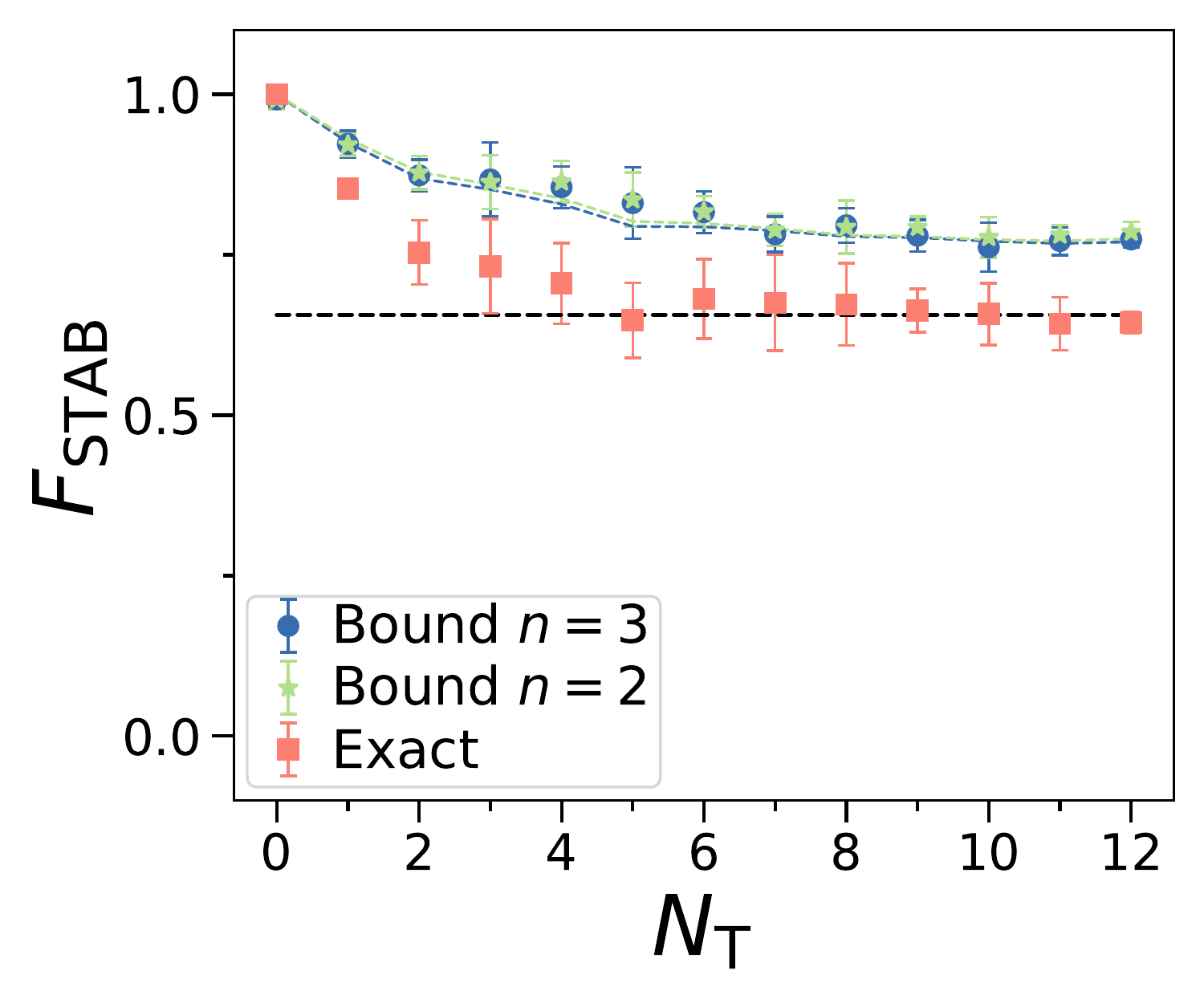}
	\subfigimg[width=0.3\textwidth]{b}{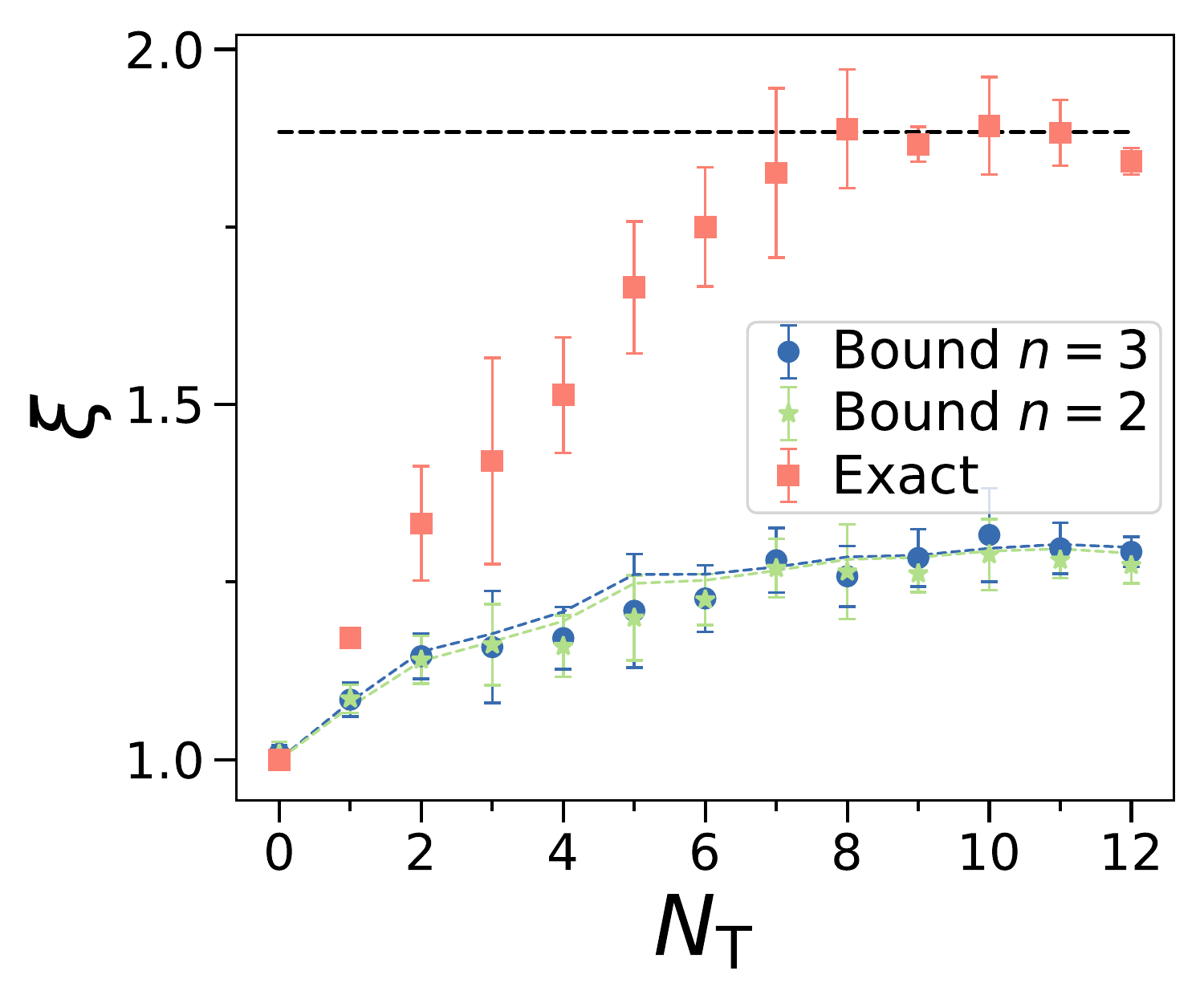}
 	\subfigimg[width=0.3\textwidth]{c}{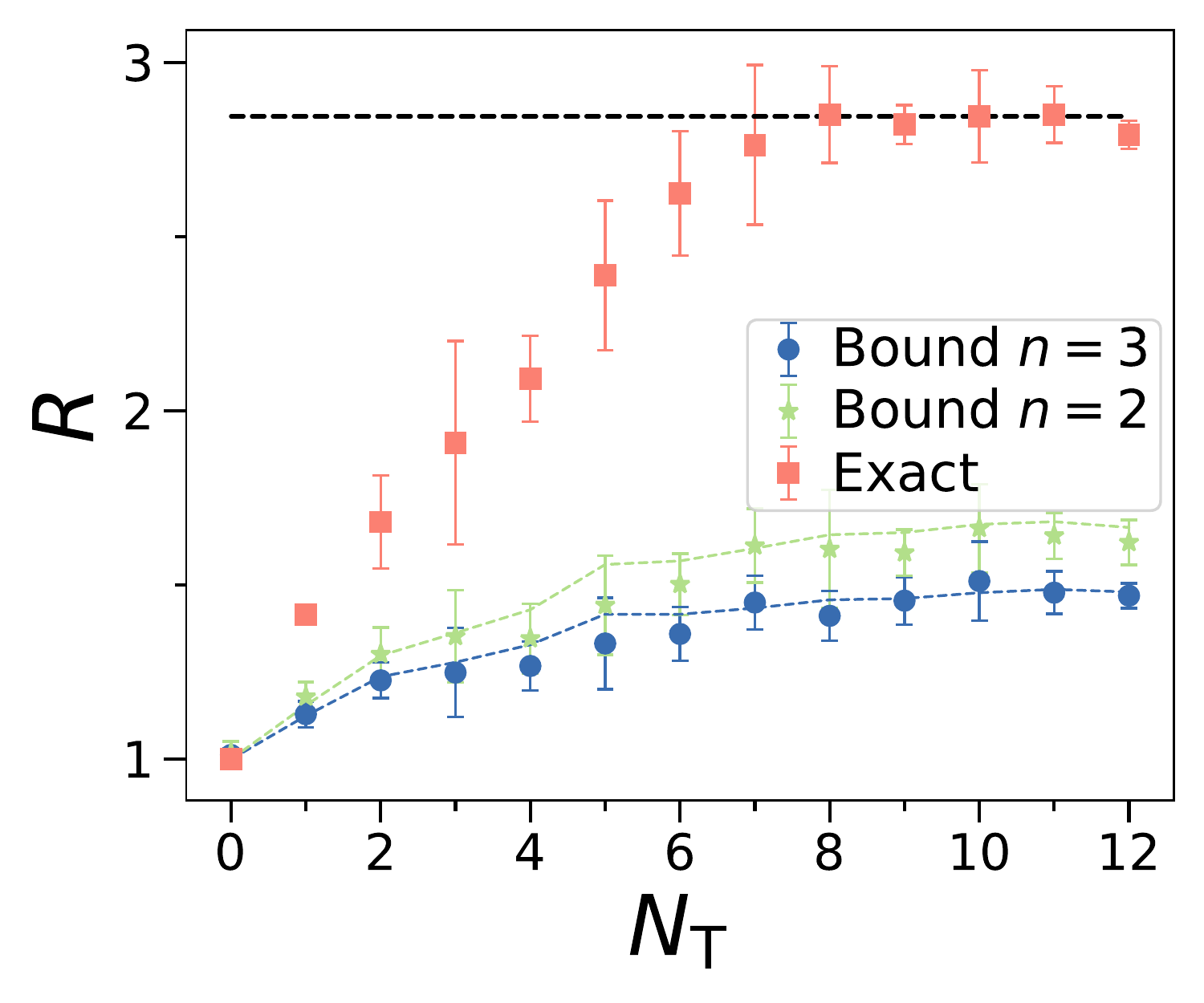}
	\caption{Compute bounds for other nonstabilizerness monotones by using $A_2$ and $A_3$. 
 \idg{a} Stabilizer fidelity $F_\text{STAB}$, \idg{b} Stabilizer extent $\xi$ and \idg{c}
 robustness of magic $R$. 
 Orange dots indicate exact simulation of the measure. Blue an green dots are bounds from error mitigated measurements, while dashed lines indicate bounds computed using the simulated $A_n$.
 We show quantum states generated by random Clifford circuits doped with $N_\text{T}$ T-gates on the IonQ quantum computer, where we compute 6 random instances of the state. 
We have $N=3$ qubits, $L=10^3$ Bell measurements and a measured depolarization error of $p\approx0.1$.
	}
	\label{fig:exp_other}
\end{figure*}

In Fig.~\ref{fig:exp_other} we demonstrate our lower bounds for robustness of magic $R$ and stabilizer extent $\xi$ using our Algorithm 1 for the Pauli spectrum moments $A_n=2^{-N}\sum_{\sigma\in\mathcal{P}}\bra{\psi}\sigma\ket{\psi}^{2n}$. We compute the bounds using the error mitigated measured $A_2$ and $A_3$ using Algorithm 1 of the main text. We show the upper bound $F_\text{STAB}\le A_n^{\frac{1}{2n}}$ and the lower bounds  $\xi \ge A_n^{-\frac{1}{2n}}$. Further, we show $\text{R}\ge A_n^{\frac{1}{2(1-n)}}$ which is the improved bound only valid for $R$.
For $A_2$, Algorithm 1 is not efficient, however for our chosen qubit number and measurement samples we are able to achieve sufficient accuracy.

We find that the exactly simulated bounds closely match the result from the IonQ quantum computer. We find the bounds for $A_2$ and $A_3$ show similar result, except for $R$ where we find that $A_2$ gives a better bound. 

\begin{figure}[htbp]
	\centering	
\subfigimg[width=0.3\textwidth]{}{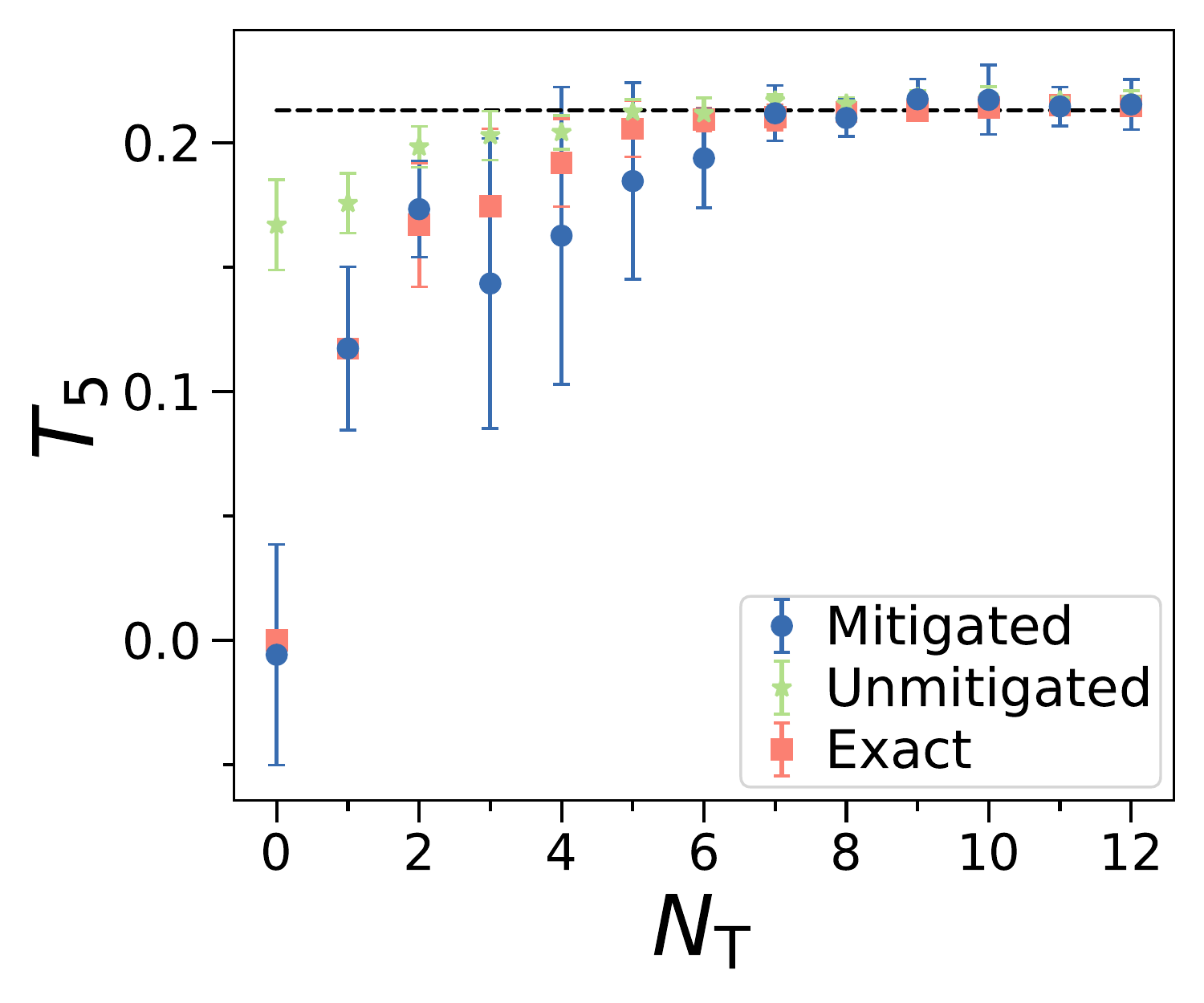}
	\caption{Measurement of $T_5$ with and without error mitigation. Dashed line is average value of $T_5$ for Haar random states.
	}
	\label{fig:exp_T}
\end{figure}

Finally, we also demonstrate measurement of the Tsallis entropy $T_5$ in Fig.~\ref{fig:exp_T}. We find similar behavior to $T_3$ as measured in the main text.

\revA{\section{Error mitigation for different noise models}\label{sec:noise}

\begin{figure*}[htbp]
	\centering	
 \subfigimg[width=0.3\textwidth]{a}{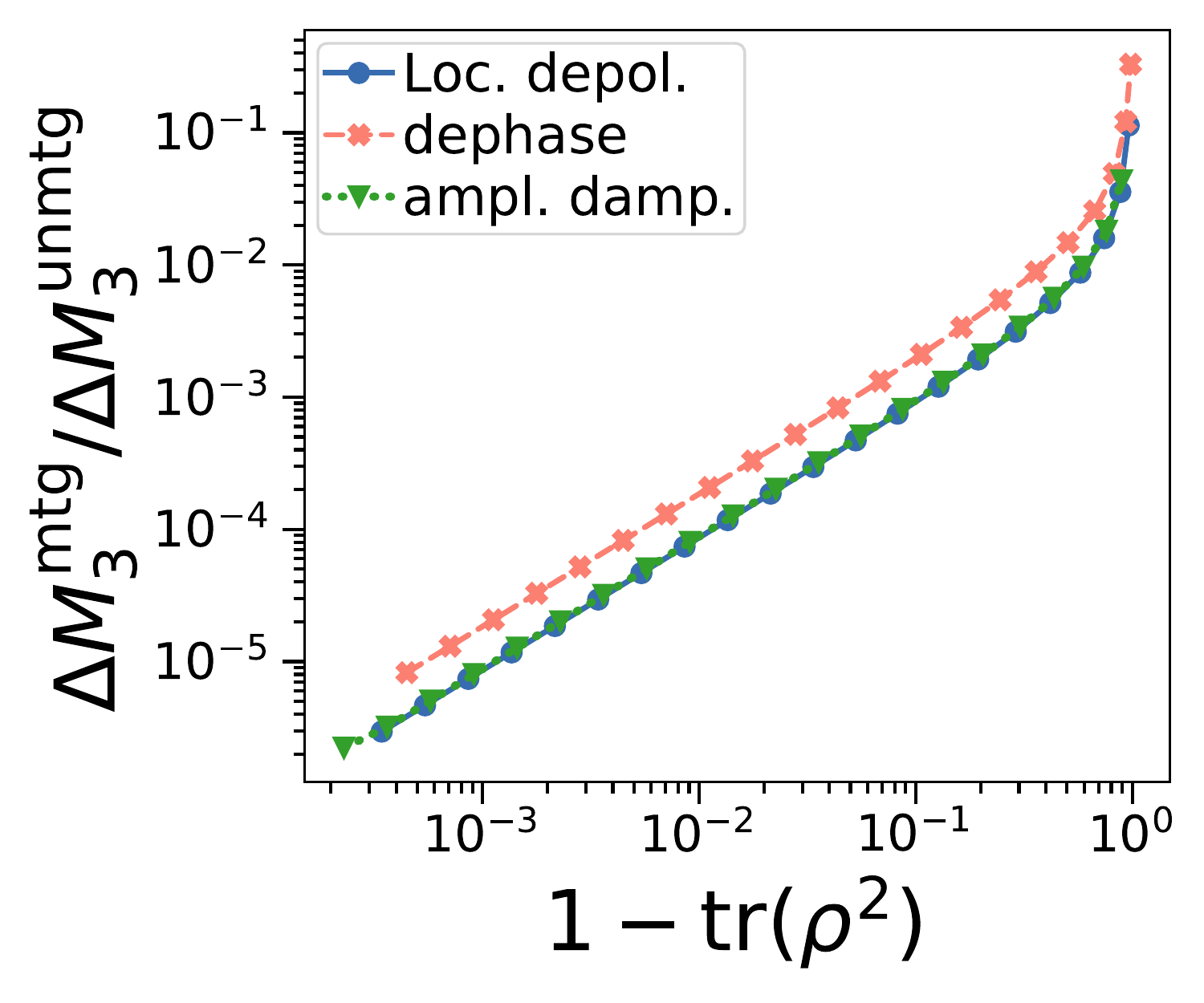}
  \subfigimg[width=0.3\textwidth]{b}{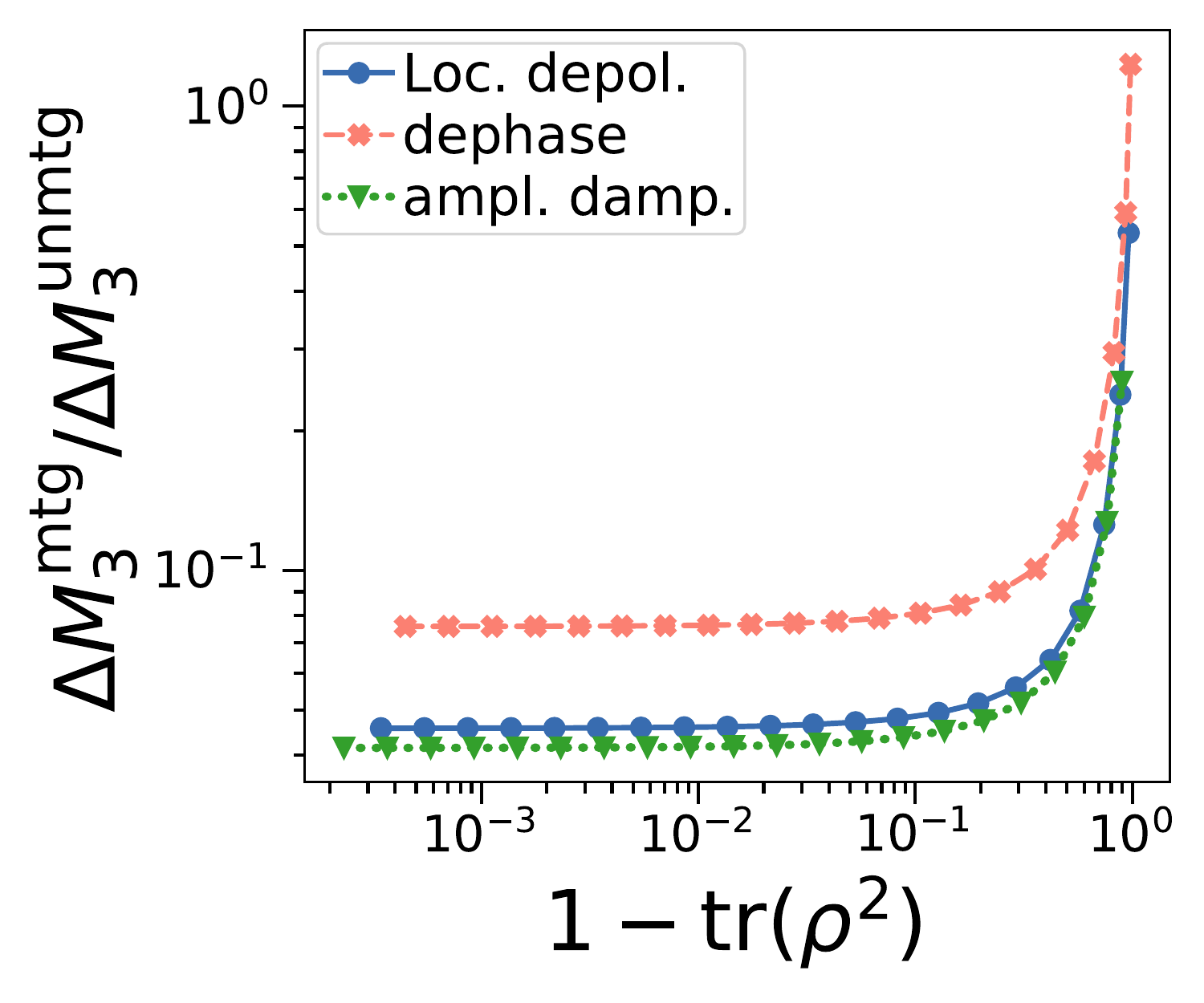}\\
\subfigimg[width=0.3\textwidth]{c}{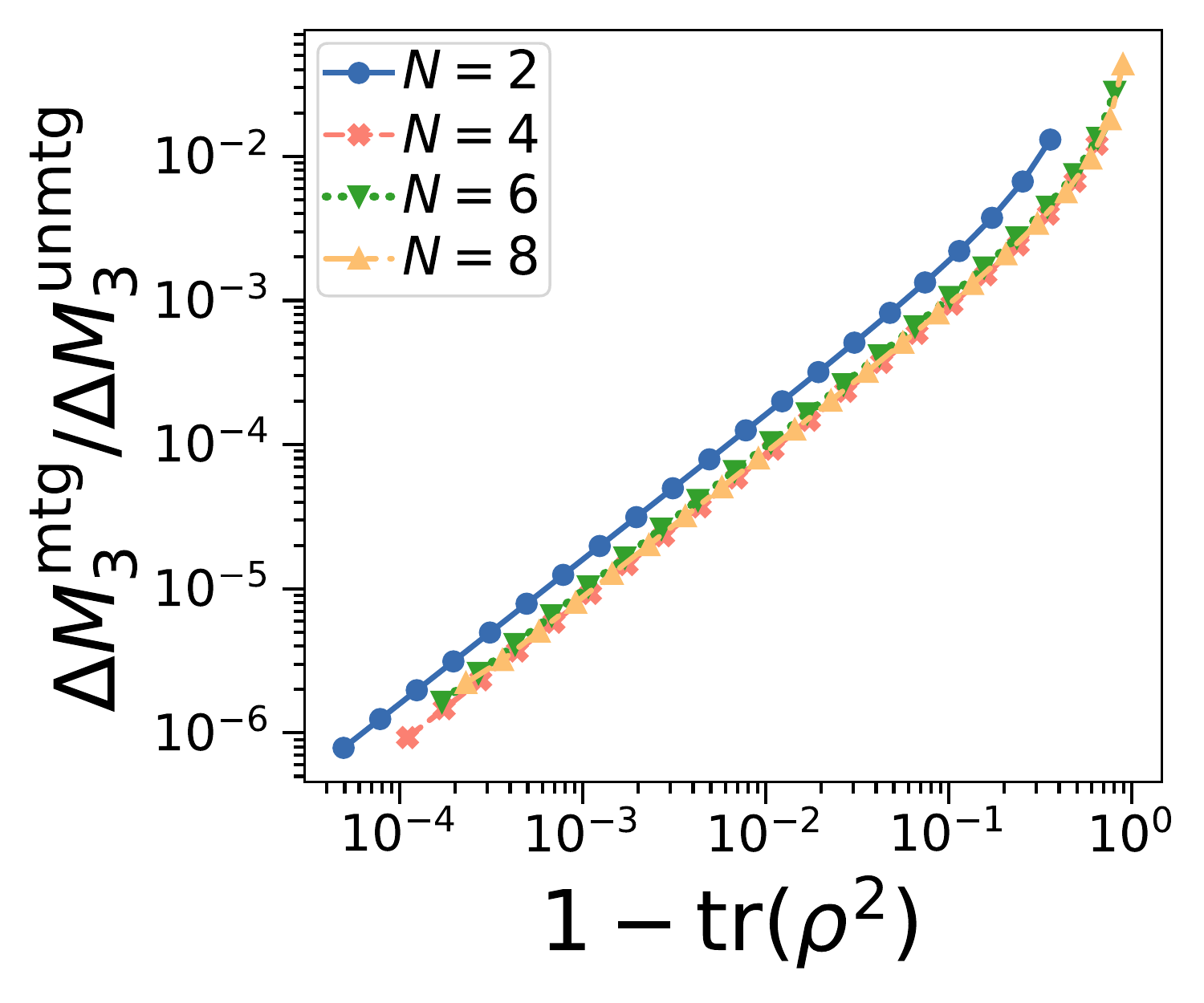}
\subfigimg[width=0.3\textwidth]{d}{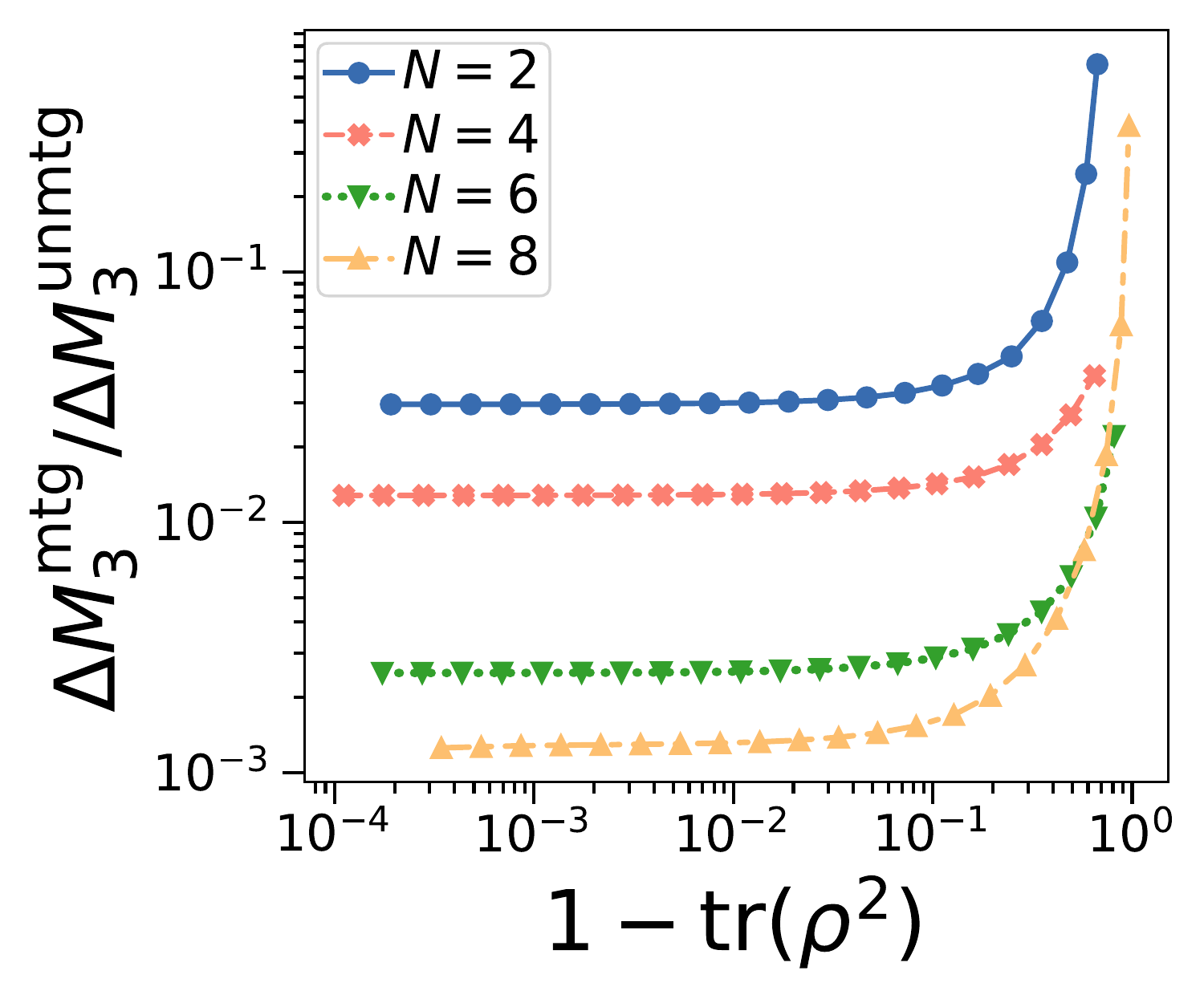}
	\caption{We simulate our error mitigation strategy for local depolarizing, dephasing and amplitude dampening noise. The noise channel is applied after each gate, where we consider layered single qubit and CNOT gates forming random Clifford circuits.
    We show relative error $\Delta M_\alpha^{\text{mtg}}/\Delta M_\alpha^{\text{unmtg}}$  against impurity $1-\text{tr}(\rho^2)$. Each point is averaged over $20$ random instances.
    \idg{a,c} We show random Clifford circuits, while in \idg{b,d} we show Clifford circuits doped with $N_{\text{T}}=N$ T-gates. 
    We show \idg{a,b} different noise models for $N=8$ qubits, while in \idg{c,d} we show different number of qubits $N$ for amplitude dampening noise.
	}
	\label{fig:noise_sim}
\end{figure*}

We now study the performance of our error mitigation strategy for different noise models. Our error mitigation strategy assumes global depolarization noise, which is a simple noise model. Our experiment indicate that it works well even for the noise of the IonQ quantum computer which is known to have a more complex noise model. Here, we provide further simulations with various noise models, finding that our model performs surprisingly well, both for unital and non-unital noise models.

We simulate random Clifford circuits constructed from layers of single-qubit rotations and CNOT gates arranged in a nearest-neighbor configuration, where we can dope the circuit with $N_\text{T}$ T-gates. After every gate, we apply a noise channel with noise strength $p$ on the qubits the gate acted on. 

We study the following three noise models, which have the following channel description for each qubit: 
For unital noise, we consider local depolarizing noise
\begin{equation}
    \Lambda_\text{ldp}(\rho)=(1-p)\rho +p/2 I_1\,,
\end{equation}
local dephasing noise
\begin{equation}
    \Lambda_\text{dephase}(\rho)=(1-p)\rho +p\sigma^z\rho\sigma^z\,.
\end{equation}
Further, we consider the amplitude dampening channel as a non-unital noise channel
\begin{equation}
    \Lambda_\text{a}(\rho)=(1-p)\rho +p\sigma^-\rho\sigma^+\,.
\end{equation}
To quantify the performance of the error mitigation, we compute the error in SE without error mitigation 
\begin{equation}
    \Delta M_\alpha^{\text{unmtg}}=\vert M_\alpha^{\text{unmtg}}-M_\alpha^{\text{pure}}\vert
\end{equation}
and with error mitigation
\begin{equation}
    \Delta M_\alpha^{\text{mtg}}=\vert M_\alpha^{\text{mtg}}-M_\alpha^{\text{pure}}\vert
\end{equation}
where $M_\alpha^{\text{pure}}$ is the noise-free SE, $M_\alpha^{\text{unmtg}}$ the unmitigated SE and $M_\alpha^{\text{mtg}}$ the mitigated SE. 
We consider now the ratio $\Delta M_\alpha^{\text{mtg}}/\Delta M_\alpha^{\text{unmtg}}$, which indicates the relative error reduction with error mitigation, which is smaller $1$ when error mitigation yields an improvement.

We show the relative error for different noise models against impurity $1-\text{tr}(\rho^2)$ in Fig.~\ref{fig:noise_sim}.
We show simulation results for random Clifford circuits in Fig.~\ref{fig:noise_sim}a,c and Clifford circuits doped with $N_\text{T}=N$ T-gates in Fig.~\ref{fig:noise_sim}b,d. In Fig.~\ref{fig:noise_sim}a,b we find that error mitigation provides a substantial improvement in error for all considered noise models, both for unital and non-unital noise. For Clifford circuits, we find empirically a linear relationship $\Delta M_\alpha^{\text{mtg}}/\Delta M_\alpha^{\text{unmtg}}\propto 1-\text{tr}(\rho^2)$ between relative error and impurity. Whereas for doped Clifford circuits the improvement converges to a constant for small $1-\text{tr}(\rho^2)$.

In Fig.~\ref{fig:noise_sim}c,d, we show amplitude dampening noise for different number of qubits $N$. We find that the reduction in error due to error mitigation  is independent of qubit number for Clifford circuits, while for doped Clifford circuits the error decreases even further with $N$. This indicates that our error mitigation strategy can work well even for more qubits.

}

\section{Nonstabilizerness, scrambling, multifractal flatness and OTOCs}\label{sec:MagicOTOC}
Here, we study the relationship of SEs, scrambling, multifractal flatness and OTOCs for various circuit and Hamiltonian evolution models.

First, we review the properties as introduced in the main text. First, we have the $n$-R\'enyi SE of state $\ket{\psi}$ as
\begin{equation}
    M_n(\ket{\psi})=(1-n)^{-1} \ln (A_n(\ket{\psi}))
\end{equation}
with the $n$-th moment of the Pauli spectrum
\begin{equation}
A_n(\ket{\psi})=2^{-N}\sum_{\sigma \in \mathcal{P}}\braket{\psi|\sigma|\psi}^{2n}\,,
\end{equation}
where $\mathcal{P}$ is the set of all Pauli strings with sign $+1$.
We can also define the $n$-R\'enyi SE $M_n(\ket{U})$ for unitary $U$ via its unique Choi state $\ket{U}=I_N \otimes U \ket{\Phi}$ with the maximally entangled state $\ket{\Phi}$.

The multifractal flatness is given by
\begin{equation}
    \mathcal{F}(\ket{\psi})=\mathcal{I}_3(\ket{\psi})-\mathcal{I}_2^2(\ket{\psi})
    \end{equation}
where $\mathcal{I}_q(\ket{\psi})=\sum_{k} \vert\braket{k\vert \psi}\vert^{2q}$ is the participation entropy,  $\ket{k}$ are computational basis states, $q>0$ and $0\le \mathcal{I}_q\le 1$ ~\cite{castellani1986multifractal}. 
Then, the multifractal flatness averaged over Clifford unitaries $\bar{\mathcal{F}}(\ket{\psi})$ is related to SE via
\begin{equation}\label{eq:flatnessCliff}
\bar{\mathcal{F}}(\ket{\psi})=\underset{U_\text{C} \in \mathcal{C}_N}{\mathbb{E}}[\mathcal{F}(U_\text{C}\ket{\psi})]=\frac{2(1-A_2(\ket{\psi}))}{(2^N+1)(2^N+2)}\,.
\end{equation}

Finally, we have the $4n$-point OTOC
\begin{align*}
&\text{otoc}_{4n}(U,\sigma,\sigma')=\left(2^{-N}\text{tr}(\sigma U\sigma' U^\dagger)\right)^{2n}\numberthis=2^{-N}\text{tr}(\langle \sigma^{(2n)} \prod_{i=1}^{2n}( U \sigma U^\dagger \sigma' \sigma^{(i-1)} \sigma^{(i)})\rangle_{\sigma^{(1)},\dots,\sigma^{(2n)}})\,,
\end{align*}
where $\sigma^{(0)}=I_N$ and $\langle X \rangle_{\sigma^{(1)},\sigma^{(2)},\dots,\sigma^{(2n)}}= 4^{-2nN}\sum_{\sigma^{(1)} ,\dots,\sigma^{(2n)} \in \mathcal{P}} X$ denotes the average over the Pauli strings $\sigma^{(1)},\dots,\sigma^{(2n)}$.
The $4n$-point OTOC averaged over all Pauli strings $\sigma$, $\sigma'$ can be related to SE via~\cite{leone2023nonstabilizerness}
\begin{equation}
4^{-N}\sum_{\sigma,\sigma'\in \mathcal{P}}\text{otoc}_{4n}(U,\sigma,\sigma')=A_n(\ket{U})\,.\label{eq:OTOC_Pauliavg_sup}
\end{equation}

\subsection{Derivation of Clifford-averaged OTOC}
While \eqref{eq:OTOC_Pauliavg_sup} relates OTOCs to the average over random Pauli operators to SEs, it would be good to connect them also to Clifford averages. 
We now derive the relationship between SEs and OTOCs averaged over random Cliffords.

First, we note that any Clifford unitary $U_\text{C}$ applied to a Pauli string $\sigma\in \mathcal{P}/I_N$ (except identity $I_N$) yields another Pauli string $\sigma'=U_\text{C}\sigma U_\text{C}^\dagger \in \mathcal{P}/I_N$. This mapping is bijective, i.e. each Pauli is mapped to exactly one other Pauli string. Further, the identity is mapped back onto itself, i.e. $U_\text{C} I_N U_\text{C}^\dagger=I_N$.

Now, we define the OTOC averaged over random Clifford unitaries ${\mathbb{E}}_{U_\text{C},U_\text{C}'\in \mathcal{C}_N}[\text{otoc}_{4n}(U_\text{C}U U_\text{C}',\sigma_{\text{a}},\sigma_{\text{b}})]$ where we have any $\sigma_{\text{a}},\sigma_{\text{b}}\in \mathcal{P}/I_N$. We find
\begin{align*}
\overline{\text{OTOC}}_{4n}(U)\equiv&\underset{U_\text{C},U_\text{C}'\in \mathcal{C}_N}{\mathbb{E}}[\text{otoc}_{4n}(U_\text{C}U U_\text{C}',\sigma_{\text{a}},\sigma_{\text{b}})]\\
=&\underset{U_\text{C},U_\text{C}' \in \mathcal{C}_N}{\mathbb{E}}[\left(2^{-N}\text{tr}(\sigma_{\text{a}}  U_\text{C} U U_\text{C}'  \sigma_{\text{b}} U_\text{C}'^\dagger U^\dagger U_\text{C}^\dagger)\right)^{2n}]\\
=&\frac{1}{(4^{N}-1)^2}\sum_{\sigma,\sigma'\in \mathcal{P}/I_N}\left(2^{-N}\text{tr}(\sigma 
 U  \sigma'  U^\dagger)\right)^{2n}=\frac{1}{(4^{N}-1)^2}(\sum_{\sigma,\sigma'\in \mathcal{P}}\left(2^{-N}\text{tr}(\sigma 
 U  \sigma'  U^\dagger)\right)^{2n}  -1)\\
 =&\frac{1}{(4^{N}-1)^2}(\sum_{\sigma,\sigma'\in \mathcal{P}}\text{otoc}_{4n}(U,\sigma,\sigma')  -1)\\
 =& \frac{A_n(\ket{U})4^{N}-1}{(4^{N}-1)^2}\numberthis\label{eq:OTOC_avg_sup}
\end{align*}
where from second to third line we used the fact that  random Clifford transformations of the form $U_\text{C}  \sigma_{\text{a}} U_\text{C}^\dagger$ map any Pauli string $\sigma_{\text{a}}\in \mathcal{P}/I_N$   to other Pauli strings (excluding identity $I_N$) with the same frequency, as well as  $\vert \mathcal{P}/I_N \vert=4^N-1$. In the third line we used $\text{otoc}_{4n}(U,\sigma,I_N)=\text{otoc}_{4n}(U,I_N,\sigma)=\delta_{\sigma, I_N}$, and in the final line we used~\eqref{eq:OTOC_avg_sup}.

We now study the 2-R\'enyi SE $M_2$, the 8-point OTOC $\text{otoc}_{8}(U,\sigma,\sigma')$ and the multifractal flatness $\mathcal{F}$ for various models. For $\mathcal{F}$, we study the flatness in regard to the state $U\ket{0}$.

\subsection{Random circuits}
First, we study random circuits composed of $d$ layers of random single-qubit rotations and nearest-neighbor CNOT gates arranged in a nearest-neighbor chain. These circuits are known to approximate Haar random unitaries in linear circuit depth~\cite{brandao2016local}.

\begin{figure*}[htbp]
	\centering	
 \subfigimg[width=0.3\textwidth]{a}{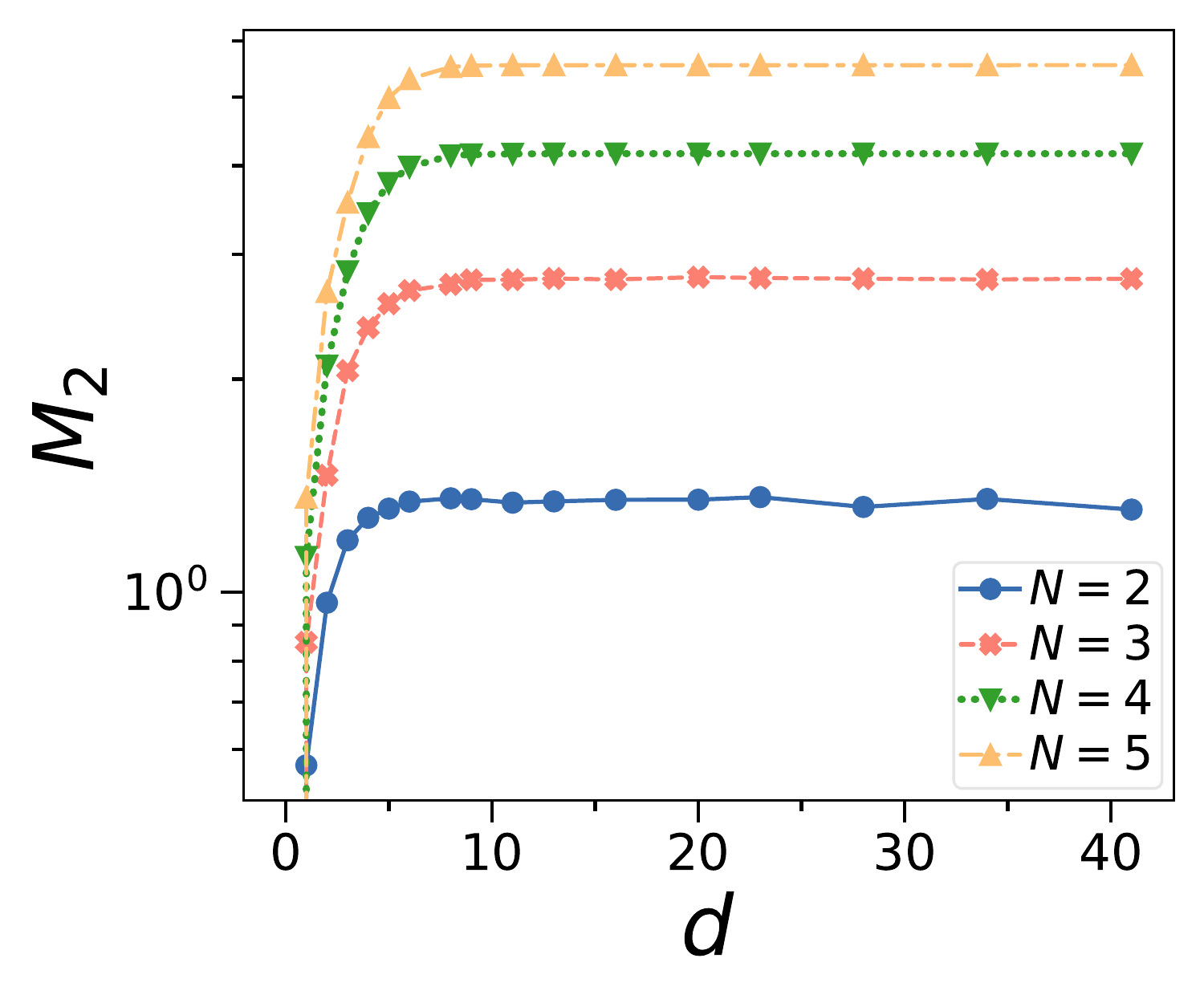}
  \subfigimg[width=0.3\textwidth]{b}{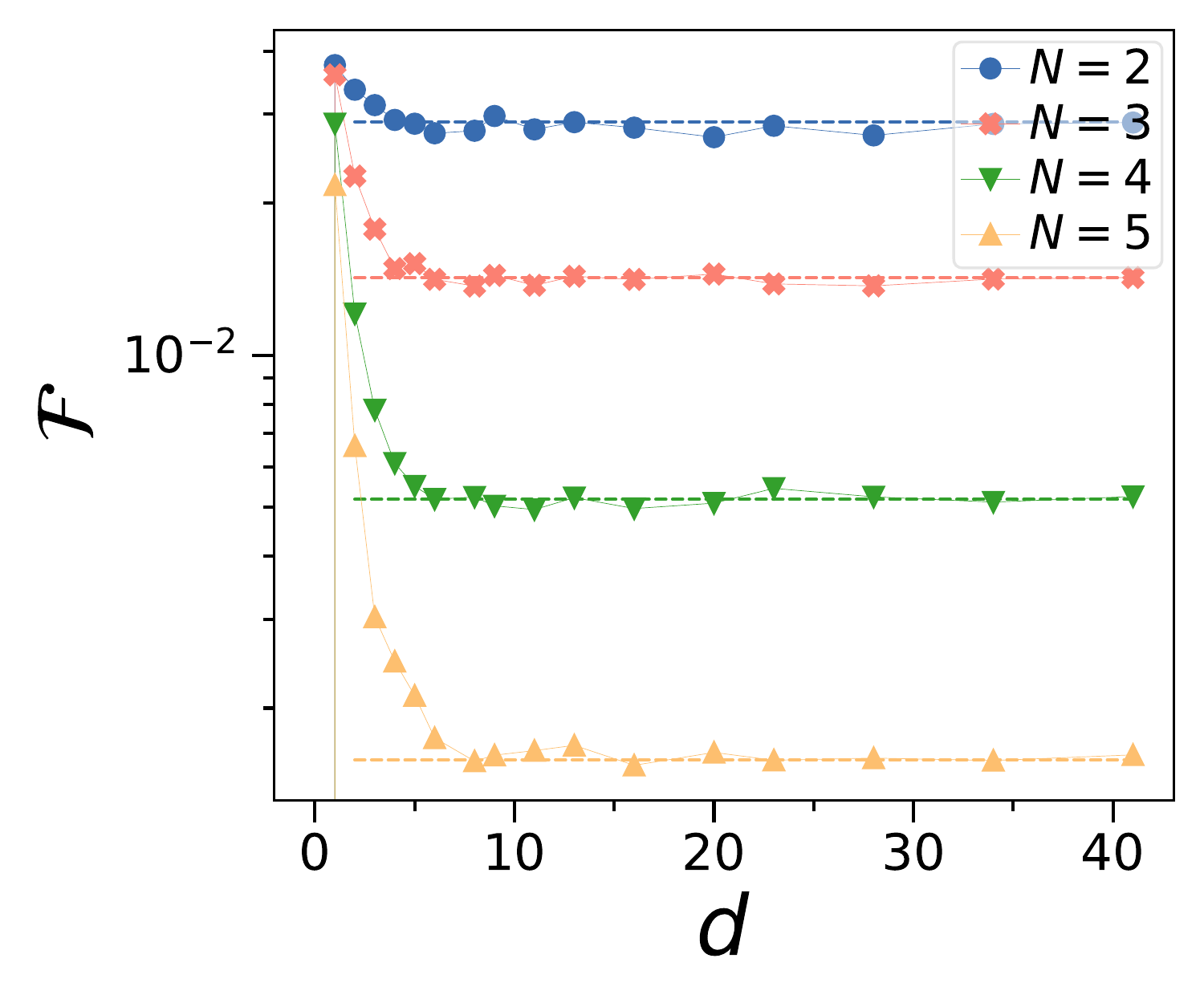}
\subfigimg[width=0.3\textwidth]{c}{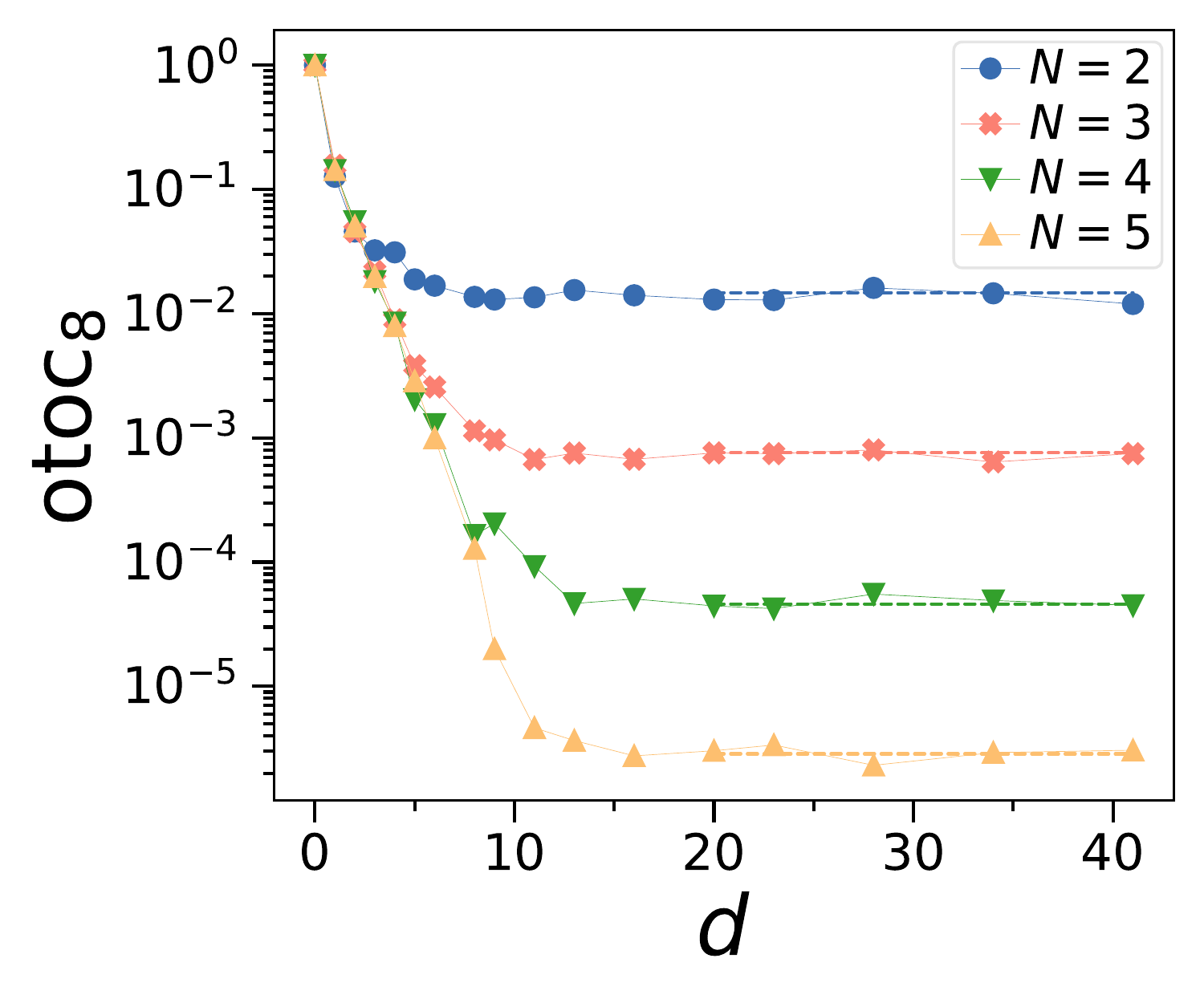}
\subfigimg[width=0.3\textwidth]{c}{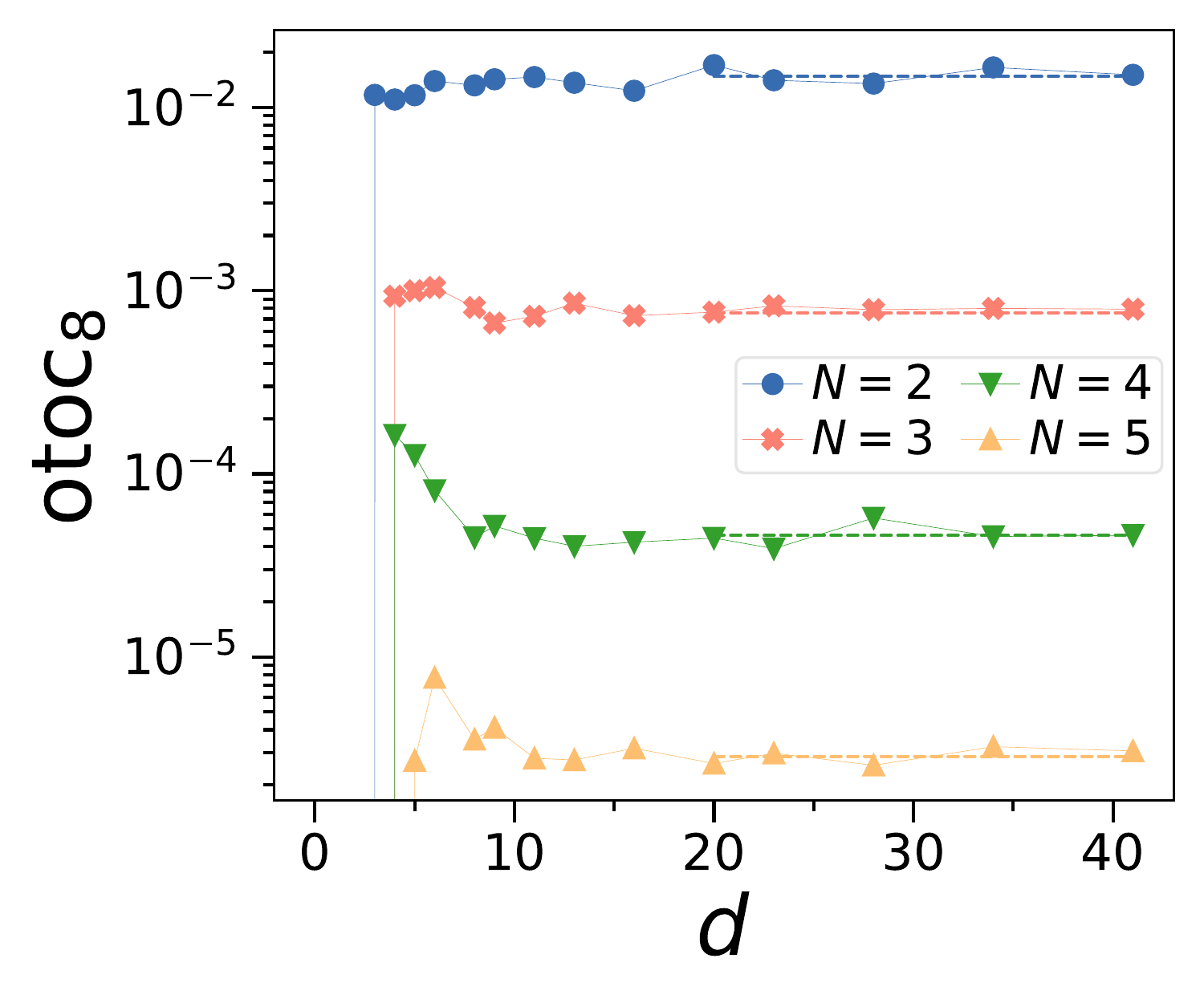}
	\caption{ We study random circuits of $d$ layers. We regard 2-R\'enyi SE $M_2(\ket{U})$, the 8-point OTOC $\text{otoc}_{8}$ and the multifractal flatness $\mathcal{F}(U\ket{0})$. We show data as function of circuit depth $d$ for different number of qubits $N$.
    \idg{a} $M_2$ against $d$.
    \idg{b} $\mathcal{F}$ against $d$.
 Dashed line is relationship between SE and Clifford averaged multifractal flatness~\eqref{eq:flatnessCliff}. 
 \idg{c} $\text{otoc}_{8}(U,\sigma^x_1,\sigma^x_1)$ against $d$, where dashed line is the Clifford-averaged OTOC given by~\eqref{eq:OTOC_avg_sup}. 
  \idg{d} $\text{otoc}_{8}(U,\sigma^x_1,\sigma^z_N)$ against $d$.
 All results are averaged over $1000$ random instances.
	}
	\label{fig:randomCircuit}
\end{figure*}

We show $M_2$, $\mathcal{F}$ and $\text{otoc}_{8}$ against $d$ in Fig.~\ref{fig:randomCircuit}. We show $M_2$ against $d$ in Fig.~\ref{fig:randomCircuit}a, observing a rapid increase with $d$ which converges approximately at $d\propto N$ depth. We observe that the multifractal flatness in Fig.~\ref{fig:randomCircuit}b converges to its Clifford-averaged value with $d$.
Similarly, we find in Fig.~\ref{fig:randomCircuit}c,d that same-site OTOC $\text{otoc}_{8}(U,\sigma^x_1,\sigma^x_1)$ and non-local $\text{otoc}_{8}(U,\sigma^x_1,\sigma^z_N)$ converge to its Clifford-average values. Curiously, we find that OTOCs require higher $d$ to converge than multifractal flatness or SE, indicating that they are more sensitive to scrambling.

\subsection{Random clifford + T}

\begin{figure*}[htbp]
	\centering	
 \subfigimg[width=0.3\textwidth]{a}{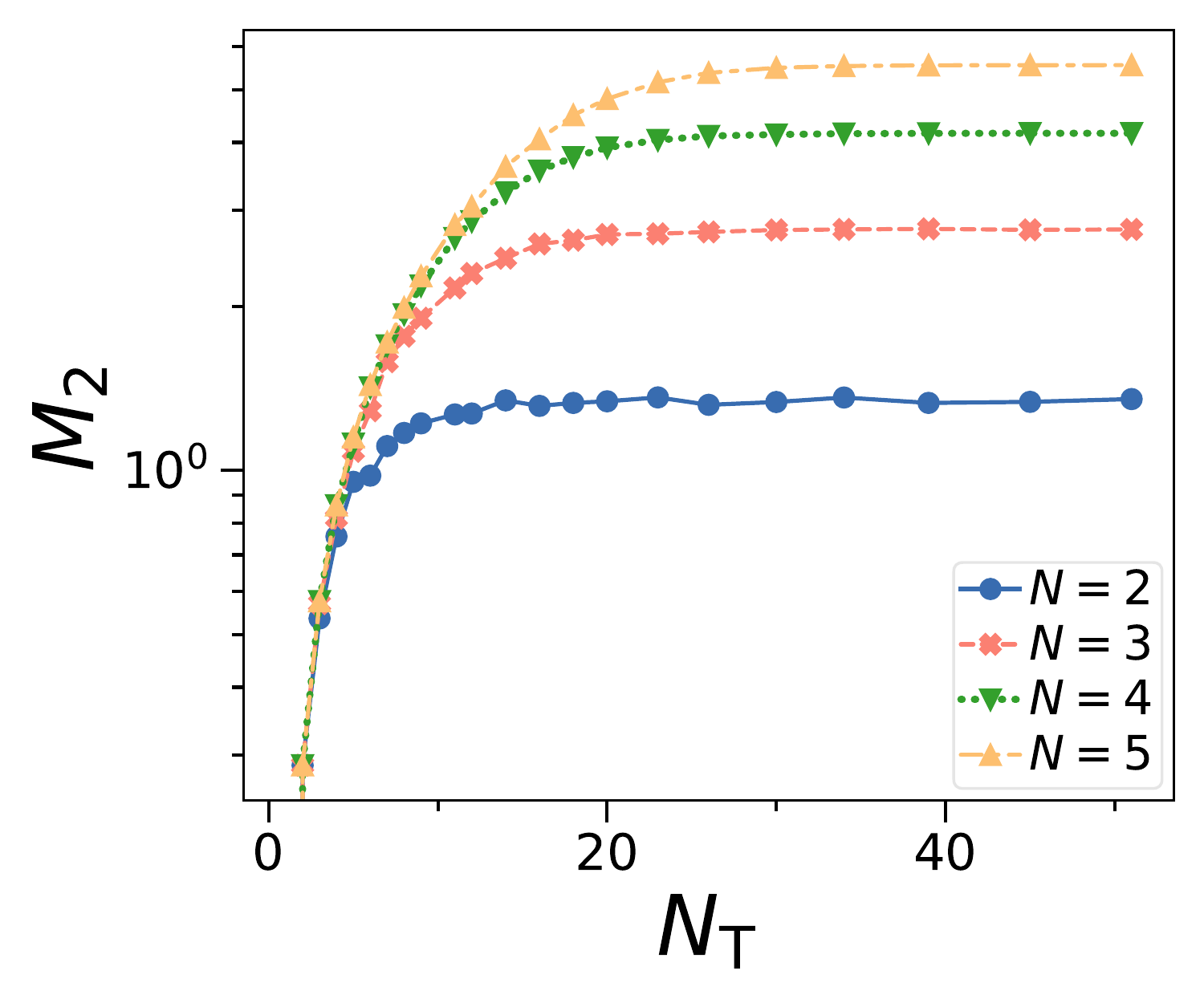}
  \subfigimg[width=0.3\textwidth]{b}{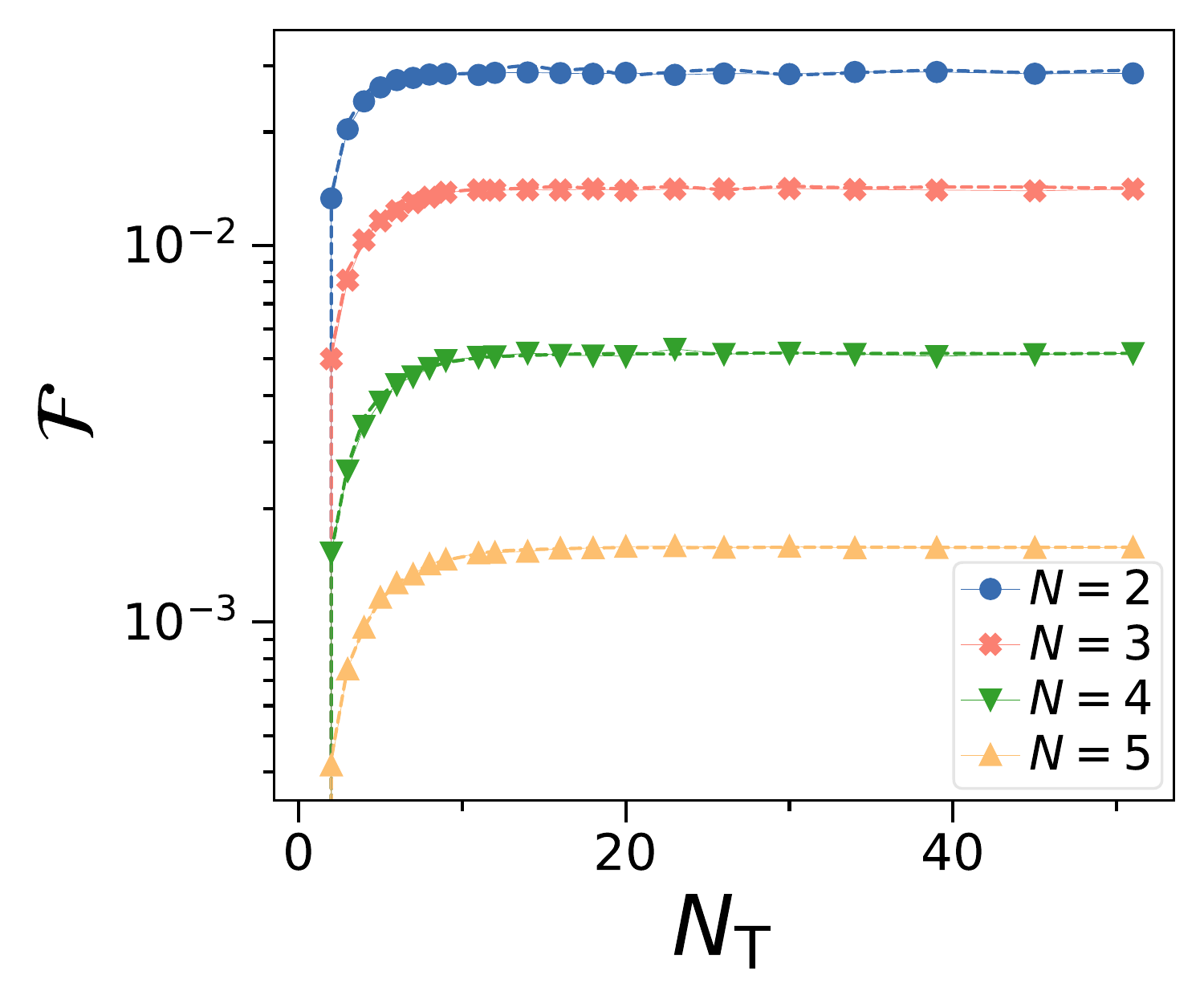}
\subfigimg[width=0.3\textwidth]{c}{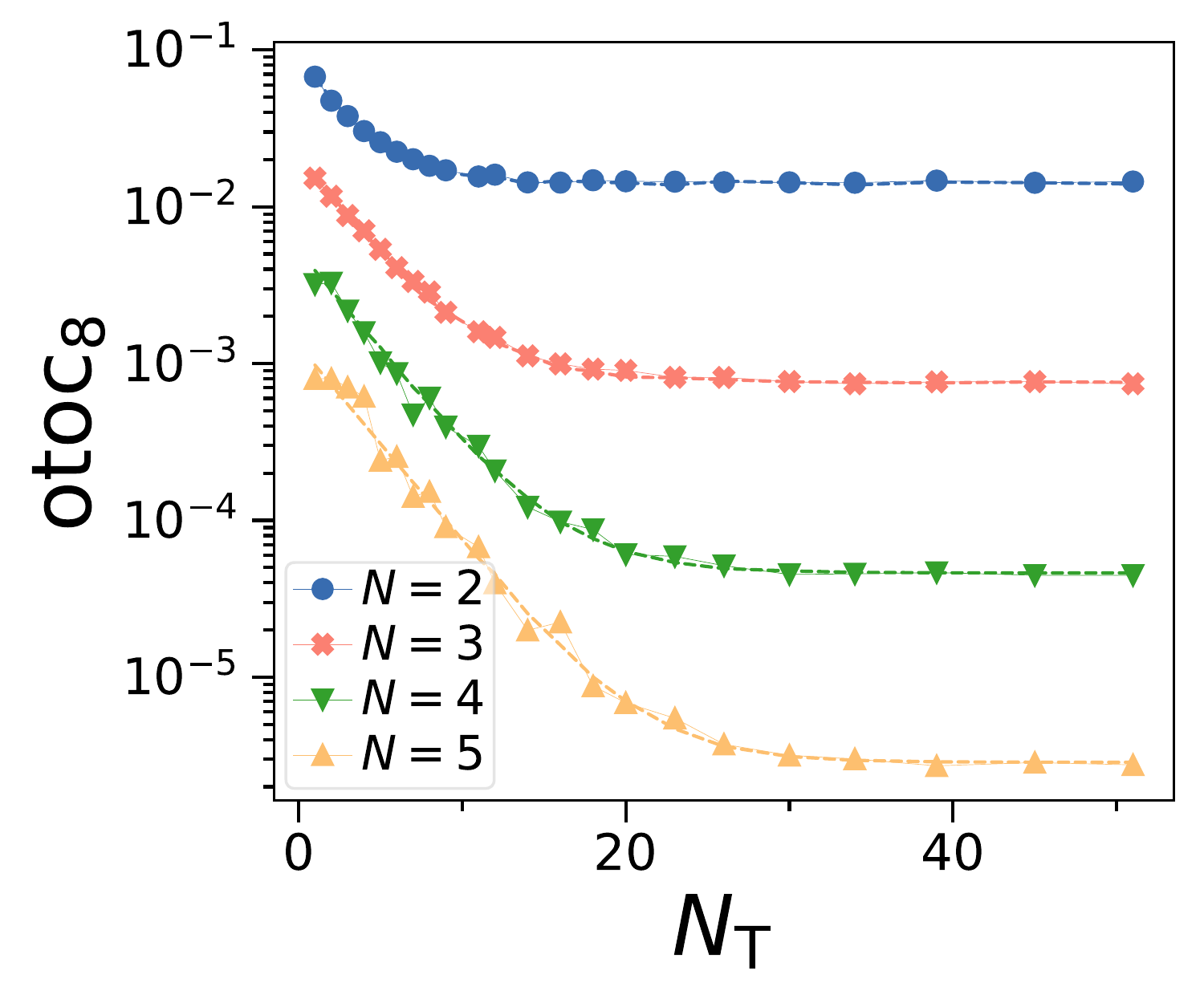}
	\caption{ We study unitary $U$ chosen as random Clifford unitaries from the Clifford group doped with T-gates~\eqref{eq:random_CliffTexp_sup}. We regard 2-R\'enyi SE $M_2(\ket{U})$, the 8-point OTOC $\text{otoc}_{8}(U,\sigma,\sigma')$ and the multifractal flatness $\mathcal{F}(U\ket{0})$. We show data as function of number of T-gates $N_{\text{T}}$, where we show different number of qubits $N$.
    \idg{a} $M_2$ against $N_{\text{T}}$.
    \idg{b} $\mathcal{F}$ against $N_{\text{T}}$.
 Dashed line is relationship between SE and Clifford averaged multifractal flatness~\eqref{eq:flatnessCliff}. 
 \idg{c} $\text{otoc}_{8}$ against $N_{\text{T}}$, where dashed line is the Clifford-averaged OTOC given by~\eqref{eq:OTOC_avg_sup}. All results are averaged over $10000$ random instances.
	}
	\label{fig:cliffordT}
\end{figure*}

Next, we study random Clifford circuits  drawn from the Clifford group $\mathcal{C}_N$ doped with $N_{\text{T}}$ T-gates which are given as
\begin{equation}\label{eq:random_CliffTexp_sup}
U(N_{\text{T}})=U_\text{C}^{(0)}\prod_{k=1}^{N_{\text{T}}} V_\text{T}^{(k)}U_\text{C}^{(k)}\,.
\end{equation}
We show $M_2$, $\mathcal{F}$ and $\text{otoc}_{8}$ against $N_{\text{T}}$ in Fig.~\ref{fig:cliffordT}. The dashed line is $\mathcal{F}$ and $\text{otoc}_{8}$ averaged over random Clifford unitaries computed using the SE $M_2$. We find that the observed values match the values predicted via~\eqref{eq:OTOC_avg_sup} and~\eqref{eq:flatnessCliff}. Note that all $\text{otoc}_{8}$ for any pair $\sigma$, $\sigma'\in\mathcal{P}/{I_N}$ yield the same value due to the scrambling with $U_\text{C}$.

\subsection{Fixed depth clifford + T}

\begin{figure*}[htbp]
	\centering	
 \subfigimg[width=0.3\textwidth]{a}{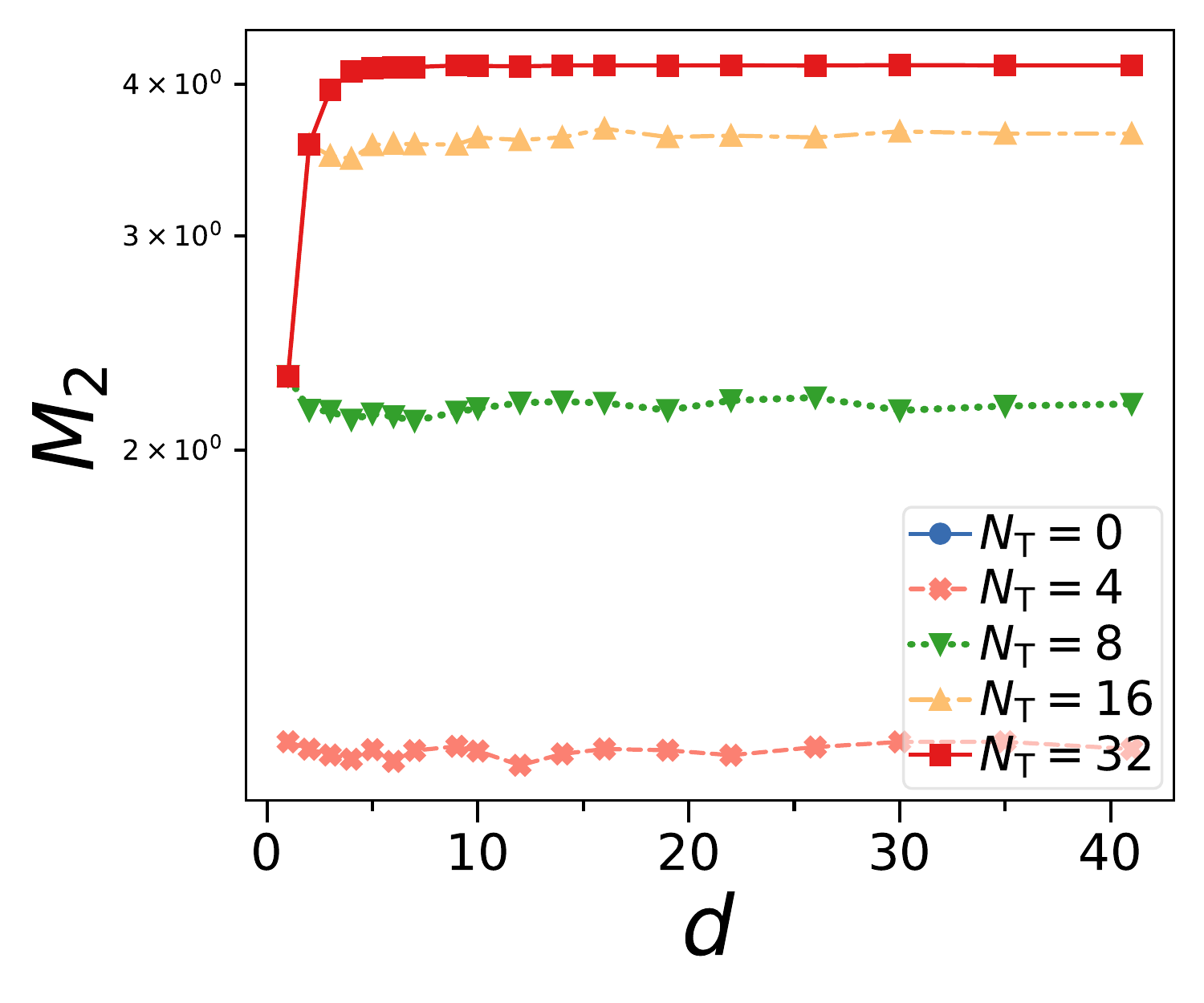}
  \subfigimg[width=0.3\textwidth]{b}{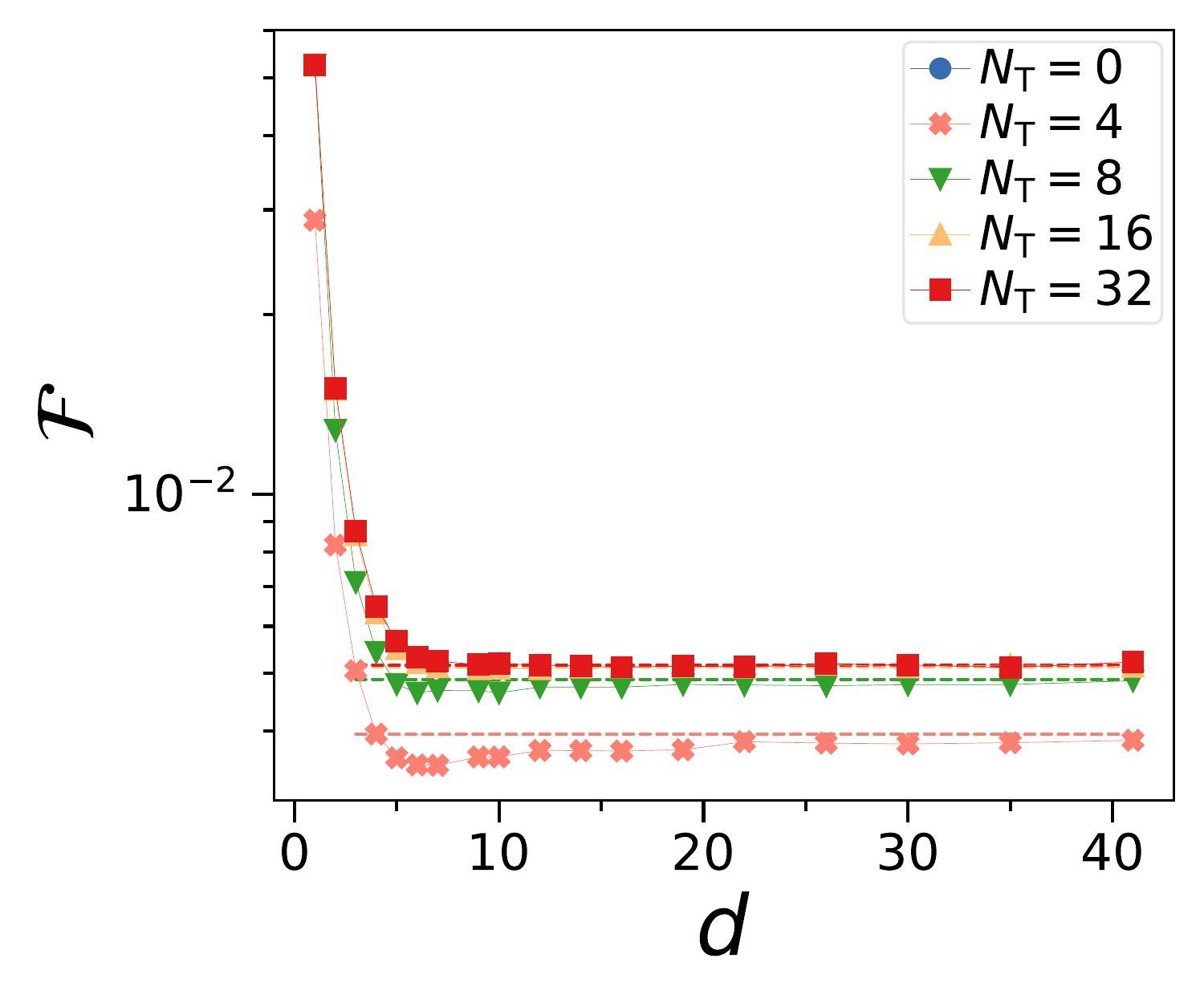}
\subfigimg[width=0.3\textwidth]{c}{otocLocConvSAN4d40m1c32r30000z1W1p25R.pdf}
\subfigimg[width=0.3\textwidth]{d}{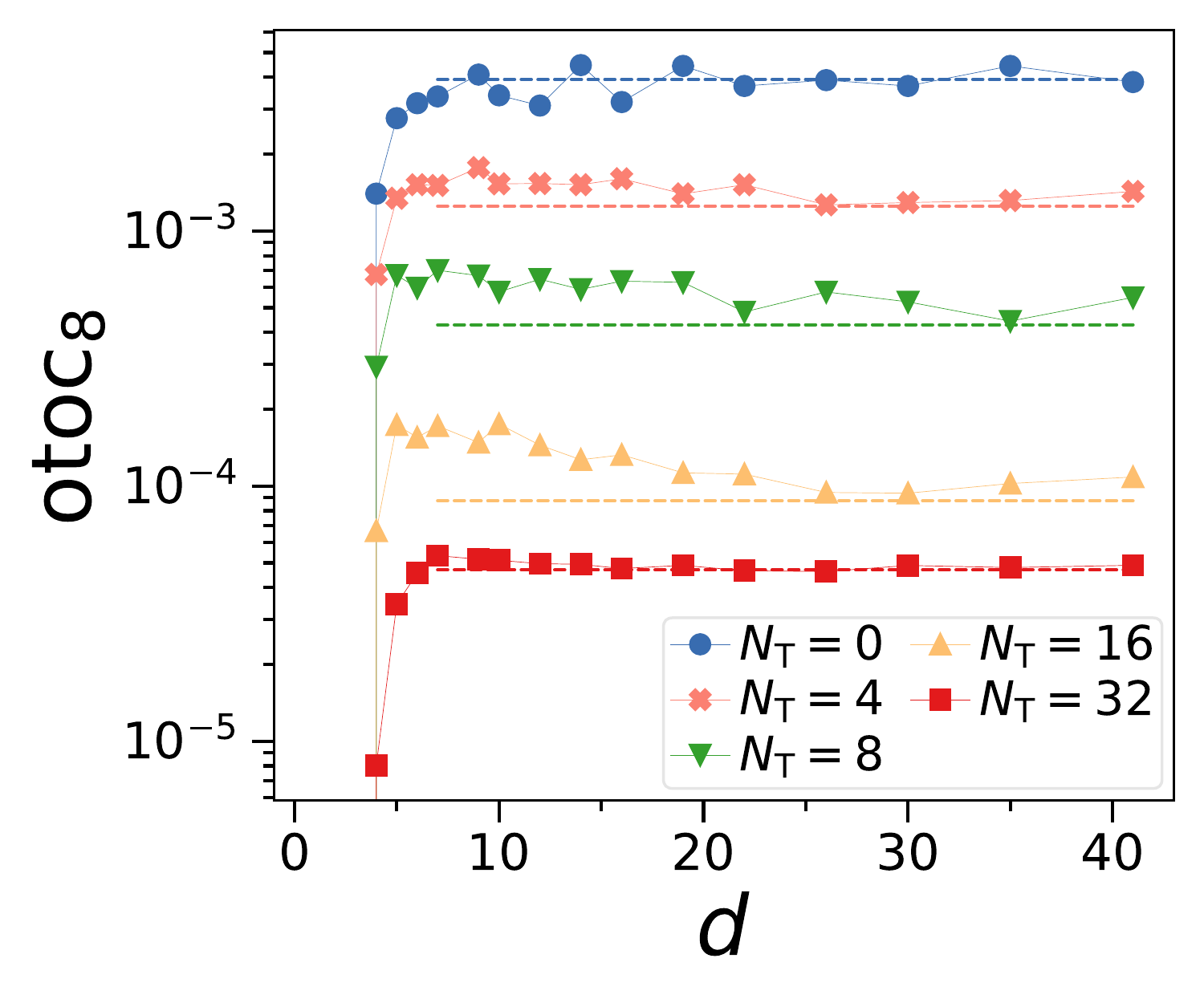}
	\caption{Circuit with $d$ layers of Clifford gates with $N_{\text{T}}$ T-gates injected. We study  2-R\'enyi SE $M_2(\ket{U})$, the 8-point OTOC $\text{otoc}_{8}(U,\sigma^x_1,\sigma^x_1)$ and the multifractal flatness $\mathcal{F}(U\ket{0})$. We show data as function of number of number of Clifford gate layers $d$, with varying number of T-gates $N_{\text{T}}$.
    \idg{a} $M_2$ against $d$.
    \idg{b} $\mathcal{F}$ against $d$.
 Dashed line is relationship between SE and Clifford averaged multifractal flatness~\eqref{eq:flatnessCliff} for large $d$. 
 \idg{c} $\text{otoc}_{8}(U,\sigma^x_1,\sigma^x_1)$ against $d$, where dashed line is the Clifford-averaged OTOC given by~\eqref{eq:OTOC_avg_sup} for large $d$. 
  \idg{d} $\text{otoc}_{8}(U,\sigma^x_1,\sigma^z_N)$ against $d$.
 All results are for $N=4$ qubits averaged over $30000$ random instances.
	}
	\label{fig:cliffordTd}
\end{figure*}

Next, we study in Fig.~\ref{fig:cliffordTd} fixed depth Clifford circuits doped with T-gates. In particular, we simulate a $d$ layer circuit, where each layer is composed of random single-qubit Clifford gates on each qubit and CNOT gates arranged in a nearest-neighbor chain. This layered structure is repeated $d$ times. Now, we randomly insert $N_{\text{T}}$ T-gates in this circuit at random positions. 
In this circuit, the degree of scrambling depends on the number of Clifford layers $d$, while the nonstabilizerness depends on $N_{\text{T}}$. 
In Fig.~\ref{fig:cliffordTd}a, we find that $M_2$ converges very fast with $d$, where the final value depends only on $N_{\text{T}}$. 
In Fig.~\ref{fig:cliffordTd}b, we study flatness $\mathcal{F}$. We find convergence to the minimal value given by the Clifford averaged value already for $d\sim 5$ layers. We observe that for small $N_{\text{T}}$ and intermediate $d$, $\mathcal{F}$ is smaller than its Clifford average value (shown by dashed line), while for large $N_{\text{T}}$ we find that $\mathcal{F}$ is always strictly larger. 
We regard two different OTOCs, in particular $\text{otoc}_{8}(U,\sigma^x_1,\sigma^x_1)$ in Fig.~\ref{fig:cliffordTd}c  
 and $\text{otoc}_{8}(U,\sigma^x_1,\sigma^z_N)$ in Fig.~\ref{fig:cliffordTd}d. $\text{otoc}_{8}(U,\sigma^x_1,\sigma^x_1)$ is initially $1$ and then decreases monotonously with more layers $d$, where the minimal value is given by its Clifford averaged value~\eqref{eq:OTOC_avg_sup}. 
 For $\text{otoc}_{8}(U,\sigma^x_1,\sigma^z_N)$, we find a fast increase for $d\ge N$ when information has propagated across $N$ qubits. The $d$ needed to converge to the Clifford average value depends on $N_{\text{T}}$. Thus, determining whether an OTOC as converged requires knowledge of its stabilizer entropy. We find that for intermediate $N_{\text{T}}$ the OTOC requires significantly larger values to converge, while for small $N_\text{T}$ and very large $N_\text{T}\ge32$ the convergence requires  significantly lower $d$.
 
 We note that both OTOCs take much higher $d$ to converge compared to flatness $\mathcal{F}$. This indicates that although OTOC and flatness of the participation ratio relate to scrambling, it appears that OTOCs are more sensitive as a measure of scrambling.

Our fixed depth Clifford + T circuits can be implemented directly in experiment. This can be realized with current noisy quantum computers~\cite{arute2019quantum}, and for a low number of qubits even with recently developed fault-tolerant quantum computers~\cite{bluvstein2023logical}.

\subsection{Evolution with GUE Hamiltonian}

\begin{figure*}[htbp]
	\centering	
 \subfigimg[width=0.3\textwidth]{a}{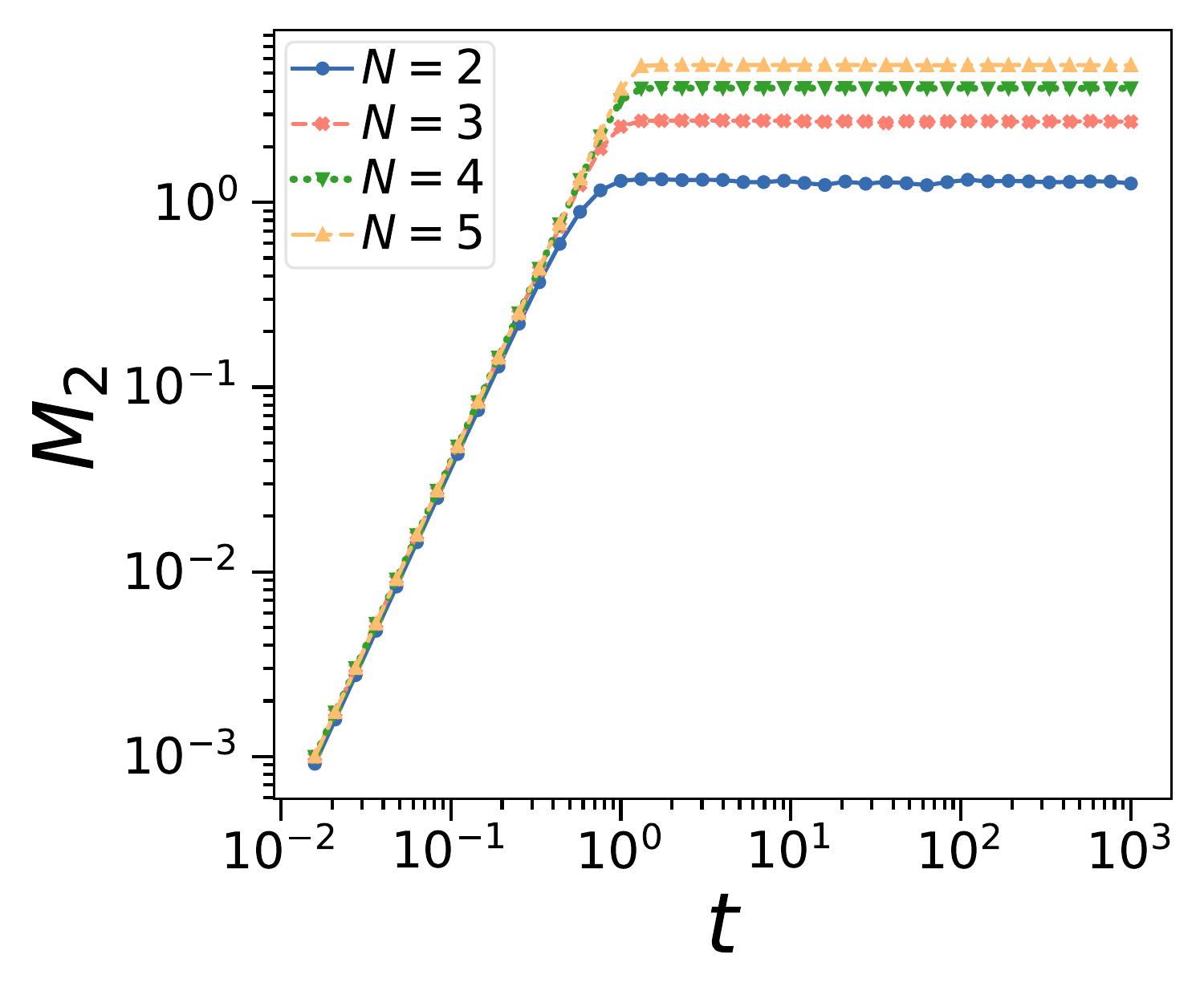}
  \subfigimg[width=0.3\textwidth]{b}{flatnessConvScrN5d1000m4c100r10000z0_2W5p51.pdf}
\subfigimg[width=0.3\textwidth]{c}{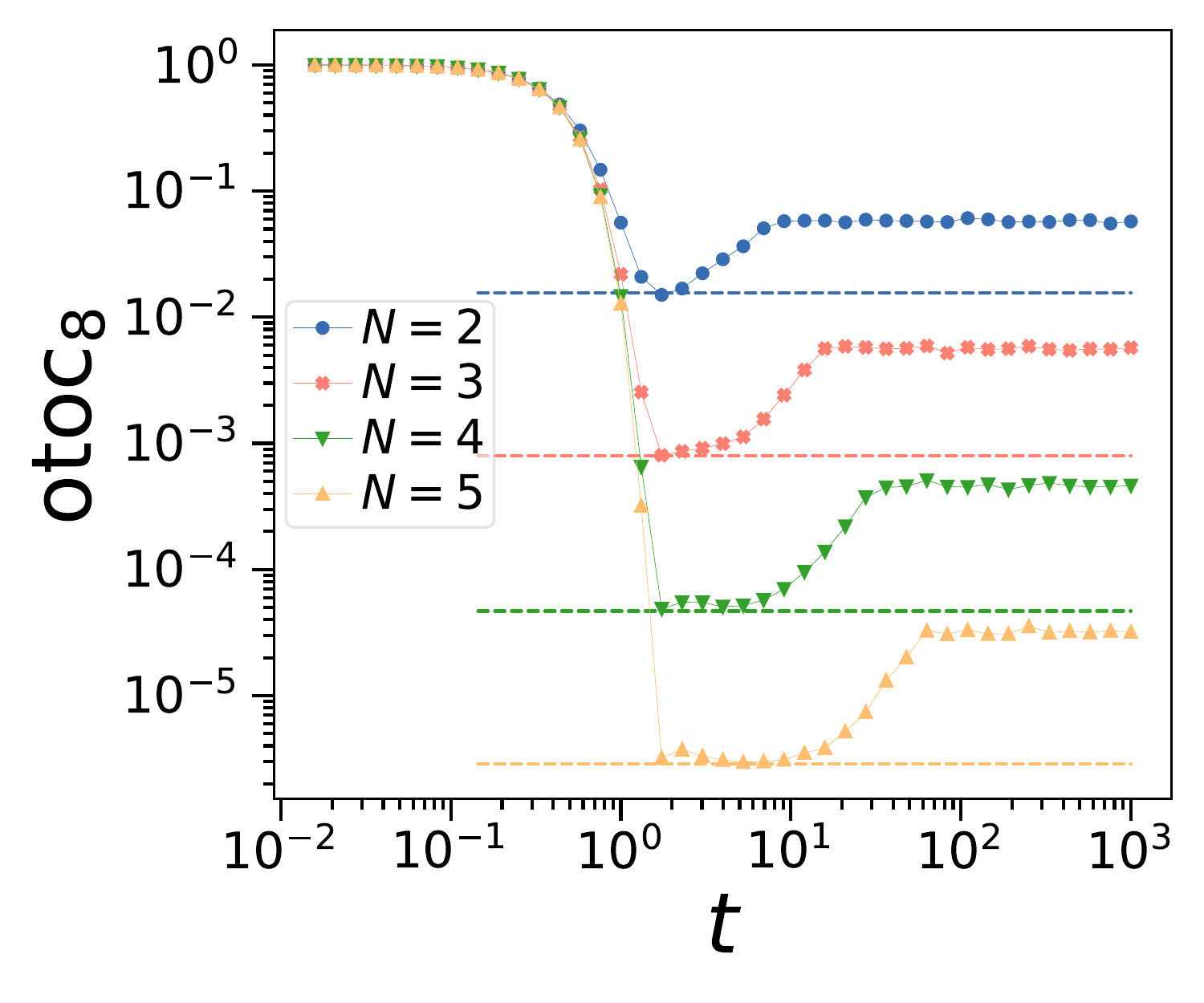}
\subfigimg[width=0.3\textwidth]{d}{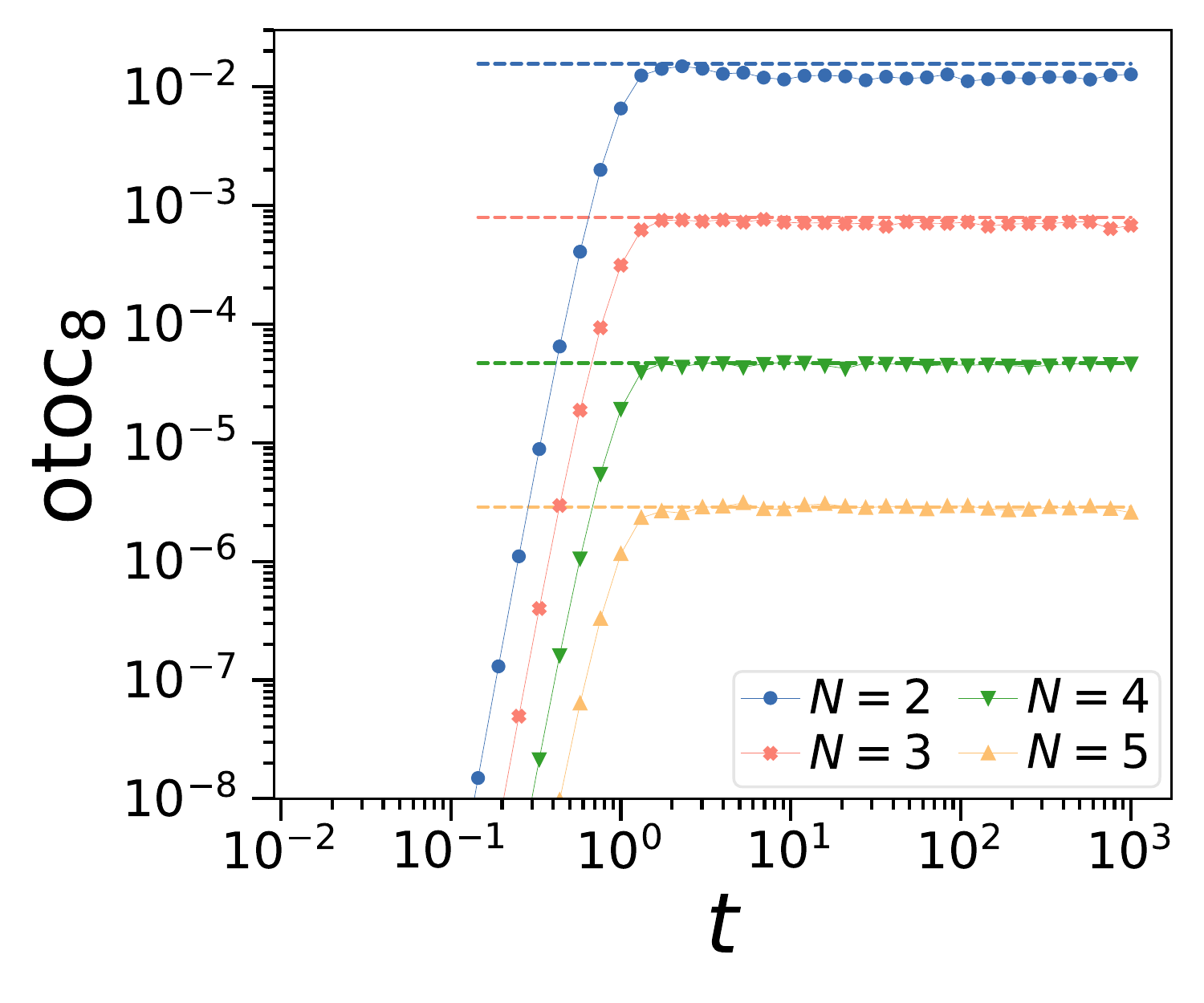}
	\caption{  2-R\'enyi SE $M_2(\ket{U})$, the 8-point OTOC $\text{otoc}_{8}(U,\sigma^x_1,\sigma^x_1)$ and the multifractal flatness $\mathcal{F}(U\ket{0})$ for evolution with random Hamiltonians $U(t)=\exp(-iH_\text{GUE}t)$ (\eqref{eq:GUE}). We show data as function of time $t$ for different number of qubits $N$.
    \idg{a} $M_2$ against $t$.
    \idg{b} $\mathcal{F}$ against $t$.
 Dashed line is Clifford averaged multifractal flatness~\eqref{eq:flatnessCliff} computed using $M_2(t\gg1)$.
 \idg{c} $\text{otoc}_{8}(U,\sigma^x_1,\sigma^x_1)$ against $t$, where dashed line is the Clifford-averaged OTOC given by~\eqref{eq:OTOC_avg_sup} computed using $M_2(t\gg1)$.
  \idg{d} $\text{otoc}_{8}(U,\sigma^x_1,\sigma^z_N)$ against $t$.
 All results are averaged over $10000$ random instances.
	}
	\label{fig:GUE}
\end{figure*}

Next, in Fig.~\ref{fig:GUE} we study the evolution 
\begin{equation}\label{eq:GUE}
    U(t)=\exp(-iH_\text{GUE}t)
\end{equation} 
with Hamiltonians $H_\text{GUE}$ sampled randomly from the Gaussian Unitary ensemble (GUE). We normalize the Hamiltonian with a prefactor $2^{-N/2}$ to restrict the eigenvalues of the Hamiltonian between $[-2,2]$ to get dynamics that is independent of $N$~\cite{cotler2017chaos}. 
We observe that SE in Fig.~\ref{fig:GUE}a increases until converging to its maximum around $t\sim \sqrt{N}$. 
However, the dynamics of $\mathcal{F}$ and $\text{otoc}_{8}$ shows more structure. 
In Fig.~\ref{fig:GUE}b, an initial increase with $t$ finds a characteristic peak around $t\sim1$. This is followed by a sudden dip to a low value. This dip value is given by the Clifford-averaged multifractal flatness. After longer times $\mathcal{F}$ increases again, converging to a larger value above the dip value. 
We observe similar dip behavior also for the $\text{otoc}_{8}(U,\sigma^x_1,\sigma^x_1)$ in Fig.~\ref{fig:GUE}c, reaching a minimal value which is given by the Clifford-averaged OTOC. At larger times the OTOC ramps again and converges to a larger value. Curiously, the duration of the dip is shorter for the OTOC compared to $\mathcal{F}$, again indicating that OTOCs are more sensitive to the degree of scrambling.
In Fig.~\ref{fig:GUE}d we regard $\text{otoc}_{8}(U,\sigma^x_1,\sigma^z_N)$ the OTOC of observables at different points in space. Here, we find that the dip-ramp behavior is not clearly visible.

For GUE evolution, the dip behavior has been first noted in Ref.~\cite{cotler2017chaos} for the frame potential and $2$ and $4$ point OTOCs. In Ref.~\cite{cotler2017chaos} it has been shown that at the dip the ensemble of GUE evolved Hamiltonians matches a $k$-design. Here, $k$-design refers that the ensemble of evolution unitaries is indistinguishable up to the $k$th moment to Haar random unitaries. However, at longer times it is not a $k$-design anymore. In particular, there is a characteristic ramp time where the OTOC increases at which the evolution stops being a $k$-design. This has been attributed due to the conservation of energy in the GUE evolution, where at exponential times the energy  gap between different eigenvalues can be resolved, leading to a divergence from being a $k$ design.

Here, we find that the dip and ramp behavior is also visible in multifractal flatness and in particular $8$-point OTOCs, where the dip value can be computed from the SE using our Clifford-averaged formulas for $\text{otoc}_{4n}$ and $\mathcal{F}$. 
Curiously, the dip and ramp behavior are not clearly visible in SE and non-local OTOCs, while the multifractal flatness and same-site OTOCs can resolve it clearly. This highlights the usefulness of multifractal flatness and same-site OTOCs as witness for $k$-designs. 
Both multifractal flatness and 8-point OTOCs can be experimental measured, allowing one to compute both the evolution and their expected converged value in experiment.

While the assumed value of the flatness and OTOC are exponentially small, we note that for small number of qubits $N$ these values still can be resolved in experiment. We highlight that these features are visible even for $N=2$ qubits, making experimental demonstration straightforward with available noisy quantum computers and simulators.

\subsection{Evolution with Hamiltonians of random Paulis}
\begin{figure*}[htbp]
	\centering	
 \subfigimg[width=0.3\textwidth]{a}{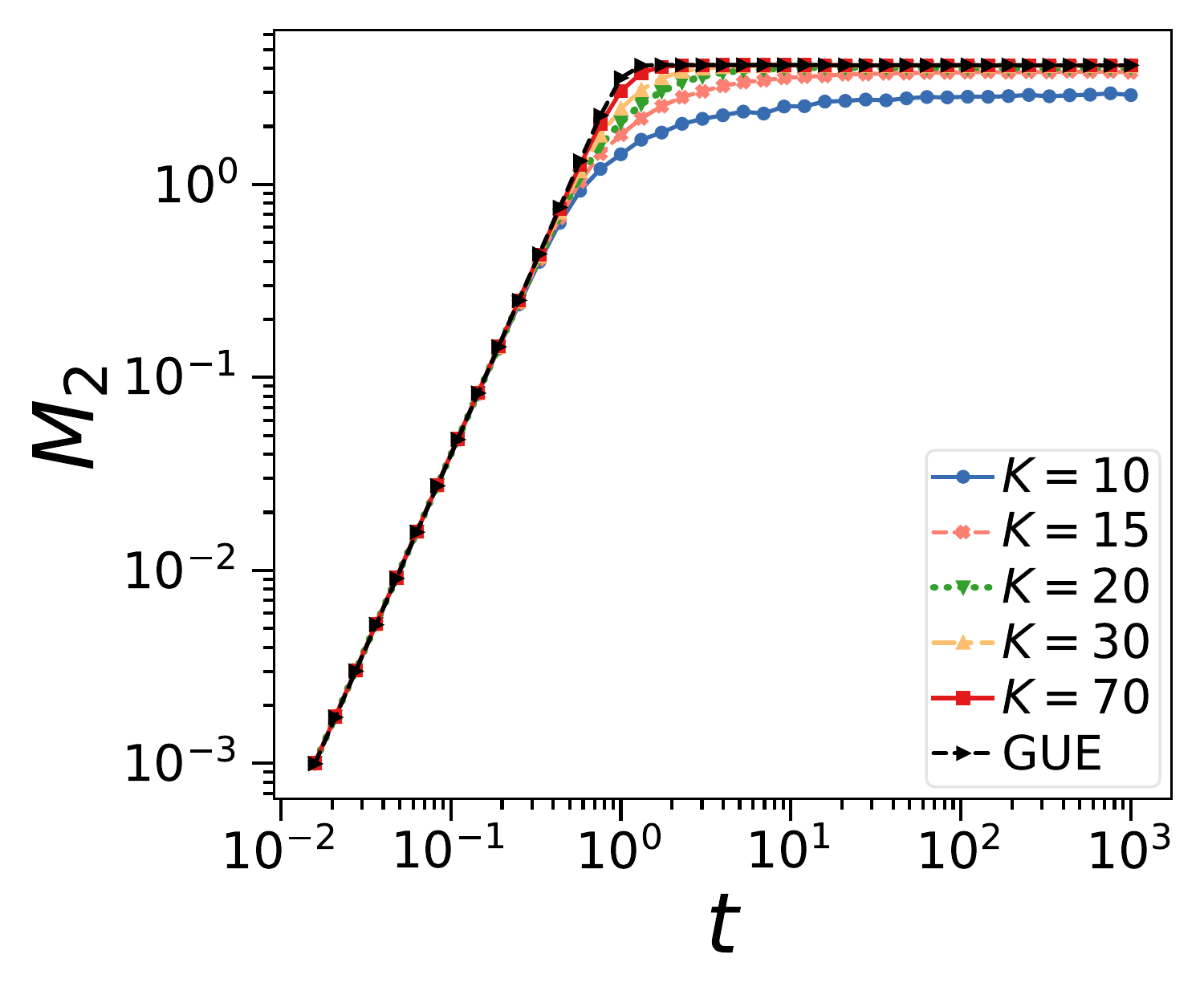}
  \subfigimg[width=0.3\textwidth]{b}{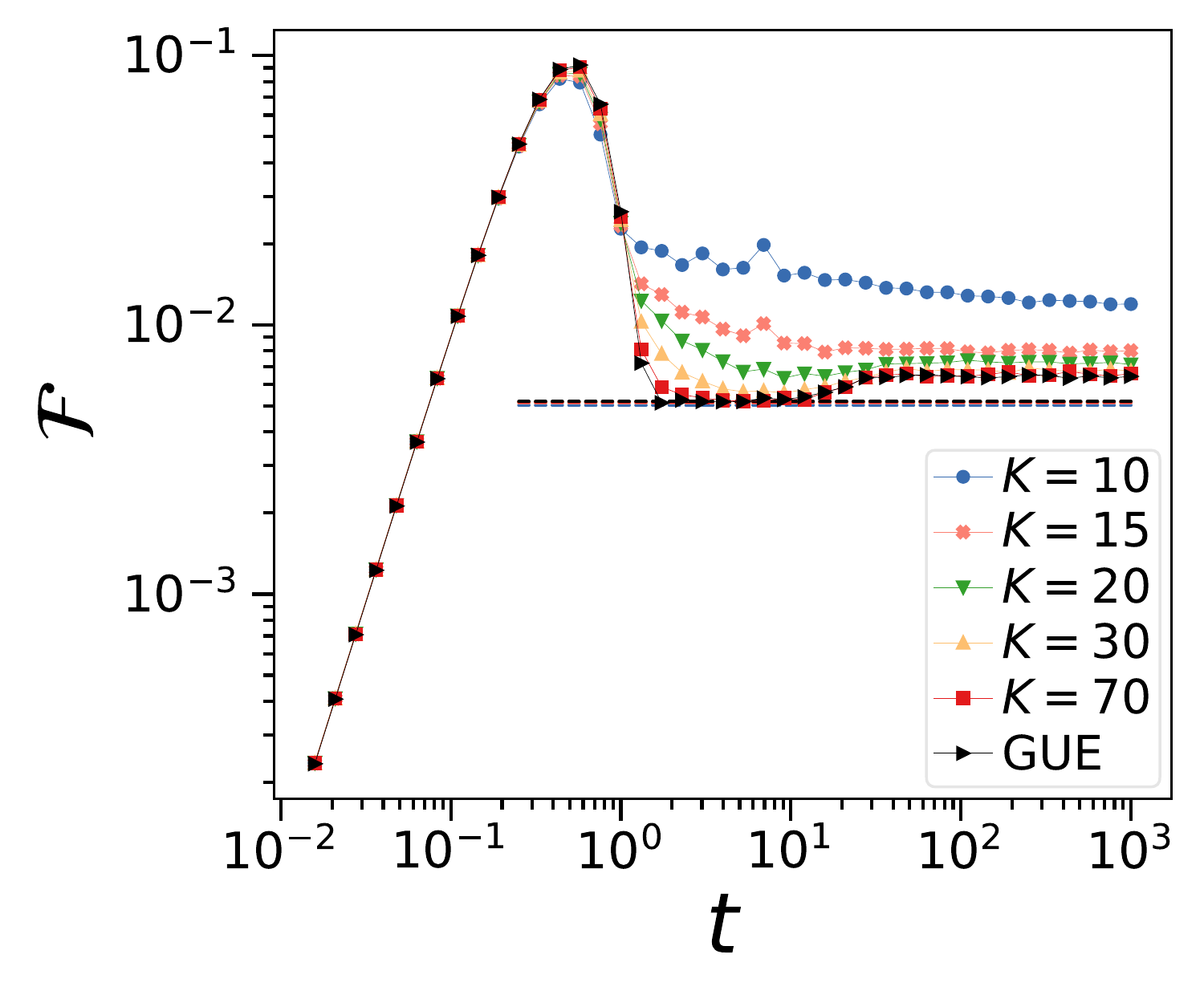}
\subfigimg[width=0.3\textwidth]{c}{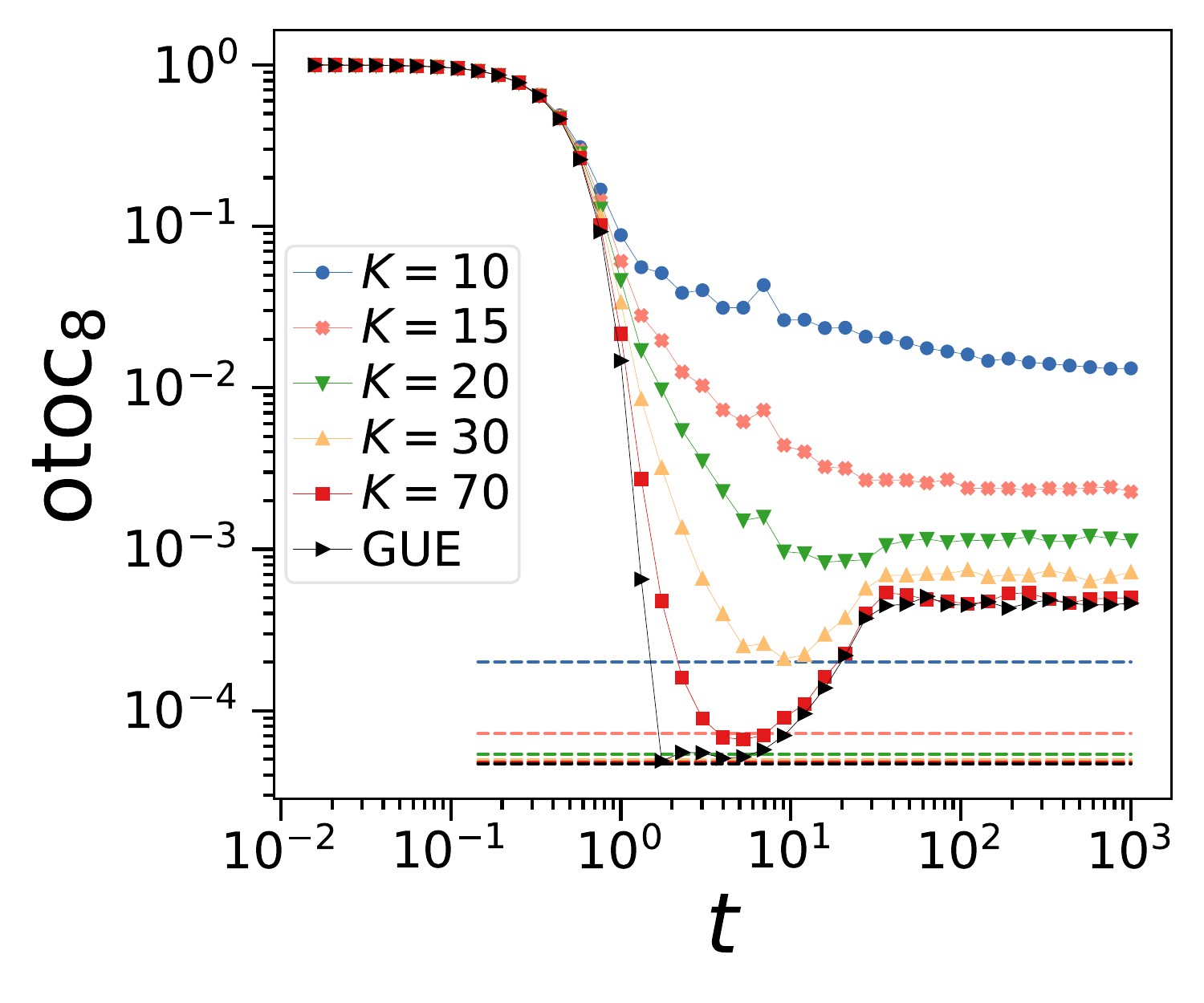}
\subfigimg[width=0.3\textwidth]{d}{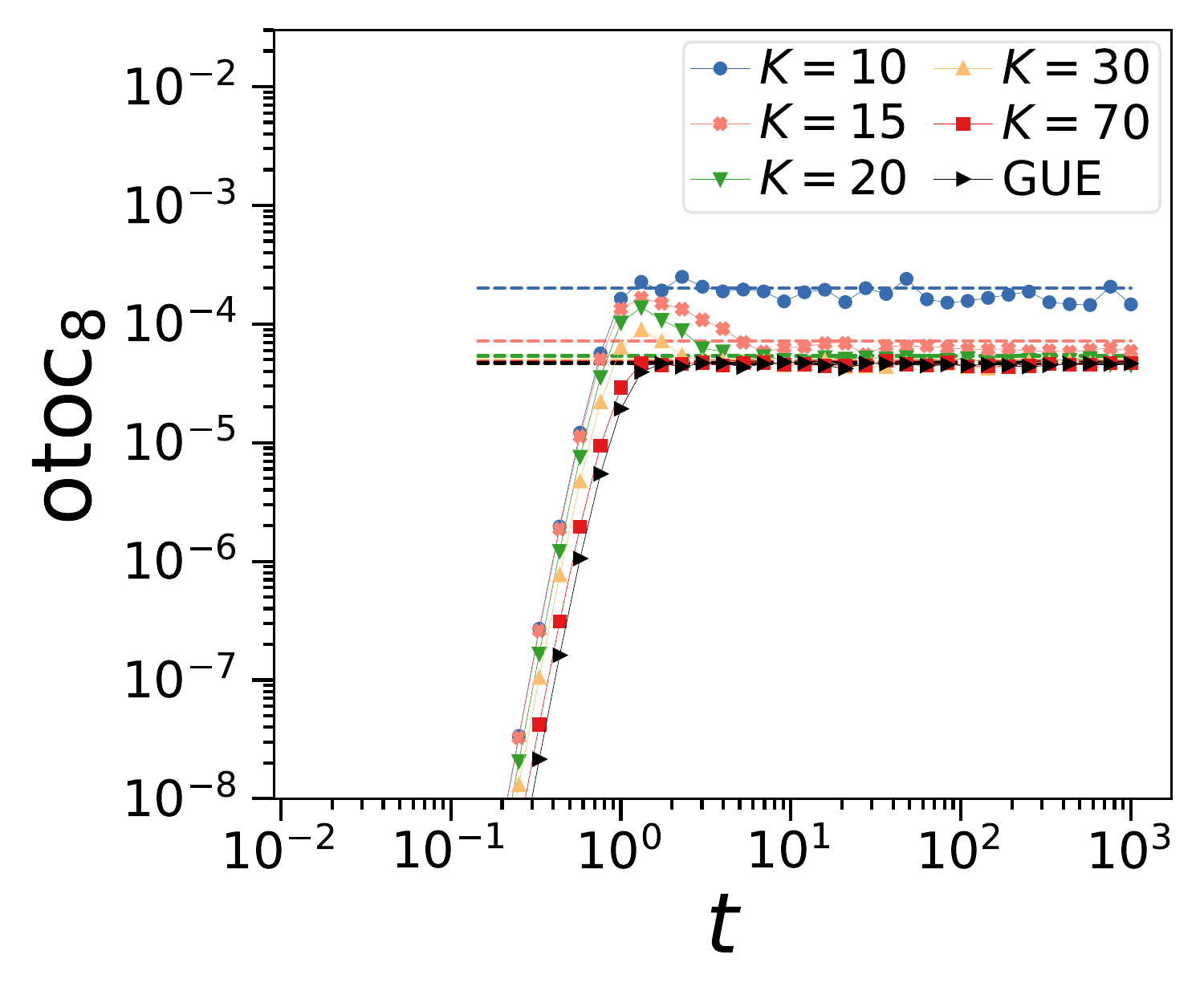}
	\caption{ Evolution $U(t)=\exp(-iHt)$ with Hamiltonian composed of $K$ random Pauli strings~\eqref{eq:randomHamiltonian}. We show 2-R\'enyi SE $M_2(\ket{U})$, the 8-point OTOC $\text{otoc}_{8}(U,\sigma^x_1,\sigma^x_1)$ and the multifractal flatness $\mathcal{F}(U\ket{0})$. We show data as function of time $t$ for different number of Pauli strings $K$, where we show the GUE evolution in black as reference
    \idg{a} $M_2$ against $t$.
    \idg{b} $\mathcal{F}$ against $t$.
 Dashed line is Clifford averaged multifractal flatness~\eqref{eq:flatnessCliff} computed using $M_2(t\gg1)$.
 \idg{c} $\text{otoc}_{8}(U,\sigma^x_1,\sigma^x_1)$ against $t$, where dashed line is the Clifford-averaged OTOC given by~\eqref{eq:OTOC_avg_sup} computed using $M_2(t\gg1)$.
  \idg{d} $\text{otoc}_{8}(U,\sigma^x_1,\sigma^z_N)$ against $t$.
 We show $N=4$ qubits averaged over $10000$ random instances.
	}
	\label{fig:randomHamilton}
\end{figure*}

We now study in Fig.~\ref{fig:randomHamilton} the evolution with a Hamiltonian composed of $K$ randomly chosen Pauli strings
\begin{equation}\label{eq:randomHamiltonian}
    H=\sum_{k=1}^K g_k  \sigma_k
\end{equation}
with $g_k$ chosen randomly from the normal distribution and random Pauli string $\sigma_k$. Here, we normalize $H$ as $H\rightarrow H/\sqrt{\text{tr}(H^2)/K}$. This normalization matches our normalization of the GUE Hamiltonian for $K$ large.

In Fig.~\ref{fig:randomHamilton}a, we study $M_2$ as function of evolution time $t$. $M_2$ quickly increases and converges to the maximum, where the behavior has little dependence on $K$.

In Fig.~\ref{fig:randomHamilton}b, we show the multifractal flatness. We find an initial increase in time which does not depend on $K$. At around $t\sim1$, there is a sudden dip, which for $K\ge30$ reaches the Clifford-averaged $\mathcal{F}$. At longer times $t$, $\mathcal{F}$ ramps up again for larger $K$, converging to a value larger than the Clifford-average. For $K\sim 70$, we dynamics matches the GUE nearly exactly.

In Fig.~\ref{fig:randomHamilton}c, we study the local $\text{otoc}_{8}(U,\sigma^x_1,\sigma^x_1)$. For smaller $K<20$, we find a montonous decrease. For larger $K$, there is a a characteristic dip behavior towards the Clifford-averaged OTOC, followed by a ramp which converges to a constant value. 

In Fig.~\ref{fig:randomHamilton}d, we study a nonlocal OTOC. We observe an initial increase that quickly converges to the Clifford-average value. Surprisingly, we find this convergence for any $K$, in contrast to the same-site OTOC. Thus, it seems that same-site OTOCs are more sensitive to convergence to a Clifford scrambled system.

While the GUE Hamiltonian is hard to implement, our random Pauli model has a straightforward implementation on quantum computers. 
In particular, the evolution can be realized by grouping the Pauli strings $\sigma_k$ of the Hamiltonian into commuting sub-Hamiltonians $H_j$, with $H=\sum_j H_j$. The evolution can now be implemented via trotterization by decomposing the evolution into timesteps $t=L\Delta t$ with $U(t)=\exp(-iHt)\approx\prod_{\ell=1}^L (\prod_{j} \exp(-i H_j \Delta t))$.

\subsection{Evolution with Ising Hamiltonian}
As reference, we also study the evolution with a one-dimensional Heisenberg Hamiltonian
\begin{equation}
    H_\text{ising}=\sum_{k=1}^N (\sigma_k^x\sigma_{k+1}^x+\sigma_k^y\sigma_{k+1}^y)+\Delta\sigma_k^z\sigma_{k+1}^z + h_k \sigma^z_k\,,
\end{equation}
where $\Delta$ is the interaction strength and random disorder $h_k\in[-W,W]$. 
This model is chaotic for small $W$, while it features a many-body localization transition for larger $W$~\cite{vznidarivc2008many}. Note that for the small system size considered here the MBL transition is obscured by finite-size effects.

\begin{figure*}[htbp]
	\centering	
 \subfigimg[width=0.3\textwidth]{a}{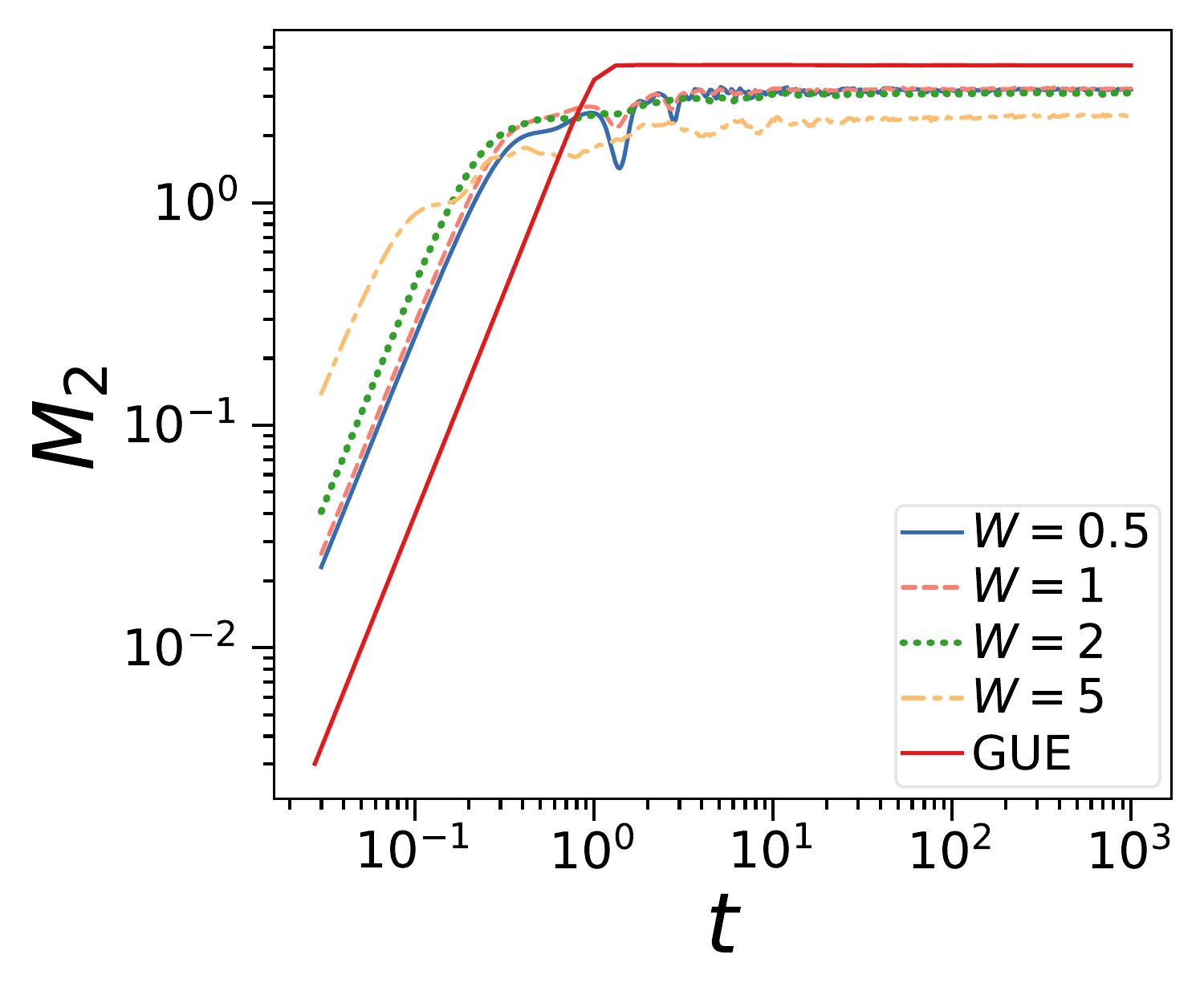}
  \subfigimg[width=0.3\textwidth]{b}{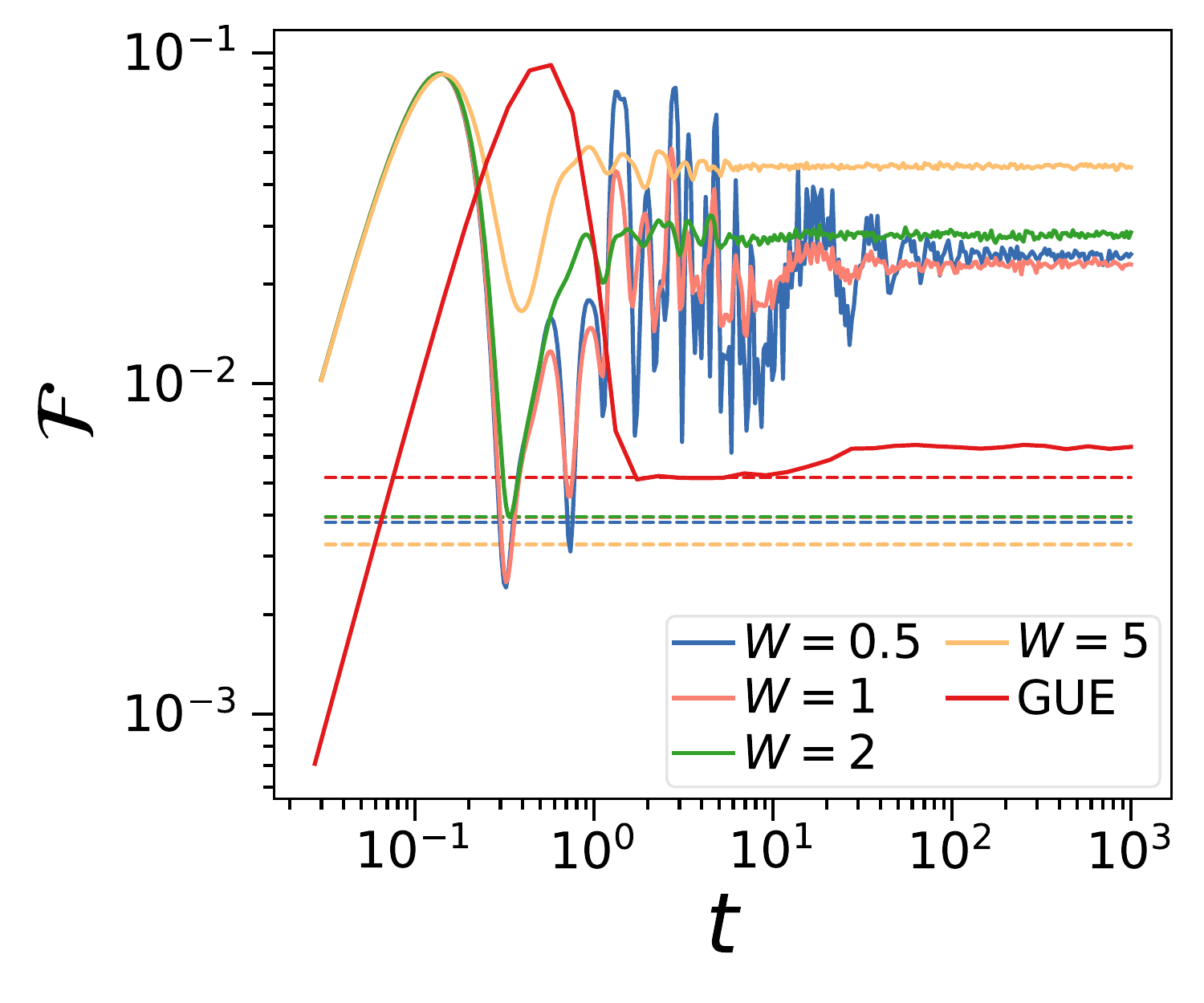}
\subfigimg[width=0.3\textwidth]{c}{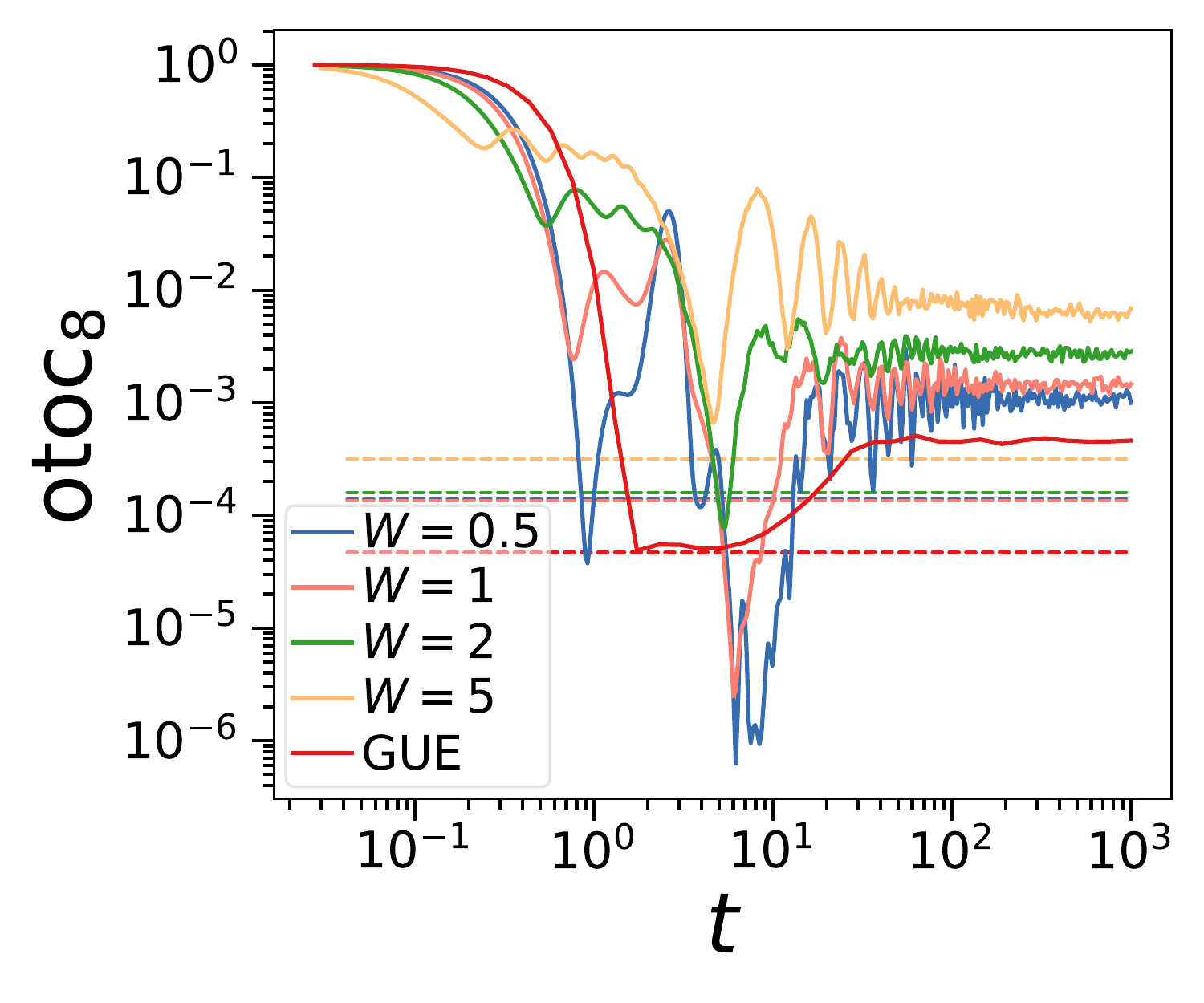}
\subfigimg[width=0.3\textwidth]{d}{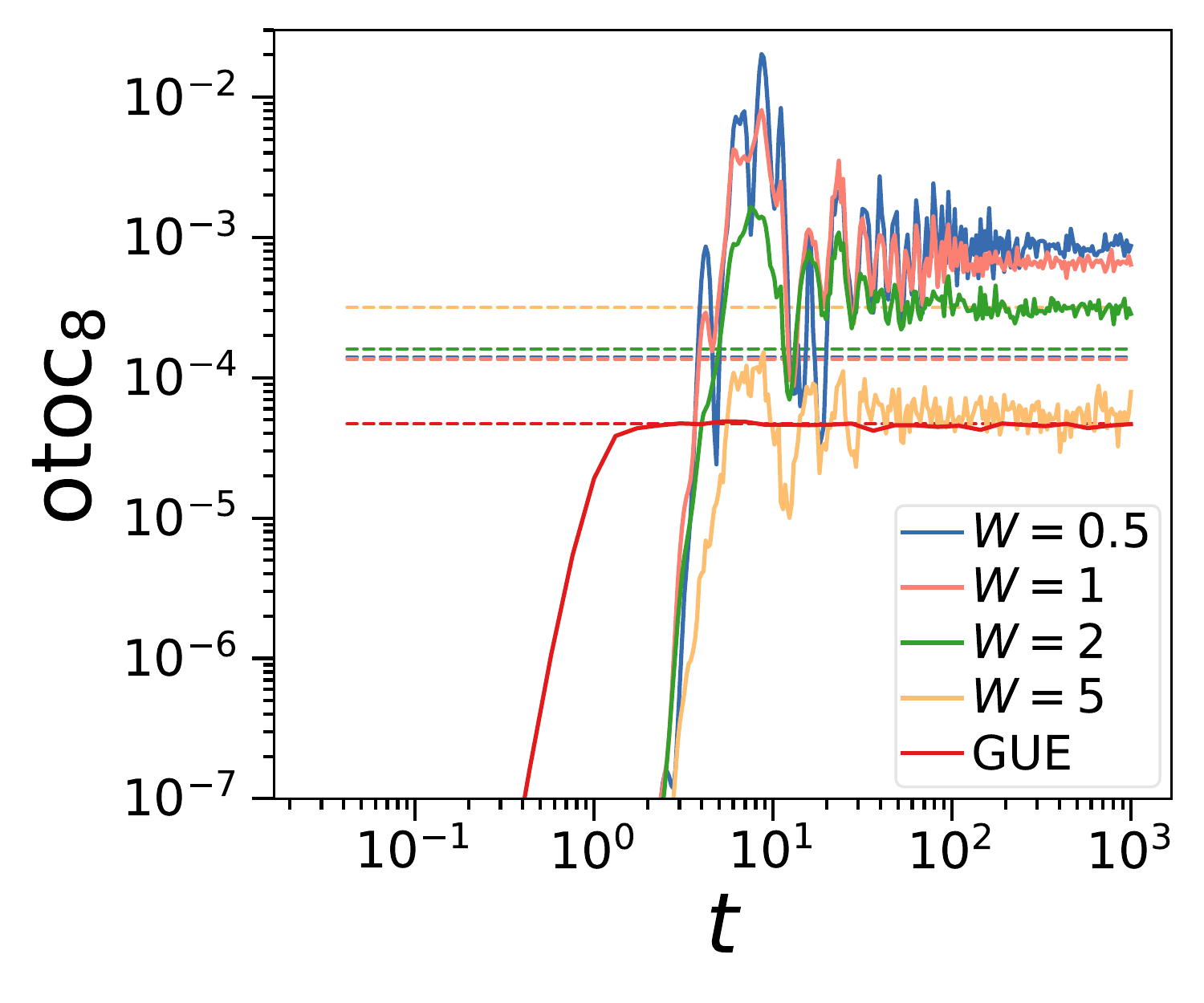}
	\caption{ Evolution $U(t)=\exp(-iH_\text{ising}t)$ with Ising Hamiltonian with $\Delta=0.2$. We show 2-R\'enyi SE $M_2(\ket{U})$, the 8-point OTOC and the multifractal flatness $\mathcal{F}(U\ket{0})$. We show data as function of time $t$ for different disorder strength $W$.
    \idg{a} $M_2$ against $t$.
    \idg{b} $\mathcal{F}$ against $t$.
 Dashed line is Clifford averaged multifractal flatness~\eqref{eq:flatnessCliff} computed using $M_2(t\gg1)$.
 \idg{c} $\text{otoc}_{8}(U,\sigma^x_1,\sigma^x_1)$ against $t$, where dashed line is the Clifford-averaged OTOC given by~\eqref{eq:OTOC_avg_sup} computed using $M_2(t\gg1)$.
  \idg{d} $\text{otoc}_{8}(U,\sigma^x_1,\sigma^x_N)$ against $t$.
 We show $N=4$ qubits averaged over $2000$ random instances.
	}
	\label{fig:MBL}
\end{figure*}

We show in Fig.~\ref{fig:MBL}a $M_2$. Notably, GUE evolution shows the same scaling for small times, however converged to a larger SE.
In Fig.~\ref{fig:MBL}b, we show the multifractal flatness against time $t$. The Ising evolution exhibits oscillations which decay over long times $t$. 
We find that for short times, the initial increase, peak and following decrease of $\mathcal{F}$ are universal features of both Heisenberg evolution and GUE evolution. However, the dip assumes in general not the Clifford-averaged multifractal flatness, and converges to much larger $\mathcal{F}$ at long times.
In Fig.~\ref{fig:MBL}c,d we study OTOCs as function of $t$. We note the pronounced oscillations, which are absent in the GUE.
While the Heisenberg model is ergodic for small $W$, it still shows widely different behavior from the GUE evolution, and does not converge to its Clifford-average values.

\subsection{Measuring multifractal flatness and OTOCs}
Here, we discuss the measurement of multifractal flatness and OTOCs in experiment.

Multifractal flatness can be directly computed from the inverse participation ratio $\mathcal{I}_q(\ket{\psi})=\sum_{k} \vert\braket{k\vert \psi}\vert^{2q}$ for $q=2$ and $q=3$. $\mathcal{I}_q$ can be efficiently measured by sampling the state in the computational basis, and counting whether $q$ sampled bitstrings being correlated.
This can be seen by using the replica trick to get 
\begin{equation}
    \mathcal{I}_q(\ket{\psi})=\sum_k \text{tr}(\Pi_q \ket{\psi\bra{\psi}^{\otimes q}})
\end{equation}
with the projector $\Pi_q= \sum_k(\ket{k}\bra{k})^{\otimes q}$. $\Pi_q$ has eigenvalues $\{0,1\}$. Thus, Hoeffding's inequality shows that one can measure the expectation value of this projector using $O(q \epsilon^{-2})$ samples, where $\epsilon$ is the additive precision. While the projector involves $q$ copies, it is sufficient to prepare and measure single copies of the state on a $N$-qubit quantum computer, and compute the expectation value via post-processing of outcomes. In particular, to compute one simply draws $q$ sampled bitstrings, and check whether they are identical (corresponding to eigenvalue 1) or not (eigenvalue 0).
Note the case $q=2$ corresponds to coherence of pure states which is detailed in Ref.~\cite{haug2023pseudorandom}. 
We conclude that the measurement setup for multifractal flatness is simple and amenable to current quantum computers.

Measuring OTOCs requires in general access to the inverse of the unitary which has been implemented in multiple experimental platforms successfully~\cite{garttner2017measuring,mi2021information}. Note that experimental protocols without inverse exist~\cite{blocher2022measuring}. However, in general the protocols to measure OTOC require deeper circuits than measuring the multifractal flatness due to the need of forward and backward evolution.

\revA{
\section{Bell measurements, stabilizer entropy and Bell magic}\label{sec:Bell}

Bell measurements are a powerful measurement scheme to estimate non-linear properties of quantum states. One takes two copies $\ket{\psi}\ket{\psi}$ and transforms them into the Bell basis via $U_\text{Bell}$ as described in the main text. Sampling in the Bell basis gives an $2N$ bit outcome $\boldsymbol{r}$ with a probability~\cite{montanaro2017learning}
\begin{equation}\label{eq:probBell}
P(\boldsymbol{r})=2^{-N}\vert\bra{\psi}\sigma_{\boldsymbol{r}}\ket{\psi^*}\vert^2\,,
\end{equation} 
where $\sigma_{\boldsymbol{r}}$ is a Pauli string. 
Bell measurements can be used to measure the purity, fidelity (as they realize a destructive SWAP test~\cite{garcia2013swap}), entanglement~\cite{garcia2013swap}, coherence, imaginarity~\cite{haug2023pseudorandom} and estimate the absolute values of any Pauli expectation value as a type of shadow tomography~\cite{huang2021demonstrating}. Further, Bell measurements can be used to learn stabilizer states~\cite{montanaro2017learning}. All these properties can be extracted at the same time by using different post-processing steps. 

For example to measure purity $\text{tr}(\rho^2)$, one performs a logical AND on outcome $\boldsymbol{r}$ between each qubit pair corresponding to the first and second copy, then compute its parity~\cite{garcia2013swap}.

For learning stabilizers, the core routine is Bell difference sampling~\cite{gross2021schur}. Here, one performs two Bell measurements with outcomes $\boldsymbol{r}$, $\boldsymbol{r}'$ and takes their difference $\boldsymbol{q}=\boldsymbol{r}-\boldsymbol{r}'$. 

The probability of outcome $\boldsymbol{q}$ for Bell difference sampling is given by~\cite{gross2021schur}
\begin{align*}
Q(\boldsymbol{q})=&\sum_{\boldsymbol{r}\in\{0,1\}^{2N}}P(\boldsymbol{r})P(\boldsymbol{r}\oplus\boldsymbol{q})=4^{-N}\sum_{\boldsymbol{r}\in\{0,1\}^{2N}}\vert\bra{\psi}\sigma_{\boldsymbol{r}}\ket{\psi^\ast}\vert^2\vert\bra{\psi}\sigma_{\boldsymbol{r}\oplus\boldsymbol{q}}\ket{\psi^\ast}\vert^2\\
=&4^{-N}\sum_{\boldsymbol{r}\in\{0,1\}^{2N}}\vert\bra{\psi}\sigma_{\boldsymbol{r}}\ket{\psi}\vert^2\vert\bra{\psi}\sigma_{\boldsymbol{r}\oplus\boldsymbol{q}}\ket{\psi}\vert^2=4^{-N}\sum_{\boldsymbol{r}\in\{0,1\}^{2N}}\vert\bra{\psi}\sigma_{\boldsymbol{r}}\ket{\psi}\vert^2\vert\bra{\psi}\sigma_{\boldsymbol{r}}\sigma_{\boldsymbol{q}}\ket{\psi}\vert^2\,.
\end{align*}
The last step follows from the commutation rules of Pauli operators which gives $\vert\bra{\psi}\sigma_{\boldsymbol{r}\oplus\boldsymbol{q}}\ket{\psi}\vert=\vert\bra{\psi}\sigma_{\boldsymbol{r}}\sigma_{\boldsymbol{q}}\ket{\psi}\vert$, while the derivation for the second last step can be found in~\cite{gross2021schur,haug2022scalable}.

When $\ket{\psi}$ is a stabilizer state, then it is described by a commuting group of Pauli $G$ with expectation value $\vert\bra{\psi}\sigma\ket{\psi}\vert^2=1$, $\forall \sigma \in G$. In the commuting group $\sigma,\sigma'\in G$, the commutator is zero with $[\sigma,\sigma']=0$. For all other Pauli strings $\sigma\notin G$ we have $\vert\bra{\psi}\sigma\ket{\psi}\vert^2=0$. 
Thus, only when $\ket{\psi}$ is a stabilizer state, one can show that $Q(\boldsymbol{q})=2^{-N}\vert\bra{\psi}\sigma_{\boldsymbol{q}}\ket{\psi}\vert^2$. Or in other words, for stabilizer states, Bell difference sampling samples Pauli strings $\sigma_{\boldsymbol{q}}$ according to the Pauli spectrum $\Xi(\sigma_{\boldsymbol{q}})=2^{-N}\vert\bra{\psi}\sigma_{\boldsymbol{q}}\ket{\psi}\vert^2$ and all sampled Pauli strings must commute.

This commuting property is used in Bell magic $\mathcal{B}$ to generate a measure of nonstabilizerness~\cite{haug2022scalable} 
\begin{equation}\label{eq:BellMagic}
\mathcal{B}(\ket{\psi})=\sum_{\boldsymbol{r},\boldsymbol{q}\in\{0,1\}^{2N}}Q(\boldsymbol{r})Q(\boldsymbol{q})\left\lVert[\sigma_{\boldsymbol{r}},\sigma_{\boldsymbol{q}}]\right\rVert_{\infty}
\end{equation}
where the infinity norm is zero $\lVert[\sigma_{\boldsymbol{r}},\sigma_{\boldsymbol{q}}]\rVert_{\infty}=0$ when the two Pauli strings commute $[\sigma_{\boldsymbol{r}}$, $\sigma_{\boldsymbol{q}}]=0$, and $\lVert[\sigma_{\boldsymbol{r}},\sigma_{\boldsymbol{q}}]\rVert_{\infty}=2$ otherwise. Correspondingly, the additive Bell magic is defined as $\mathcal{B}_\text{a}=-\log_2(1-\mathcal{B})$.  Due to the commuting property, only for pure stabilizer states $\ket{\psi_\text{C}}$ we have $\mathcal{B}(\ket{\psi_\text{C}})=0$, else it is greater zero. One can also check that it is invariant under Clifford unitaries. We also note that the complexity of the error mitigation scheme for Bell magic scales as $1/(1-p)^8$  where $p$ is the depolarization probability. In contrast, for stabilizer entropy it scales as $1/(1-p)^{2n}$, e.g. for $n=3$ we have $1/(1-p)^{6}$.

The measurement protocols for stabilizer entropy uses a different method to measure nonstabilizerness. In particular, one has to measure the operator $\Gamma^{\otimes N}=\sum_{\sigma\in\mathcal{P}}\sigma^{\otimes 2n}$ which measures the $n$-th moment of the Pauli spectrum. This operator is diagonalized by Bell transformations as shown in SM~\ref{sec:parity}. By appropriately post-processing of Algorithm 1 in the main text one can then measure
\begin{equation}
    A_n=2^{-N}\sum_{\sigma \in \mathcal{P}} \braket{\psi|\sigma|\psi}^{2n}=\bra{\psi}^{\otimes 2n}\Gamma_{n}^{\otimes N} \ket{\psi}^{\otimes 2n}\,.
\end{equation} 
Note that the fact that Pauli strings of stabilizer states commute is not used at all, in contrast to Bell magic. Instead, stabilizer entropy is a measure of the moment of the Pauli spectrum determine nonstabilizerness. In particular, stabilizer states have a peaked Pauli distribution with either $\vert\bra{\psi}\sigma\ket{\psi}\vert^2=0$ or  $\vert\bra{\psi}\sigma\ket{\psi}\vert^2=1$. In contrast, highly magical states have a broad  distribution $\vert\bra{\psi}\sigma\ket{\psi}\vert^2\approx4^{-N}$.

Our Algorithm 2 uses yet another approach to measure the stabilizer entropy. It samples directly from the Pauli spectrum (via the complex conjugate $\ket{\psi^\ast}$ in a Monte-Carlo fashion and then averages over the sampled Pauli strings via 
\begin{equation}
A_n=\underset{\sigma\sim \Xi(\sigma)}{\mathbb{E}}[\bra{\psi}\sigma\ket{\psi}^{2n-2}]\,.
\end{equation}
}

\end{document}